\documentclass[useAMS,usenatbib]{mn2e}
\usepackage{graphicx}
\usepackage{latexsym}
\usepackage{natbib}
\usepackage{amssymb}
\usepackage{amsmath}
\usepackage{lscape}

\title{Exploring an Alternative Channel of Evolution Towards SNa Ia Explosion }

\author[E. Chiosi,  C. Chiosi,  P. Trevisan, L. Piovan, \& M. Orio]{E. Chiosi$^{1}$,  C. Chiosi$^{1,2}$, P. Trevisan$^{2}$, L. Piovan$^{2}$, \& M. Orio$^{1,3}$\\
 $^{1}$ Astronomical Observatory of Padova, INAF, Vicolo dell'Osservatorio 5, 35122 Padova, Italy\\
 $^{2}$ Department of Physics and Astronomy, University of Padova,
       Vicolo dell'Osservatorio 2, 35122 Padova, Italy\\
 $^{3}$ Department of Astronomy, University of Wisconsin Madison,
       475 N. Charter Str., Madison WI 53706, USA\\
E-mail: {\tt emanuela.chiosi@oapd.inaf.it (EC); cesare.chiosi@unipd.it (CC); patrizia.trevisan@unipd.it (PT);}\\
E-mail: {\tt lorenzo.piovan@gmail.com (LP); marina.orio@oapd.inaf.it (MO) }
 }

\date{\tt Submitted: September 2014; Accepted: **** }

\pubyear{2014}

\begin{document}
\maketitle
\title{Exploring an Alternative Channel of Evolution Towards SNa Ia Explosion}

\begin{abstract}

In this paper
we explore the possibility that isolated CO-WDs  with mass smaller than  the Chandrasekhar limit may undergo nuclear runaway and SNa explosion, triggered by the energy produced by under-barrier pycno-nuclear reactions between carbon and light elements. Such reactions would be due to left over impurities of the light elements, which   would remain inactive until the WDs transit from  the liquid to the solid state. We devise  a simple formulation for the coulombian potential and the local density in a ionic lattice affected by impurities and introduce it in the known rates of pycno-nuclear reactions  for  multi-component plasmas. Our semi-analytical results indicate that the energy generated by these pycno-nuclear reactions  exceeds the WD luminosity and provides enough energy to elementary cells of matter to balance the energy cost for C-ignition at much younger ages than the age of the Universe, even for WDs with  masses as low as $\simeq 0.85\, M_\odot$. A thermonuclear runaway may thus be triggered in isolated WDs. The explosion would occur  from  few hundred thousand to a few million years after the WD formation in the mass interval    $0.85 - 1.2\, M_\odot$.

\end{abstract}


\section{Introduction}\label{intro}

Carbon-Oxygen White Dwarfs (CO-WDs) originate from  low and intermediate mass progenitors in the mass interval $0.7-0.8 M_\odot$ to $6 M_\odot$ \citep{weidemann1967,weidemann1977,weidemann1990,ChiBerBre1992,weidemann2000}, which after the central H- and He-burning phases develop a highly electron degenerate CO core (electrons are fully degenerate and nearly relativistic). After the thermally pulsing AGB phase, these stars  eject the whole envelope, baring the CO core. A WD is thus formed by a CO core surrounded by thin (if any) layers of lighter elements.

Given any reasonable initial mass function, about 97\% of the stars  of any generation  with lifetime shorter than the age of the Universe ($13.7\pm 0.2$\,Gyr according to \citet{spergel2003}) end  as WDs.
\citet{kepler2007} examined a sample of 7755 DA-WDs  from the SDSS catalog and concluded that
the vast majority of the 616 DBs and DOs listed in \citet{eisenstein2006}, fall in the mass interval  $0.5-0.7 M_\odot$.  The  peak value is  $0.6 M_\odot$ and  the range of masses extends  from about 0.4$M_\odot$ toward the low mass end  to $1.2-1.3 M_\odot$ on the opposite side.
According to theoretical models, there also is  a clear relationship between the progenitor and WD masses, $M_{i}$ vs $M_{WD}$, \citep[][]{weidemann1967,weidemann1977,weidemann1990,ChiBerBre1992,weidemann2000,marigo2001,catalan2008} for which
there is a plausible explanation based on current theory of stellar structure and evolution (see below for more details).

The most important facts of the WD theory are: (i) Except for a thin surface layer, the equation of state (EOS) can be approximated as fully degenerate electrons ($P\equiv P_e$) with kinematic conditions changing from  non relativistic ($P_e= K_{5/3} \rho^{5/3}$)  to fully relativistic ($P_e= K_{4/3} \rho^{4/3}$)  with increasing central and mean density. (ii) The structure of a WD of given chemical composition (mean molecular weight of ions  and/or electrons, $\mu_i$ and $\mu_e$ respectively) is fully determined by its central density. (iii) Assuming Newtonian hydrostatic equilibrium, the WD mass has a maximum value, called the Chandrasekhar mass $M_{\mathrm{Ch}}$,  for which the central density  is infinite so that the EOS is fully relativistic.  In this case,  $M_{\mathrm{Ch}}= 5.85 / \mu_e^2$ with $\mu_e$ the molecular weight of electrons.  For a typical CO-WD, $\mu_e \simeq 2$ so   $M_{\mathrm{Ch}} \simeq 1.46 M_\odot$. However,  the radius of  the Chandrasekhar mass is zero. Clearly such non physical situation means that no WD can be born with the   Chandrasekhar mass, which therefore  is a mere ideal value. (iv) Along the sequence toward the  Chandrasekhar value at increasing central density, two important physical processes can intervene, overall instability driven by General Relativity (GR) effects, and pycno-nuclear burning.   WDs more massive than about $1.3\,M_\odot$, i.e. denser than about $\rm 3\times 10^9\, g\, cm^{-3}$, become dynamically  unstable because of GR \citep{shapiro1983};
C-O WDs denser  than about $6\rm \times 10^9\, g\, cm^{-3}$, may start pycno-nuclear burning  during the liquid-solid regimes \citep{shapiro1983}.  The onset of these phenomena  may lead to  a thermo-nuclear runaway. Therefore, all stable WDs we observe must have formed with masses lower than $1.2 - 1.3\, M_\odot$.

The current theory of type Ia SNe  assumes that the Chandrasekhar mass can be reached by accretion or merging and although the details are not fully known, carbon is ignited via the pycno-nuclear channel. This is followed by C-detonation or
C-deflagration (the latter is more likely) and by a thermal runaway, because  the
gravo-thermal specific heat of the star is positive. The liberated nuclear energy exceeds  the gravitational binding
energy and the star is thorn apart   \citep[see the recent review by][and references therein]{Nomoto2013a}.
Binary  models for type Ia SNe are classified as  \emph{double-degenerate}, i.e.  the merger of two gravitationally bounded WDs, and \emph{single-degenerate}, i.e. the evolution to the explosive phases is due to the
accreting material from a companion star  \citep[see e.g.,][]{Napi03,Trimble04,Orio2013}.

Owing to the important role of the pycno-nuclear reactions in destabilizing a WD close to the Chandrasekhar mass and triggering  type Ia SNa explosions, it is worth examining in some detail the condition under which pycno-nuclear reactions can occur.
The pycno-nuclear regime starts in  very dense and cool environments, i.e. in the liquid/solid state. While both the central and mean densities are determined by the WD mass and remain nearly constant if the mass does not change, the temperature decreases because the WD is radiating energy from the surface. Because there are no nuclear sources and the electrons are highly degenerate, while the WD radiates,  the  ions must cool down and  transit from gaseous to liquid, and eventually solid (crystallised) conditions, i.e.  carbon and oxygen ions  form a lattice which is  pervaded by a gas of electrons. As the temperature decreases, the ions eventually reach the fundamental energy state  under a coulombian potential that approximately has the form of a harmonic oscillator. In  quantum physics, the energy of   the fundamental state of a harmonic oscillator is $1/2 \hbar \omega$ ($\omega$ is the plasma frequency). This  means that even at extremely low temperatures there is a finite probability for the C and O ions to penetrate the repulsive coulombian barrier.  In this scheme, the  rates of pycno-nuclear reactions between C and/or O ions  were first calculated by \citet{SalpeterVanHorn69} and then refined over the years.

In their pioneering study,  \citet{SalpeterVanHorn69} first noticed that  impurities may enhance the pycno-nuclear reactions among the nuclei of the lattice. The enhancement is due to local over-densities in the sites of the impurities.  Furthermore, even the contaminant nuclei themselves may  react with the lattice nuclei,  thus contributing to the total energy generation.  If this happens,    WDs that are less dense and/or less massive  than the  limits  we discussed above become unstable to the ignition of pycno-nuclear reactions triggered  by  impurities.  Since at  the distance scales corresponding to the densities of old WDs,  electric forces dominate the scene, impurities due to  light elements, such as hydrogen or helium, may induce higher  local over-densities so that light elements can more easily cross  the coulombian barriers.

In order to explore this idea, first we   evaluate the change induced by the light element contamination on the typical inter-ion  distance  $R_o$.   Second, we  present the reaction rates in the pycno-nuclear channel for reactions like H+C, He+C, etc.
Finally,  we explore the possibility that even isolated WDs  with mass significantly smaller than  the Chandrasekhar limit and relatively low density (i.e. in the ranges  $0.85 - 1.2\ M_\odot$  and $\rm  10^7 - 10^8\,  g\, cm^{-3}$, respectively) may undergo nuclear runaway, because of the energy produced by under-barrier nuclear reactions by contaminant elements, when the WDs  reach the liquid/solid state.

The paper is organized as follows. In Section \ref{general} we first  shortly review the  history of a WD progenitor and the cooling and crystallization processes of ions in the WD,  and  introduce the relationships between the  progenitor mass and the WD mass and between the WD mass and its central density. In Section \ref{nuclear} we describe the fundamentals of nuclear burnings in WDs, summarize two current sources for the reaction rates in both  the thermal and pycno-nuclear regimes for single and multi-component fluids, and present a preliminary comparison of the reaction rates.    In Section \ref{impurities} we  evaluate the changes in the rates due to  impurities, first in the transmission probability of the coulombian  potential barrier penetration and then in the local density. Since impurities by light elements are more efficient than those by heavy elements, in Section \ref{evaluation} we estimate the abundances of hydrogen and helium left over by previous evolutionary phases. After an initial phase at the beginning of the cooling sequence in which part of the energy may be of nuclear origin, for a long time the only source of energy is the thermal energy of the ions and all nuclear sources are turned off.  Therefore the light  elements remain inactive for a long time until the WD reaches the conditions for the activation of the  pycno-nuclear regime. In Section \ref{results_Kita_Yako}  we estimate and compare the energy production by reactions like $\rm ^{1}H+^{12}C$ and $\rm ^{4}He+^{12}C$ showing that they can produce the typical luminosity of an old WD in the pycno-nuclear stage.  In Section \ref{evolution}, we follow the evolution of  WDs of different mass along their cooling sequences  towards the pycno-nuclear regime and explore the effect of different abundances of light elements (H in particular) on the energy release by nuclear reactions. We estimate the abundance of contaminant elements and the epoch at which the nuclear energy generation first equals and then exceeds  the WD luminosity. In addition to this we formulate a new condition for C-burning ignition in a liquid,  semi-solid medium and propose a two-steps mechanism (named \textit{"the fuse  C-ignition"}) for igniting carbon  in elementary volumes defined by the mean free path of thermal conduction.  For plausible values of H and He abundances, WDs with masses as low as $\simeq 0.85 - 0.9\, M_\odot$ (and above) reach the critical condition for rekindling the nuclear energy production and initiate a thermal runaway. Finally, in Section \ref{conclusions} we draw some conclusions about the possible implications of these results for the progenitors of  type Ia SNe.

\section{General properties of WDs}\label{general}
\subsection{Evolution of the progenitors}

The structure and evolution of low and intermediate mass stars,  the progenitors of CO-WDs, can be described with the aid of three milestone masses \citep[for all details see][]{IbenRenzini1983,kippenhahn1990,ChiBerBre1992}.  We define low mass stars those  that develop an electron degenerate helium core, shortly after leaving the main sequence toward the red giant branch (RGB). When the mass  of the He-core has grown to a critical value (0.45-0.50 $M_{\odot}$, the precise value depends  on the composition, star mass, and input physics),  a He-burning  runaway (called He flash) starts  in the core and continues until electron degeneracy is removed.  Then nuclear burning proceeds quietly. The maximum initial mass of the star  for this to occur is  $M_{HeF}$.  Stars more massive than $M_{HeF}$ are classified  as intermediate-mass or massive depending on the physics of carbon ignition in the core.  After core He-exhaustion,  intermediate mass stars develop a highly degenerate CO core, and undergo helium shell  flashes or thermal pulses as asymptotic giant branch (AGB) stars. The AGB phase is terminated either  by
envelope  ejection  and formation of a CO-WD (with initial mass $M_i$ in the range $M_{HeF} \leq M_{i} \leq M_{w}$)  or by carbon ignition and deflagration in a highly degenerate core, once it has grown  to  the  Chandrasekhar  limit of 1.4 $M_{\odot}$.

The  limit  mass $M_{w}$  is regulated by the efficiency of mass loss by stellar wind during the  RGB and AGB phases.  The minimum mass of the CO  core, below  which carbon ignition in non degenerate condition fails  and  the above scheme holds, is 1.06 $M_{\odot}$. The initial mass to reach a core of 1.06 $M_\odot$ is called  $M_{up}$.
The exact value of $M_{HeF}$,  $M_{up}$, and $M_w$ depends on many details of stellar physics.
$M_w$ is mainly controlled by mass loss during the AGB phase and is about 6 $M_\odot$. The values of $M_{HeF}$ and $M_{up}$ are  $\simeq 1.8-2.2 M_\odot$ and  $\simeq 7-9 M_\odot$, respectively, in absence of convective overshooting. These ranges  become $M_{HeF} \simeq 1.7- 1.8M_\odot$ and  $M_{up}\simeq 6 M_\odot$ when convective overshooting
is included \citep{ChiBerBre1992}. Since the ignition mass for a fully degenerate fully relativistic CO core is 1.46 $M_\odot$, the possibility that C-ignition may occur in single CO-WDs is definitely ruled out.

\subsection{Physical state of WD interiors} \label{phys_pro}
\textbf{Generalities}.
The interiors of CO-WDs are made of ions of C, O,  traces of other elements, and free electrons. Ions are fully ionized and electrons form a uniform background. In other words, there is a multi-component mixture (customarily named multi-component plasma, MCP)   of ion species $i=1,2,...$ with mass numbers $A_i$, atomic number $Z_i$, and number densities $n_i$. The total number density is $n=\sum_i n_i$. For one component plasma (OCP) the suffix $i$ is omitted.
A two components medium is defined a binary ionic medium or BIM.

The number density is related to the mass density $\rho$ of the matter by
\begin{equation}\label{density}
    n_i = {X_i \rho \over A_i m_u}
\end{equation}
where $X_i$ is the mass fraction or abundance of ions $i$, and $m_u$ the atomic mass unit ($m_u=1.660 \times 10^{-24} g$). If the density is not very high,  the total mass fraction contained in the nuclei is $X_N=\sum_i X_i  \approx 1$, whereas for a density higher than $\sim 4 \times 10^{11} \, {\rm g\, cm^{-3}}$ above which  neutrons may be free,  $X_N < 1$. Introducing the  fractional number $x_i=n_i/n$, with $\sum_i x_i = 1$, we use two groups of  useful relationships:

\begin{equation}\label{mass_atomic}
    \langle Z \rangle = \sum_i x_i Z_i, \qquad \qquad \langle A \rangle = \sum_i x_i A_i
\end{equation}
where $\langle Z \rangle$ $\langle A \rangle$ are the mean atomic and mass  number of ions and
\begin{equation}\label{ne_rho}
    n_e=n \langle Z \rangle,\qquad \rho = {m_u n \langle A \rangle \over X_N}, \qquad x_i= {X_i / A_i \over \sum_j X_j/A_j}\, .
\end{equation}

The physical state is best described by the Coulomb coupling parameter $\Gamma_i$ for ions $i$:

\begin{equation}
 \Gamma_i = \frac{(Z_i e)^2}{a_i k_B T} = \frac{Z_i^2 e^{5/3} }{a_e k_B T}
\end{equation}
\begin{equation}
a_e   = \left[\frac {3} {4 \pi n_e}\right]^{1/3},  \qquad a_i = Z_i^{1/3} a_e,\qquad
            a_i= \left[\frac {3} {4 \pi n_i}\right]^{1/3}
\end{equation}
where $T$ is the temperature, $k_B$ the Boltzmann constant, $a_e$ the electron-sphere radius, and $a_i$ the ion -sphere radius (a radius of a sphere around a given ion, where the electron and ion charge balance each other). The coupling parameter
$\Gamma_i$ is  the ratio  between the Coulomb energy, $E_C= \frac{(Z_i e)^2}{a_i}$,  to the thermal energy $E_{th}=k_BT$ of the ions.
If $\Gamma_i << 1$  the ions constitute an almost ideal Boltzmann gas. If $\Gamma_i \geq 1$ the ions  are strongly coupled by the Coulomb forces and constitute  a Coulomb liquid or a solid. The transition  gas to  liquid occurs smoothly at $\Gamma_i \sim 1$ with no phase transition. According to  \citet{lindemann1910} and
recent accurate Monte Carlo simulations by \citet{DeWitt2001}, a classical OCP of ions solidifies at $\Gamma_i \simeq 175$ via a  weak second-order phase transition.

In most cases, BIMs or MCPs are supposed to represent the composition of  WDs, therefore it may be useful to introduce the mean ion Coulombian parameter $\langle \Gamma \rangle = \sum_j x_j \Gamma_j$. Strongly coupling occurs if
$\langle \Gamma \rangle \geq 1$, causing a transition from plasma to liquid. The temperature $T^L$ at which this occurs is given by
\begin{equation}\label{trans_liquid}
    k_B T^L = \sum_j \left[{ Z_j^2 e^2 \over a_j} \right] x_j
    \equiv k_B T \langle \Gamma \rangle \, .
\end{equation}
\noindent
With decreasing temperature, the ion motions  can no longer be considered as classical,  but must be quantized.  The nuclei form a Debye plasma with temperature $T^P$ associated with the ion plasma frequency $\omega^P$

\begin{equation}\label{debye}
    T^P = { \hbar \omega^P \over k_B}, \qquad \quad
       \omega^P = \sqrt{ \sum_j{ {4 \pi Z_j^2 e^2 n_j \over A_j m_u } } } \, .
\end{equation}
This is the critical stage at which the specific heat of the material is determined by nuclei oscillations  of frequency $\omega_P$ rather than by free thermal motions.

\noindent
Increasing $\langle \Gamma \rangle$ further, by either lowering the temperature or increasing the density or both, the matter crystallizes into a rigid  Coulomb lattice. The solidification (or melting) temperature  $T^M$ is given by

\begin{equation}\label{melting}
T^M = {1 \over k_B} \sum_j \left[{ Z_j^2 e^2 \over a_j} \right] x_j = {T^L \over  \langle \Gamma \rangle_M}
\end{equation}
where $\langle \Gamma \rangle_M$ = 175 \citep{DeWitt2001}.
At such high densities, even the small zero point oscillation  allow the neighbouring nuclear wave functions  to overlap, inducing  nuclear reactions that depend on density and not on temperature. This is the  pycno-nuclear regime.

\textbf{WD cooling}.
After a short lived initial phase, during  which the energy supply is sustained by some  nuclear burning in the two progressively extinguishing surface shells  and the large  energy losses by  neutrinos emission, the evolutionary rate of the WD is driven only by  the internal energy of the ions and electrons. Owing to the very different specific heat at constant volume of ions and electrons,  the ions are the dominating source. While the ions cool down,  the WD   undergoes several phase transitions. When the temperature is $\simeq 10^8 K$,  the WD is gaseous and the ions behave like a perfect Maxwell-Boltzmann gas. As the temperature decreases, first the ions become liquid and eventually they form a solid lattice. Because the  electrons are fully degenerate, they cannot cool down.

At this stage, in the energy conservation equation  ${L_r \over dM_r}=\epsilon_n + \epsilon_g + \epsilon_\nu$ (with obvious meaning of all the symbols)  $\epsilon_n$ and   $\epsilon_\nu$ can be neglected and the gravitational term
$\epsilon_g = -C_v \dot{T} + {T\over \rho^2}({\partial P \over \partial T})_v \dot{\rho}$ is approximated as $\epsilon_g \simeq -C_v \dot{T}$. Therefore the luminosity of a WD is given by

\begin{equation}\label{cool}
L= - \int_o^M C_v \dot T dM_r
\end{equation}
where $C_v$ and $\dot T$ are function of the position and time.

As the interior of a WD cools down, the ion specific heat $C_v^{ion}$  per gram  gradually changes from

$$C_v^{ion}\simeq \frac{3}{2} \frac{k_B}{A m_u}  \quad \qquad {\rm to} \qquad \quad C_v^{ion} \simeq {3}\frac{k_B}{A m_u} $$

\noindent (with the usual meaning of all the symbols) whereby the first relation is  for the hot gaseous phase and the second one is for the phase in which    the temperature has decreased but it is still far from the Debye value (whereby quantum effects become important).  The increase by a factor of 2 is due to increasing  correlations of the ion positions driven by the growing importance of the Coulomb interaction energies as compared to the thermal energies of the ions. In other words, the spatial scale of the coulombian interactions is comparable to the inter-ion spacing determined by the density. When the temperature is close to the Debye value, $C_v^{ion}$ decreases dramatically and becomes proportional to $T^3$.

As far as  electrons are concerned,  the specific heat of the electrons can be neglected with respect to that of ions and their contribution to the cooling rate of the WD can be ignored
\citep{kippenhahn1990}. To conclude,  as the ions are the main contributors to the luminosity of a WD,  the above relationships are used to calculate the cooling sequence of a WD (see Section \ref{existing_models} below).

\textbf{Liquefaction and crystallization.}\label{crystal}
The liquefaction and crystallization theory \citep{VanHorn68} predicts that in WDs ions start to liquefy and  freeze in an ordered cristalline lattice from the center to the outer layers when the temperature falls below $T^L$ and $T^M$.
The phase transition from an isotropic Coulomb liquid to a crystalline solid implies a discontinuity in the distribution of the plasma ions. Because symmetry  cannot be achieved  instantaneously, the transition is a first-order phase change. Therefore, for  $\langle\Gamma\rangle=\langle\Gamma\rangle_{\mathrm{m}}=172-175$
\citep[][]{kitamura2000} latent heat is released \citep{VanHorn68}.

For a ionic mixture with more than one species of ions, chemical separation  may occur, either at solidification or in the fluid phase. The mixture behaviour, in this case, derives from the peculiar shape of the state diagram. Because a phase separation,  with the companion stratification  of elements,  is a source of gravitational energy (caused by sinking  of the heavier ions), able to deeply modify the WD cooling time, it is of fundamental importance to obtain detailed phase diagrams for the BIM or MCP of interest. Moreover, for an accreting WD, chemical separation may produce  chemical stratification, thus affecting  the electron capture, opacity and fusion rates.

The most difficult problem with a strongly coupled MCP at low temperatures is  understanding its actual state. MonteCarlo simulations of the freezing of classical OCP by \citet{DeWitt1992} indicate that it freezes into imperfect body-centered cubic (BCC) or faced-centered cubic (FCC) micro-crystals. Unfortunately there are no reliable simulations of freezing for MCPs. Cold MCPs are much more complex  than OCPs; they can be  regular lattices with impurities or an amorphous uniformly mixed structure or a lattice of one phase with admixtures of other ions; or even a mosaic of phase separated regions. Fortunately, these extreme conditions are seldom reached.  In  a typical CO-WD (with $\rm X_C =X_O = 0.5$), the temperature and density plane is confined in the ranges
$7.5 \leq \log \rho \leq 10.5$ and $7.0 \leq  \log T \leq  9.5$. At $\rho > 4\times 10^{10}$ g cm$^{-3}$ the carbon nuclei cannot survive in dense matter because of beta captures. At $\rm \rho > 2 \times 10^{10} g cm^{-3}$, the oxygen nuclei are also destroyed by beta captures. In this plane, the loci of $T^L$, $T^P$ and $T^M$ are straight lines whose terminal points
[$\log \rho$, $\log T$] are: $T^L$[8.4, 9.5], $T^P$[10.5, 8.8], and $T^M$[10.5, 7.9]. There is some marginal effect that depends on the fractional abundances $x_j$ \citep[see][for more details]{yakovlev2006}.

\begin{table}
\footnotesize
\caption{Masses $M_G$(in units of $M_\odot$) and radii  R (in units of $R_\odot$) of CO WDs at varying the central density $\rho_c$ (in $\rm g\, cm^{-3}$). The mass is the gravitational mass; it  tends to the Chandrasekhar limit as the density goes to infinity.    }	
\begin{tabular}{|r l c|  r l c|}  
\hline
$\log\rho_c$   & $M_G$    &    $R$                 & $\log\rho_c$   & $M_G$   &   $R$     \\
    \hline
   5.29        & 0.20    &  2.10$\times10^{-2}$   &   8.29         & 1.24    & 5.33$\times10^{-3}$   \\
   5.57        & 0.27    &  1.87$\times10^{-2}$   &   8.57         & 1.30    & 4.54$\times10^{-3}$   \\
   5.71        & 0.31    &  1.77$\times10^{-2}$   &   8.71         & 1.32    & 4.19$\times10^{-3}$   \\
   5.85        & 0.35    &  1.67$\times10^{-2}$   &   8.85         & 1.34    & 3.85$\times10^{-3}$   \\
   6.00        & 0.40    &  1.57$\times10^{-2}$   &   9.00         & 1.36    & 3.54$\times10^{-3}$   \\
   \hline
   6.29        & 0.51    &  1.39$\times10^{-2}$   &   9.29         & 1.38    & 2.98$\times10^{-3}$   \\
   6.57        & 0.63    &  1.23$\times10^{-2}$   &   9.57         & 1.40    & 2.49$\times10^{-3}$   \\
   6.71        & 0.69    &  1.16$\times10^{-2}$   &   9.71         & 1.41    & 2.27$\times10^{-3}$   \\
   6.85        & 0.75    &  1.08$\times10^{-2}$   &   9.85         & 1.41    & 2.07$\times10^{-3}$   \\
   7.00        & 0.81    &  1.02$\times10^{-2}$   &  10.00         & 1.42    & 1.88$\times10^{-3}$   \\
   \hline
   7.29        & 0.93    &  8.89$\times10^{-3}$   &  10.29         & 1.42    & 1.55$\times10^{-3}$   \\
   7.57        & 1.04    &  7.73$\times10^{-3}$   &  10.57         & 1.42    & 1.28$\times10^{-3}$   \\
   7.71        & 1.09    &  7.20$\times10^{-3}$   &  10.71         & 1.42    & 1.15$\times10^{-3}$   \\
   7.85        & 1.14    &  6.69$\times10^{-3}$   &  10.85         & 1.42    & 1.04$\times10^{-3}$   \\
   8.00        & 1.18    &  6.21$\times10^{-3}$   &  11.00         & 1.42    & 9.44$\times10^{-4}$   \\
\hline
\end{tabular}
\label{tabrho_mass.tab}
\end{table}

\begin{table}
\footnotesize
\caption{Properties of the progenitor star at the start of the TP-AGB phase prior to the formation of the WD. $M_i$, $M_{co}$, $M_{WD}$ are in $M_\odot$, $\rho_c$ is in $\rm g\, cm^{-3}$, the age is in years.
$M_{WD} $ is derived from eqn. (\ref{mass_ini_mass_wd}). $\rho_c$ is interpolated from Table \ref{tabrho_mass.tab}.  }	
\begin{tabular}{|l  l l l  l|}
\hline
      &    Y=0.26 & Z=0.017  &          &            \\
$M_i$ & $M_{co}$ & $M_{WD}$ &$ \log\rho_c$ & $ Age (yr) $ \\
    \hline
6.0   & 0.956    & 1.14    &   7.95  & 7.881$\times 10^{ 7}$  \\
5.0   & 0.869    & 1.04    &   7.60  & 1.233$\times 10^{ 8}$  \\
4.0   & 0.802    & 0.87    &   7.15  & 2.216$\times 10^{ 8}$  \\
3.0   & 0.725    & 0.75    &   6.84  & 4.999$\times 10^{ 8}$  \\
2.0   & 0.625    & 0.64    &   6.56  & 1.464$\times 10^{ 9}$  \\
1.0   & 0.524    & 0.53    &   6.35  & 1.269$\times 10^{10}$  \\
0.8   & 0.517    & 0.52    &   6.25  & 2.899$\times 10^{10}$  \\
\hline
\end{tabular}
\label{mi_mco_mwd.tab}
\end{table}

\subsection{Current models of WDs}

Because in first approximation, the EOS of WDs does not  depend on the temperature but only on the density, the polytropic description can be used, i.e.  the mechanical and thermal structure of the WDs can be treated separately \citep{chandrasekhar1939}.  However, whenever the thermal history of a WD is required to estimate correctly the nuclear energy release or the  luminosity as a function of time,  or other details of the cooling sequence, complete models of WDs  are required. Both types of models  have been calculated by many authors and have been made available in the literature. To mention a few, recent state-of-the-art models of WDs have been calculated by \citet{althaus1997,althaus1998}, \citet{althaus2009,althaus2012,althaus2013}, \citet{millerbertolami2013}, \citet{panei2007}, \citet{renedo2010}, and \citet{salaris2010,salaris2013}. Such models  will be used in our analysis.

In concluding this section, we need to present and shortly discuss a few   relationships between important parameters of WDs such as  (i) the initial mass $M_i$ of the progenitor star, (ii) the  CO core mass, $M_{CO}$, at the beginning of the TP-AGB phase, (iii) the central density and the age of the progenitor at the end of the AGB phase, (iv) the mass $M_{WD}$ of the descendent WD (which  is also named the gravitational mass $M_G\equiv M_{WD}$),  and finally (v) the central density,  radius, and age.

A recent empirical estimate of the relationship between $M_{WD}$ and $M_i$ has been given  by \citet{catalan2008}
to whom we refer whenever necessary
\begin{eqnarray}
M_{WD} &=& (0.096 \pm 0.005)M_i + (0.429 \pm 0.015) \nonumber \\
       && \quad\quad\quad\quad\quad\quad\quad\quad\quad\quad {\rm for \quad M_i < 2.7 M_\odot}  \nonumber \\
M_{WD} &=& (0.137 \pm 0.007)M_i + (0.318 \pm 0.018) \nonumber \\
       && \quad\quad\quad\quad\quad\quad\quad\quad\quad\quad {\rm for \quad M_i > 2.7 M_\odot} \, .
\label{mass_ini_mass_wd}
\end{eqnarray}

The relationships between the central density $\rho_c$, the mass $M_G\equiv M_{WD}$, and the total radius  are derived from \citet{althaus1997,althaus1998} and are reported in Table \ref{tabrho_mass.tab}. The cooling sequences for the same WD masses are also from \citet{althaus1997,althaus1998}.   The mass $M_{CO}$ of the CO core at the beginning of the TP-AGB phase and the age of the progenitor  are   from the Padova Library of stellar models \citep{Bertelli2008,Bertelli2009} and are listed in Table \ref{mi_mco_mwd.tab}. Owing to the very short duration of the TP-AGB phase, the stellar age at the beginning of this phase nearly coincide with the age of the progenitor star when the CO core is exposed and the WD phase begins. We neglect here the effects  of the initial chemical composition on  $M_{CO}$,  $M_{WD}$  and   ages, and  focus on the case  [Y=0.26, Z=0.017] typical of a young population. If the chemical composition is taken into account, because of the different total lifetimes of stars of the same mass but different chemical composition and other details of stellar structure, at the low mass end of the WD mass distribution not all combinations of $M_i$ and $M_{WD}$  correspond to realistic cases. The lower  mass limit is determined  by the age of the Universe, i.e.  $13.7\pm 0.2$\,Gyr according to WMAP data \citep{spergel2003}.

Finally, we would like to remind that, owing to the different mass size of the external envelope surrounding the $M_{CO}$ core of TP-AGB stars, the core increases little  during this phase in low mass stars (up to  about $3\, M_\odot$) while it increases significantly in stars with mass in the range $3\, M_\odot$ to $6\, M_\odot$. Furthermore, the upper mass limit $M_{up}$ for AGB phase to occur (and to WDs to be generated)  depends on the initial chemical composition;  the detailed stellar models by \citet{Bertelli2008, Bertelli2009} show that it can vary from $5\, M_\odot$ to $6\, M_\odot$. Therefore the $M_{WD}$ vs $M_i$ relationship of Table \ref{mi_mco_mwd.tab} has to be considered as only indicative of the overall trend.

\section{Nuclear burning in WDs}\label{nuclear}

\subsection{Generalities}\label{generalities}

In  the first stages of the WD  life, nuclear burning can occur in two  shells, close to the surface  \citep[see ][and references therein]{renedo2010,Corsico2014}.
Hydrogen may still be burning  via the CNO-cycle, because a small amount of it (mass abundances in the range $\rm 10^{-4}\leq  X_H \leq10^{-5}$) is left on the  surface, and the temperature is still sufficiently high to sustain  nuclear burning.
Similarly, somewhat deeper inside there may be also  residual He-burning in a shell surrounding the inert CO-core.

In general, as the temperature decreases, all  thermal nuclear burnings are turned off  for billions of years until the WD has cooled down to very low temperature, so that the pycno-nuclear regime is reached. This is possible only when the internal energy of ions and the WD luminosity are very low, and  the baryonic matter (mostly C and O ions) is crystallized, usually  when the WDs are quite old (several Gyrs).
An exception to the above picture are the  results of \citet{millerbertolami2013} who have recently  shown  that  for  low-mass  WDs resulting from  low-metallicity progenitors, residual  H-burning
constitutes  the  main contributor  to  the  stellar luminosities  for
luminosities as low  as $\log L/L_\odot \simeq   -3.2$.

The pycno-nuclear reactions were studied long ago by \citet{SalpeterVanHorn69}.  These authors modeled the pycno-nuclear potential with a harmonic oscillator and discussed the possible influence of electron-screening and other effects. They also suggested  that impurities may  significantly increase the nuclear burning rates (see Section \ref{impurities}). This suggestion is  the starting point of this study.

Other  formulations of nuclear reactions in  the pycno-nuclear regime are by  \citet{schramm1990}, \citet{ogata1991}, \citet{ichimaru1992}, \citet{kitamura1995}, \citet{brown1997}, \citet{ichimaru1999}, \citet{kitamura2000}, \citet{gasques2005}, \citet{yakovlev2006},  \citet{Beard2010}, and references therein. All these authors  considered and derived the pycno-nuclear reaction rates for a lattice composed of  a single element \citep[the one-component plasma OCP of][]{gasques2005}, two elements \citep[the binary ionic medium BIM of][]{kitamura2000},  or a multi-component plasma (MCP) of \citep[][]{yakovlev2006}. They also   discussed  electron screening  and  phase transitions from gaseous to liquid to solid phases. However, they did not discuss the presence of impurities (see Section \ref{existing_models}).

\subsection{Formalism for reaction rates}
We are interested in nuclear fusion reactions
$$ (A_i, Z_i) + (A_j, Z_j) \rightarrow (A_c, Z_c) $$

\noindent
where $A_c=A_i+A_j$, $Z_c = Z_i + Z_j$ refer to the compound nucleus $c$. To study these reactions we must extend the formalism presented in Sect. \ref{phys_pro} and introduce the ion-sphere quantities

\begin{equation}\label{aij}
a_{ij}= {a_i + a_j \over 2}, \quad  \mu_{ij} = m_u{A_i A_j \over A_c}, \quad  \Gamma_{ij}= {Z_i Z_j e^2 \over a_{ij} k_B T }
\end{equation}
\begin{eqnarray}\label{tempij}
    T^{L}_{ij} &=& {Z_i Z_j e^2 \over a_{ij} k_B} \nonumber \\
    T^{P}_{ij} &=& {\hbar \over k_B} \omega_{ij}\qquad  \omega^P_{ij}=\sqrt{\frac{
          4\pi Z_i Z_j e^2 n_{ij}}{2 \mu_{ij}}}  \nonumber \\
     T^{M}_{ij}&=& \frac{Z_i Z_j e^2}{a_{ij} \Gamma^M_{ij}}
\end{eqnarray}

\noindent where $\mu_{ij}$ is the reduced mass, $a_{ij}$ characterizes the equilibrium distance between neighbouring nuclei (the corresponding number density is $n_{ij} = {3 / 4\pi a_{ij}^3}$), $\Gamma_{ij}$ describes their coulomb coupling, $T^L_{ij}$ is the temperature of the onset of strong coupling (or liquefaction temperature), $T^{P}_{ij}$ is the Debye temperature for the oscillations of ions $i$ and $j$, and $T^{M}_{ij}$ is the melting or solidification temperature\footnote{In the case of an OCP all the symbols reduce to those already defined in Sect. \ref{phys_pro}}.
Finally, we need the generalized Bohr radius

\begin{equation}\label{bohr}
    r_{Bij} = {\hbar^2 \over 2 \mu_{ij} Z_i Z_j e^2}
\end{equation}
which becomes the ion Bohr radius for equal ions $i=j$, and the parameter $\lambda_{ij}$,  corresponding to the parameter $\lambda$ introduced by \citet{SalpeterVanHorn69}

\begin{eqnarray}\label{lambdaij}
    \lambda_{ij} & = & r_{Bij} \left( { n_{ij} \over 2 } \right)^{1/3}  = { 2  r_{Bij} \over (Z_{i}^{1/3} + Z_{j}^{1/3}) } \left( {\rho X_N \langle Z \rangle \over 2 \langle A \rangle m_u } \right)^{1/3}  
\end{eqnarray}

\subsection{Five regimes of nuclear burning}\label{fiveregimes}
In WDs  the central density ranges from $10^6 \rm g\, cm^{-3}  $ to $10^{10} \rm g\, cm^{-3} $ depending on the initial mass of the star, and  the density decreases from the center to the surface, so we must consider  a wide range of densities. The temperature also spans a wide range, from $10^8 K$ at the beginning of the cooling sequence to typical values of about $10^5 K$ after a few  Gyrs. In the plane $T-\rho$, the WD undergoes two phase transitions: from gas to liquid and from liquid to solid (crystallization). In parallel to this,
there are five burning regimes \citep[see][]{SalpeterVanHorn69,yakovlev2006,Beard2010}: (i) the  classical thermo-nuclear;  (ii) the thermo-nuclear with strong electron screening; (iii) the thermo-pycno-nuclear; (iv) the thermally enhanced pycno-nuclear; (v) and eventually the zero temperature pycno-nuclear. For a complete description of the regimes see Beard (2010).  The five regimes have the following characteristics:

i) The classical thermo-nuclear one takes place when $\Gamma_{ij} <<1$.  The mean inter-nuclear distance is much greater than the typical scale at which the particles feel electrostatic interaction: the nuclei are fully stripped and there is a small screening effect from the background electron gas. The bulk matter behaves like an ideal gas.

ii) The second regime is bounded by the temperatures $T^L_{ij}$ and $T^P_{ij}$
with $T^P_{ij} \leq T \leq T^L_{ij}$. For temperatures lower than $T^L_{ij}$ the ions are in the liquid phase, and for temperatures close to  $T^P_{ij}$, the ions cannot be considered  a free gas but part of a lattice with vibrations $\omega^P_{ij}$. The thermonuclear burning associated to very strong electron screening operates in this temperature range.

iii) The third regime corresponds to the temperature range  $T^M_{ij} \leq T \leq T^P_{ij}$ and $\Gamma_{ij} \geq 1$.
Nuclei are  bound to the lattice sites, so that the reactions occur between highly thermally excited bound nuclei, which oscillate with frequencies higher than the plasma frequency and have energy greater than the zero point energy of the plasma. The nuclei are also  embedded in a highly degenerate electron gas, so that the reaction rates are enhanced by the charge screening electron background.
For $\Gamma^M_{ij} \geq 175$, and   temperatures lower than $T^M_{ij}$,  the nuclei form a solid lattice.

iv) In the thermally enhanced pycno-nuclear regime, $\Gamma_{ij}  >> 1$ but the melting temperature is not  yet reached (this requires $\Gamma_{ij} \geq 175$). Most of the nuclei are bound to the lattice, but some nuclei are highly exited states, and  reactions may occur between neighboring pairs of these nuclei. Electron screening is always very strong.

v) The last domain is the one of pure pycno-nuclear behaviour ($T \simeq 0$). Almost all the ions are in the fundamental state of the lattice (now crystallized to a solid) but their energies are larger than zero
because of  the Heisenberg indetermination principle. Therefore, there is still a finite possibility of penetrating  the Coulomb barrier (tunneling effect). Reactions are possible only between closest pairs.  Because there is no longer a temperature dependence, these reactions are referred to as $T=0$ pycno-nuclear rates.

\subsection{Thermal vs pycno-nuclear burnings}\label{thermo_pycno}
There are some important differences in the hypotheses underlying thermo-nuclear and pycno-nuclear reactions that are worth highlighting. In the thermo-nuclear reactions each particle can interact with all the others,  therefore the rate contains the product of the two densities of the two interacting species. The reaction rate is

\begin{equation}
R_{ij}=N_i N_j < \sigma v>
\end{equation}

\noindent The probability of interaction is proportional to the product of two factors: the Maxwell-Boltzmann statistical factor and the probability of tunneling. Their product  gives origin to the so-called Gamow window. The first factor is proportional to

\begin{equation}
e^{-E/(K_B T)}
\end{equation}
whereas the second one is proportional  to
\begin{equation}
e^{-\sqrt{E_G/E}}
\end{equation}
where $E_G$ is the Gamow energy.
If we define $E_{ij}^{pk}=\frac{1}{2}\sqrt{E_G} k_BT$ we can write the
thermo-nuclear reaction rate as
\begin{equation}
R_{ij}^{th}=4 \frac{n_i n_j}{1+\delta_{ij}}\left(\frac{2 E_{ij}^{pk}}
{3 \mu_{ij}}\right)^{1/2} \frac{S( E^{pk})}{k_BT}e^{-\tau_{ij}}
\end{equation}
with $\tau_{ij}=\frac{ 3 E_{ij}^{pk}}{k_BT}$.

The rate for the pycno-nuclear reactions is quite different. Each nucleus in the lattice can interact only with the nearest neighbours. Therefore, the  rate is
\begin{equation}
R_{i}^{pyc}=\frac{n_i}{2}<\nu_i p_i>
\end{equation}
where $p_i$ is the reaction probability between a pair of ions and $\nu_i$ is the number of nearest neighbours. According to \citet{SalpeterVanHorn69},  the reaction rate probability can be written as

\begin{equation}
p_i=D_{pyc}\frac{\lambda_i^{3-C_{pl}}S(E_i^{pk})}{\hbar r^2_{Bi}}
exp\left(\frac{-C_{exp}}{\lambda_i^{1/2}}\right)
\end{equation}
where $\lambda_i$ is the ratio of the Bohr radius to the lattice spacing. For more details on the meaning of the various symbols  see   \citet{SalpeterVanHorn69,Beard2010} and \citet{shapiro1983}.

\subsection{Rates for pycno-nuclear reactions in mono- and multi-component fluids}\label{existing_models}
In the literature on pycno-nuclear reactions, there  are basically  three  different formulations or models:
(i) the classical one, based on the simple harmonic oscillator at zero temperature proposed long ago by \citet{SalpeterVanHorn69}, that we will shortly revisit to explore the effect of impurities (Sect. \ref{impurities}),  (ii) the model by
\citet[][and references]{kitamura2000} and  finally (iii) the models by \citet{gasques2005} and \citet{yakovlev2006}.
The last two models include also the effect of temperature, and they are mutually  consistent although they make use of different analytical expressions.

\subsubsection{Summary of the Kitamura (2000) reaction rates }\label{new_rates_kita}
\citet[][see also references therein]{kitamura2000} considerably improved the simple description based on the harmonic oscillator. He derived analytic expressions for the reaction rates,  taking into account the dielectric functions of relativistic and non-relativistic electrons, the screening potentials based on the Monte Carlo simulations, and the interaction free energies in dense electron screened BIMs.   He found that  under-barrier reaction rates  can be expressed as the product of three terms

\begin{equation}\label{gamow_under_strong}
R= R_G \, A_{ij}^{(0)} \, A_{ij}^{(e)}
\end{equation}

\noindent where $R_G$ is the so-called Gamow channel, representing the basic binary interaction between any two  particles, and is expressed as the Gamow rate. It is dominant in tenuous plasmas in which the effect of surrounding particles is negligible, so the nuclei interact via the bare Coulomb potential and the rate is expected to be strongly dependent on temperature. The other two terms take the effects of the surrounding particles into account. The so-called few-particles interactions are expressed by $A_{ij}^{(0)}$ and  occur independently of the aggregation state of nuclei. The shielding effect stems from local variations of particle  density  with respect to the background (also referred to as polarization).  The net effect is to reduce the Coulomb potential barrier and to strongly enhance the rate. Once the temperature is below a certain value, the rate is expected to   increase as the  temperature decreases.
Finally the third term $A_{ij}^{(e)}$ is due to the many-particles processes that may occur when the electrons can be considered as a uniform background. This is expected to produce a small effect at the typical temperatures and densities of liquid-solid WDs, because  the so-called screening temperature $T_S$, at which this effect is important, is much lower. Therefore, the third term is expected to be small and to become dominant only at very low temperatures, unlikely to be reached in WDs.

The most important pycno-nuclear reactions are the few-particles ones rather than the many-particles interactions. Quoting \citet[][page 2694]{IchimaruKitamura99} \textit{"Since the enhancement factors generally increase steeply with lowering of the temperature, a maximum pycno-nuclear rate may be attained in a liquid-metallic substance near the conditions of freezing transitions; reaction rates in a solid, depending on the amplitudes of atomic vibrations, increase with the temperature"}.

\citet{kitamura2000}'s rates are particularly useful to understand the effects of impurities because they include both
BIMs  and   OCPs, and    provide a better description of the physics  during the solid transition, which implies a smoother discontinuity in the reaction rates. We will apply the \citet{kitamura2000} rates   over the whole range of temperatures and densities, for any kind of reacting pairs.

In a BIM  of two elements ``\textit{i}'' and ``\textit{j}'', with charges $Z_i$ and $Z_j$, mass number $A_i$ and $A_j$, mass fraction $\rho_m$ and temperature $T$,   the  reaction rates are

\begin{equation}
\begin{split}
R_{ij}(\mbox{reactions cm}^{-3}\mbox{
s}^{-1})=&\frac{1}{1+\delta_{ij}}\,\frac{X_iX_j(A_i+A_j)}{Z_iZ_j{(A_iA_j)}^2}\\
&\cdot{\left[\rho_m\left(\mbox{g
cm}^{-3}\right)\right]}^2\\&\cdot[S_{ij}(E_{\mathrm{eff}})\,
(\mbox{MeV barns})]R_0
\end{split}
\label{rate Kitamura generale}
\end{equation}

\noindent
where $\delta_{ij}$ is the Kr\"onecker delta function ($\delta_{ij}=1$ if $i=j$ and $\delta_{ij}=0$ otherwise), $S_{ij}$ is the nuclear cross-section factor for the analyzed reaction and $R_0$ depends the aggregation state of the matter.

\noindent For the liquid phase, if $T_s$ is the critical screening temperature and
    $T \geq T_s$, we have

\begin{subequations}
\begin{align}
R_0^{\mathrm{fluid}}=&2.613\times 10^{32}\,
\tau_{ij}^2\sqrt{1-{\left[\tanh{\left(\frac{T_s}{T}\right)}^8\right]}^{1/12}}\nonumber\\
&\cdot\exp
\left(-\alpha_{ij}\pi\,\sqrt{\frac{D_s}{r_{ij}^*}}+\xi_{ij}\right)\\
{\bf \rm if \,\, instead} & \,\,\,\ T < T_s  \nonumber\\
R_0^{\mathrm{fluid}}=&1.600\times 10^{33}{\left(\frac{D_s}{r_{ij}^*}\right)}^{1/2}\left\{
1+\left[0,543{\left(\frac{D_s}{r_{ij}^*}\right)}^{1/2}-1\right]\right.\nonumber\\
&\cdot\left.{\left(\frac{T}{T_s}\right)}^3\right\}\,\exp
\left(-\alpha_{ij}\pi\,\sqrt{\frac{D_s}{r_{ij}^*}}+\xi_{ij}\right)
\end{align}
\label{rate liquid K}
\end{subequations}

\noindent
where  $T_s$ is the critical screening temperature at which the Gamow peak energy equals the electrostatic screening energy. The two expressions coincide for $T=T_s$.

The nuclear reaction rates for the solid state are  obtained by substituting in eqn.(\ref{rate Kitamura generale}), the following expression for $R_0$:

\begin{equation}
\begin{split}
R_0^{\mathrm{solid}}=&\,4.83\cdot10^{32}{(R_s^{ij})}^{1.809}\\
&\cdot\exp\Bigg\{\Bigg[-3.506+0.142\,\left(\frac{a_{ij}}{D_s}\right)\Bigg.\Bigg.\\
&\left.\left. +0.144\, {\left(\frac{a_{ij}}{D_s}\right)}^2+
F(Y_{ij}^s)\right] {(R_s^{ij})}^{1/2}\right\}\,\mbox{.}
\end{split}
\label{rate solid K}
\end{equation}

\noindent
The definition of all terms and meaning of all symbols used in eqns. (\ref{rate liquid K}) and (\ref{rate solid K}) above can be found in \citet{kitamura2000} to whom the reader should refer.
We  note  that the  first term in the exponential factor of eqns. (\ref{rate liquid K}) corresponds to the Gamow thermo-nuclear channel, while the second term corresponds to the under-barrier reaction channel.  In this paper, we examine the reactions listed in Table~\ref{reazioni}. The experimental value of the cross-section factor and $\mathcal{Q}$-value from  \citet{Fowler75} are also in Table ~\ref{reazioni}.

\begin{table*}
\footnotesize
\caption{Coefficients in the interpolation expressions for a reaction rate for the optimal model of nuclear burning and for the
models which maximize and minimize the rate. The parameters CT, $\alpha_{\lambda ij}$, $\alpha_{\omega ij}$ are different for the multi-component plasma (MCP) and one component plasma (OCP) (the values for
OCP  are given in brackets). For a MCP, the models assume a uniformly mixed state (see \citet{yakovlev2006} for details).}	 \begin{tabular*}{114.5mm}{|l| l l l l l l l|}
\hline
Model         & Cexp   & Cpyc& Cpl & CT           &$ \alpha_{\lambda ij}$&$ \alpha_{\omega ij}$ & $\Lambda$\\ \hline
Optimal       & 2.638  & 3.90& 1.25& 0.724 (0.724)& 1.00 (1)& 1.00 (1)& 0.5\\
Maximum rate  & 2.450  & 50  & 1.25& 0.840 (0.904)& 1.05 (1)& 0.95 (1)& 0.35\\
Minimum rate  & 2.650  & 0.5 & 1.25& 0.768 (0.711)& 0.95 (1)& 1.05 (1)& 0.65\\
\hline
\end{tabular*}
\label{tab2.tab}
\end{table*}

\subsubsection{Summary of the Gasques - Yakovlev reaction rates}\label{new_rates_yako}
\citet{gasques2005} elaborated a model for OCPs   to calculate the reaction rates in several regimes, from thermo-nuclear to pycno-nuclear. Subsequently \citet{yakovlev2006} and \citet{Beard2010} extended the model to MCPs.   These three papers  are based on the Sao Paulo potential, that takes into account the effect of Fermi statistic on the nucleons of the interacting nuclei.
The purposes  of these studies are: (a)  to evaluate the rate for the pycno-nuclear part $R_{ij}^{pyc}$ and (b) to apply a phenomenological expression for the temperature and density dependent part $\Delta R_{ij}(\rho, T)$. All the auxiliary quantities  contained in the expressions below have already been introduced in Sect. \ref{phys_pro} or are given in Table \ref{tab2.tab}.

The pycno-nuclear component is
\begin{eqnarray}
R_{ij}^{pyc}=10^{46} C_{pyc}\frac{8\rho X_N x_i x_j A_i A_j \langle A \rangle
Z_i^2 Z_j^2}{(1+\delta_{ij})A_c^2} S(E_{ij}^{pk})\nonumber \\
\times \tilde{\lambda}_{ij}^{3-C_{pl}}exp(-\frac{C_{exp}}{(\tilde{\lambda}_{ij})^{1/2}}) cm^{-3}s^{-1}
\end{eqnarray}
whereas  the phenomenological expression for the temperature and density dependent part of the rate that combines all the burning regimes is
\begin{eqnarray}
R_{ij}(\rho,T)=R_{ij}^{pyc}(\rho)+\Delta R_{ij}(\rho,T),  \\
\Delta R_{ij}(\rho,T)=\frac{n_i n_j}{1+\delta_{ij}}\frac{ S(E_{ij}^{pk})}{\hbar} r_{Bij} P F,  \\
F=exp(-\tilde{\tau_{ij}}+C_{sc}\tilde{\Gamma_{ij}}\phi e^{-\Lambda \tilde{T_{ij}^{(p)}}/T}-\Lambda \frac{\tilde{T}_{ij}^{(p)}}{T}), \nonumber \\
P=\frac{8 \pi^{1/3}}{\sqrt{3}2^{1/3}}
(\frac{E_{ij}^a}{k_B \tilde{T}})^{\gamma}
\end{eqnarray}
where
\begin{equation}
E_{ij}^a = 2 \mu_{ij} Z_i^2 Z_j^2e^4/\hbar^2
\end{equation}
\begin{equation}
\phi=\sqrt{\Gamma_{ij}}/[(C_{ij}^{sc}/\zeta_{ij})^4+\Gamma_{ij}^2]^{1/4}
\end{equation}
\begin{eqnarray}
\tilde{\tau_{ij}}=3(\frac{\pi}{2})^{2/3}(\frac{E_{ij}^a}{k_B T})^{1/3} \nonumber \\
\tilde{\Gamma_{ij}}=\frac{Z_i Z_j e^2}{a_{ij}k_B \tilde{T}}, \tilde{T}=\sqrt{T^2+C_T^2(T_{ij}^{(p)})^2},
\end{eqnarray}
\begin{eqnarray}
\gamma=(T^2\gamma_1+(\tilde{T_{ij}^{(p)}})^2\gamma_2)/(T^2+(\tilde{T_{ij}^{(p)}})^2))\\
E_{ij}^{pk}=\hbar \tilde{\omega}_{ij}^{(p)}+(\frac{Z_i Z_j e^2}{a_{ij}}+\frac{k_b T \tau_{ij}}{3})exp(-\frac{\Lambda \tilde{T}_{ij}^{(p)}}{T})
\end{eqnarray}
where $\gamma_1=2/3$ and $\gamma_2=(2/3)(C_{pl}+0.5)$. For all other details see \citet{yakovlev2006} and \citet{Beard2010}.

\subsubsection{Comparing the two sets of reaction rates}\label{compa_rates}
\begin{figure}
   \centering
   \resizebox{\hsize}{!}{\includegraphics{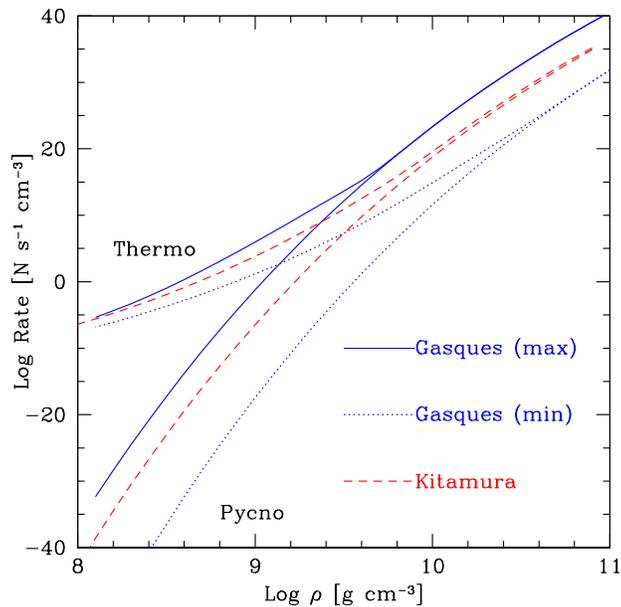}}
    \caption{Comparison  of the  \citet{gasques2005} and \citet{kitamura2000}  rates for the
    $\rm ^{12}C+^{12}C$ reaction. The coefficients of the \citet{gasques2005} rates  are those listed in Table \ref{tab2.tab}. The thermal branch and pycno-nuclear channel are calculated with  temperatures of $10^8$ K and $10^5$ K, respectively. The mass abundance of carbon is equal to $\rm X_C=1$.}
   \label{rateCC_cfr.fig}
    \end{figure}

The rates predicted by \citet[][dashed lines]{kitamura2000} and
\citet[][solid and  dotted lines ]{gasques2005} for the $\rm ^{12}C + ^{12}C$  reaction are shown as a function of density  in Fig. \ref{rateCC_cfr.fig}. For the \citet{gasques2005} rate  we adopt the numerical coefficients for  MCPs  that are listed in Table \ref{tab2.tab} and show the cases of maximum (solid lines) and minimum (dotted lines) rates. For each case we show separately  the sole pycno-nuclear component (red lines) and the pycno-nuclear plus thermal channel (blue lines).
The rates by \citet{kitamura2000} fall always in between those by \citet{gasques2005} for all regimes. In general, the three cases have a very similar trend, the only difference being the density at which the transition from thermal+pycno to pycno-nuclear regimes occurs and the absolute value of the rates. However, we note that there is a large difference between the maximum and minimum efficiency for the \citet{gasques2005} rates, amounting to about ten orders of magnitude,  and that the  \citet{kitamura2000}  rates run closer to the \citet{gasques2005} case with maximum efficiency. There are large differences in the rates  at all densities, causing a large difference in the final results.

\begin{table}
\begin{center}
\caption{Parameters for the reaction rates of most common reactions in CO-WDs.}
\begin{tabular}{|c|c|c|c|}
\hline \rule{0pt}{4ex}Reaction & Products  & $\mathcal{Q}$-value [MeV] & $S(0)$ [MeV barn]\\[6pt]
\hline\rule{0pt}{4ex}
$^{12}$C+$^{12}$C & Mg$^{24}$ & 13.931 & $8.83\times
10^{16}$\\[6pt]
$^{12}$C+$^{16}$O & Si$^{28}$ & 16.754 & $1.15\times 10^{21}$\\[6pt]
$^{16}$O+$^{16}$O & S$^{32}$  & 16.541 & $2.31\times 10^{27}$\\[6pt]
\hline
\end{tabular}
\label{reazioni}
\end{center}
\end{table}

\section{Impurities in pycno-nuclear reactions rates}\label{impurities}
When  matter becomes solid, the C and O ions are in  a fixed configuration in a crystal  whose structure is  known from solid state physics. We cannot  exclude the possibility that some impurities  of elements of any type are present in the lattice.

The dominant element with the lowest atomic number is carbon ($Z_C$=6), followed by oxygen ($Z_O=8$), so impurities can be grouped according to their atomic number:  i) heavy elements like magnesium, etc...with $Z > Z_C$; ii) light elements like hydrogen and helium with $Z < Z_C$.

\noindent
We are particularly interested here in  impurities of light elements, like hydrogen and helium. In the next section, we will see that the pycno-nuclear rates for these elements may be particularly high, so the  amount of light elements  necessary for  relevant effects is extremely low. For instance, in the case of hydrogen,   abundances  in the range  $\rm 10^{-16} \leq X_H \leq  10^{-21}$ are sufficient to produce nuclear energy in amounts comparable to the luminosity of typical WDs  at suitable ages.

This should not be confused with the  surface content of hydrogen  that can be as high as $\rm X_H \simeq 10^{-5}$ without triggering thermo-nuclear burning  \citep[see][for all details]{Fujimoto1982a,Fujimoto1982b,renedo2010}.
Higher concentrations are not allowed otherwise the associated thermo-nuclear energy generation via the  CNO-cycle would exceed the luminosity of a WD, probably inducing  a thermonuclear runaway.

In this  section we describe, with a simple formulation, how the coulombian potential and local density in a ionic lattice are modified by impurities. We adopt  the  expression of \citet{SalpeterVanHorn69} and \citet{shapiro1983} for the rate of pycno-nuclear reactions. The original theory was developed  for reactions among the same nuclei (e.g. carbon).  Since we consider at least two elements  (e.g. reactions among hydrogen and carbon), the columbian potential must be suitably modified.  We are going to present a template, to which more sophisticated formulations must be compared.

\subsection{The modified Salpeter-Van Horn model}\label{SalpHorn}
\begin{figure}
   \centering
   \resizebox{\hsize}{!}{\includegraphics{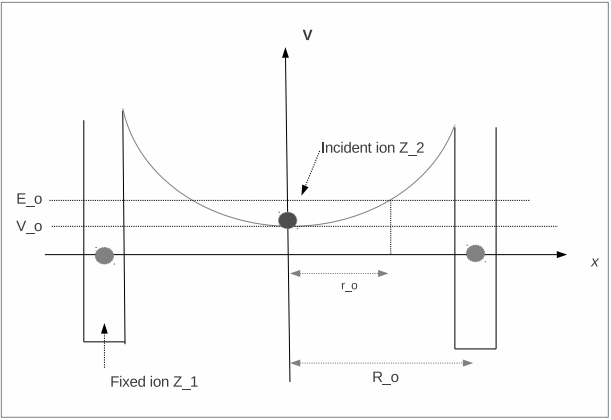}}
\caption{The potential governing the motion of one incident ion of charge $Z_2$ relative to
    an adjacent ion of charge $Z_1$. Zero point energy fluctuations in the harmonic potential well near
    the incident ion  can lead to Coulomb barrier penetration and nuclear reactions.}
\label{shapiro1.fig}
\end{figure}

Using the same formalism of \citet{shapiro1983}, we consider   an OCP made of a certain type of ions, e.g. either pure carbon or pure oxygen. In this framework, first we  consider the one-dimensional  potential of an array  of ions and   extend it to  3D. Suppose that the lattice is composed of ions of charge $Z_1$. In one cell we substitute the charge $Z_1$ with a charge $Z_2$ (for instance hydrogen or helium). If $x$ is the displacement  from the equilibrium position (see Fig.\ref{shapiro1.fig})

\noindent the coulombian potential becomes
\begin{equation}
V(x)=\frac{Z_1 Z_2 e^2}{R_0-x}+ \frac{Z_1 Z_2 e^2}{R_0+x}-2 \frac{Z_1^2 e^2}{R_0}
\label{pot_1}
\end{equation}
where $R_0$ is  the inter-ionic distance. In eqn. (\ref{pot_1}) we assume that the energy of the lattice is zero and  estimate the energy variation. Removing  a charge $Z_1$  produces the negative term (interaction of the neighbouring particles with the hole) and the addition of a  charge $Z_2$  at position $x$ introduces the positive terms.  We suppose that $x<<R_0$  and use the Taylor expansions to derive
\begin{equation}
V(x)=2\frac{Z_1 Z_2 e^2}{R_0}-2 \frac{Z_1^2 e^2}{R_0}+ \frac{2 Z_1 Z_2 e^2 x^2}{R_0^3} \,.
\label{pot_2}
\end{equation}
The constant term shifts all the energy levels. The variable  term
\begin{equation}
 + \frac{2 Z_1 Z_2 e^2 x^2}{R_0^3}
\label{pot_3}
\end{equation}
is the dominant one and  behaves like a harmonic oscillator with
\begin{equation}
\omega=(\frac{4 Z_1 Z_2 e^2}{R_0^3 \mu})^{1/2} \, .
\label{pot_freq}
\end{equation}

\noindent
Using  the equation $E_0=V(r_0)=V_0+1/2Kr_0^2=V_0+1/2 \hbar \omega$, the  turning point $r_0$ is

\begin{equation}
 r_0=( \frac{\hbar R_{0}^{3/2}}{2 (Z_1 Z_2 \mu)^{1/2} e})^{1/2}
 \label{errezero}
\end{equation}
 The solutions of the Schr\"odinger equation for a harmonic oscillator are known.
In three dimensions they become
\begin{equation}
|\psi_{SHO}|^2=\frac{\tau^3}{\pi^{3/2}}e^{-\tau^2r^2}
\label{psiquad_1}
\end{equation}
where $\tau=1/r_0$. Assuming that  the exponential value is one,
\begin{equation}
|\psi_{inc}|^2\simeq|\psi_{SHO}|^2\simeq \frac{1}{r_0^3\pi^{3/2}}  \, .
\label{psiquad_2}
\end{equation}
The transmission coefficient for an incident ion with energy $E_0$ in the WKB approximation is
\begin{equation}
T=exp[-2\int_a^b (\frac{2\mu}{\hbar^2}[E_0-V_0-1/2Kx^2])dx]
\label{transmis}
\end{equation}
where $E_0=V_0+1/2\hbar \omega$.
Following the discussion by \citet{shapiro1983}, we obtain the reaction rate per ion pair
\begin{equation}
W=v|\psi_{inc}|^2\frac{T S(E)}{E_K}
\label{rateW}
\end{equation}
where in the limit $r_0/R_0 <<1$
\begin{equation}
T=\frac{R_0}{r_0}exp(-2\frac{R_0^2}{r_0^2})  \, .
\end{equation}
This transmission coefficient is the leading term  in the reaction rate and it strongly depends on the charge $Z_2$, in fact the exponent is proportional to $(Z_1 Z_2)^{1/2}$, which means that the transmission coefficient is much higher for $Z_2<Z_1$.

We   refer to our revision of the \citet{shapiro1983} rate as the "modified harmonic oscillator" (MHO). We will later adopt the formulations of \citet{kitamura2000} and \citet{yakovlev2006}, but the MHO approximation is an efficient way to explore reactions between light  and heavier elements  when the WD reaches  the pycno-nuclear regime.
We  note that the same formalism can be applied to the case of MCPs because what matters  are the pairs of interacting ions, e.g. $\rm ^{1}H + ^{12}C $ or $\rm  ^{1}H + ^{16}O $. The lower coulombian barrier in the case of $\rm ^{12}C$ and $^{1}H$  definitely favours this ion pair, so most of discussion will be limited to these two elements.

\subsection{Local densities around impurities}\label{changeden}

\begin{figure}
   \centering
   \resizebox{\hsize}{!}{\includegraphics{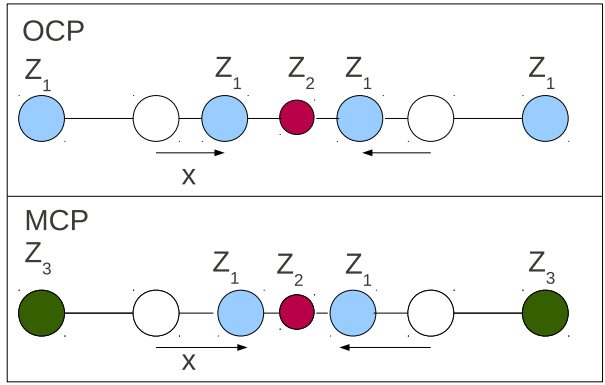}}
    \caption{\textbf{Top Panel}: OCP made of carbon in which a nucleus  of carbon ($Z_1$) is replaced by a nucleus of  hydrogen ($Z_2$).  We evaluate the forces acting on the nuclei  $Z_1$ adjacent to the impurity (charge $Z_2$) and the corresponding displacement $x$ of $Z_1$ given by the new condition of force equilibrium.  \textbf{Bottom Panel}: the same but  for a MCP made of carbon  and  oxygen  nuclei in BCC configuration in which a nucleus  of oxygen is replaced by a  nucleus of hydrogen. }
   \label{impro1.fig}
    \end{figure}

In order to evaluate  the change in the local density and consequently in the rates of energy production by pycno-nuclear reactions caused by impurities,  OCPs and MCPs must be treated separately. We can consider three cases:

\textbf{(i) OCPs, neglecting electrons in the electrostatic force}. Let us consider an array of nuclei of charge $Z_1$ (e.g. carbon) as shown in the top panel of Fig. \ref{impro1.fig}. If in one site we substitute the nucleus of charge $Z_1$ with a nucleus of charge $Z_2$, e.g.  hydrogen (the impurity), the carbon nuclei neighbouring the  hydrogen nucleus are shifted toward it because of the new equilibrium of the forces as indicated by the arrows (the situation is clearly symmetric). The  equilibrium of the forces is expressed by

\begin{equation}\label{ion_1}
\frac{Z_1 Z_2}{(R_0-x)^2}=\frac{Z_1^2}{(R_0+x)^2} \,.
\end{equation}

\noindent After developing the squares we obtain
\begin{equation}
(Z_1 Z_2-Z_1^2)x^2+2R_0(Z_1 Z_2+Z_1^2)x+(Z_1 Z_2-Z_1^2)R_0^2=0 \,.
\end{equation}\label{ion_2}
This is a second order equation with two solutions
\begin{equation}
x_{1,2}=R_0\left[-\frac{(Z_1 Z_2+Z_1^2)}{(Z_1 Z_2-Z_1^2)}\pm \sqrt{\left[\frac{(Z_1 Z_2+Z_1^2)}{(Z_1 Z_2-Z_1^2)}\right]^2-1}\right]\,.
\end{equation}\label{root_1}
We evaluate the displacement for $Z_1=6$ (carbon) and $Z_2=1$ (hydrogen) and keep only the solution $<1$ that has physical
meaning. We obtain $R_0' = R_0 - x=0.58\, R_0$, i.e. the new local inter-ion distance $R_0'$ is smaller than in the unperturbed case. For the case $Z_1=6$ (carbon) and $Z_2=2$ (helium) obtaining $R_0 - x=0.73\, R_0$. As expected, the effect is decreasing with   increasing $Z_2$.  Hereinafter $R_0'$ is  renamed $R_0$.

\textbf{(2) OCPs with  electrons in the electrostatic force}. In the above expression we neglected the contribution by electrons to the balance of electrostatic forces among ions. Taking  electrons into account,  we assume that in the space between the ions $Z_1$ and $Z_2$ electrons distribute uniformly and evaluate the field due to electrons at any distance $x$ in between the ions $Z_1$ and $Z_2$. In this case eqn. (\ref{ion_1}) becomes

\begin{eqnarray}
&-& \frac{Z_1 Z_2}{(R_0-x)^2}+\frac{Z_1^2}{(R_0+x)^2}  +  \nonumber \\
 &-& {1\over 2} \frac{(Z_1 - Z_2) Z_1 x}{R_0^3}
     + {1\over 2} \frac{(Z_1 - Z_2)Z_1 (R_0-x) }{R_0^3} = 0
\end{eqnarray}\label{mod_ion_1}

\noindent Performing easy algebraic manipulations we obtain the fifth degree equation

\begin{equation}
-2x^5 + R_0 x^4 + 4 R_0^2 x^3 - \frac{38}{5}R_0^4 x + 3R_0^5 = 0
\end{equation}\label{opc_fifth}

\noindent to be solved numerically. The solution has four complex roots and only one real with physical meaning, i.e. $x= 0.4402\, R_0$. Therefore, in the case of carbon and hydrogen the new distance between two neighbouring ions is  $R_0'= 0.56\, R_0$.

\textbf{(3) MCPs with electrons in the electrostatic force}. It is worth of interest to consider the case of MCPs, limited however to a mixture of $\rm C + O$ \citep[the BCC configuration,][]{Ichimaru1982} and including the effects of electrons in the force balance. We consider an array  of ions in the sequence O ($Z_3$) - C ($Z_1$) - H ($Z_2$) - C($Z_1$) - O($Z_2$) as shown in the bottom panel of Fig. \ref{impro1.fig}, and evaluate the displacement of the C nuclei bracketing the hydrogen nucleus  by imposing the balance of electrostatic forces.
In this case eqn. (\ref{mod_ion_1}) becomes

\begin{eqnarray}
&-& \frac{Z_1 Z_2}{(R_0-x)^2}+\frac{Z_1 Z_3}{(R_0+x)^2}  +  \nonumber \\
 &-& {1\over 2} \frac{(Z_1 - Z_2) Z_1 x}{R_0^3}
     + {1\over 2} \frac{(Z_1 - Z_2)Z_1 (R_0-x) }{R_0^3} = 0
\end{eqnarray}\label{mod_ion_2}
the associated fifth degree equation is

\begin{equation}
-10x^5 + 5 R_0 x^4 + 20 R_0^2 x^3 +4 R_0^3 x^2 -46 R_0^4 x + 19 R_0^5 = 0
\end{equation}
\label{mcp_fifth}

\noindent where $R_0$ is  the distance between  the C and O ions in the BCC lattice. Also in this case there is only one real solution, $x=0.4822 R_0$. The distance between the C and H ions is therefore $R_0'= 0.5178 R_0$.

\begin{figure}
\centering
\resizebox{\hsize}{!}{\includegraphics{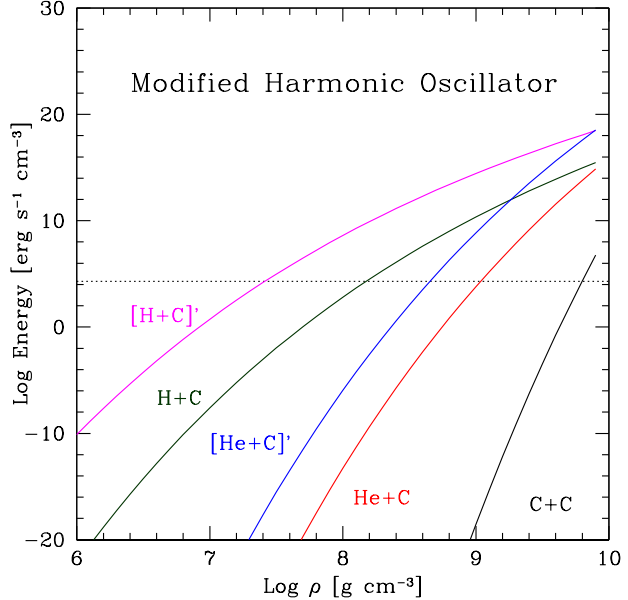}}
    \caption{The energy  generated by the reactions $\rm ^{12}C + ^{12}C$ (blue line labelled C+C), $\rm ^{1}H + ^{12}C$ (green line labelled H+C), and $\rm ^{4}He + ^{12}C$ (red line labelled He+C) according to the MHO rates. The abundances by mass of C, H and He are $\simeq 1$, $10^{-21}$, and $3\times 10^{-7}$, respectively. This corresponds to an ideal WD made of sole carbon with traces of hydrogen or helium.  The rates for the reactions $\rm ^{1}H + ^{12}C$ (magenta line labelled [H+C]') and  $\rm ^{4}He + ^{12}C$ (cyan line labelled [He+C]') have been recalculated  including also the effect on the local density caused by the presence of impurities as discussed in Section \ref{changeden}. Finally,  the dotted black line represents the mean luminosity expressed  in $\rm erg\,s^{-1}\,cm^{-3}$ of a typical WD. First  all the rates intersect the mean luminosity which is the signature of a potential nuclear runaway, second the intersection occurs at higher and higher densities with increasing atomic number. Third, the effect of impurities is conspicuous (e.g. compare the green and magenta lines for hydrogen impurities or the red and cyan lines for He impurities). }
\label{energy_sh.fig}
\end{figure}

   \begin{figure}
   \centering
   \resizebox{\hsize}{!}{\includegraphics{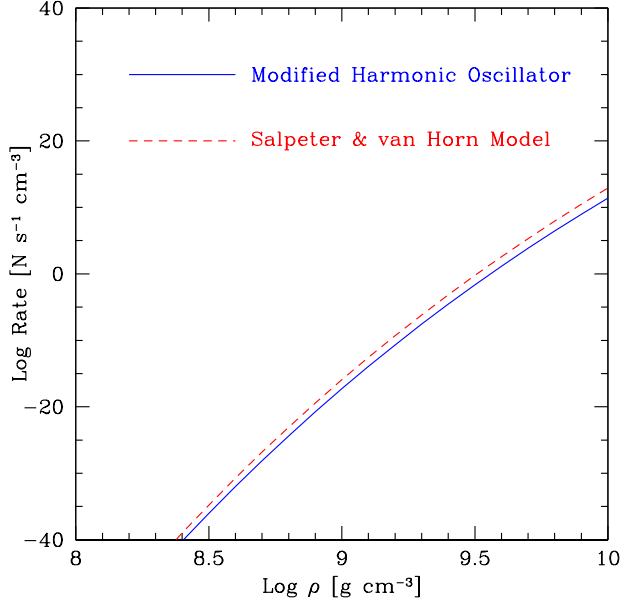}}
    \caption{ $\rm ^{12}C + ^{12}C$  pycno-nuclear reaction rates for the simple MHO model
    (blue continuous line) compared with the \citet{SalpeterVanHorn69} model (red dashed line). They are roughly similar.}
   \label{mod1.fig}
    \end{figure}

\textbf{Implementing the correction into the reaction rates}.
The effect of this correction on the interionic distance in the pycno-nuclear rates depends on the physical model for the rates. If we insert the new value of $R_0$ into  eqns. (\ref{psiquad_2}) and (\ref{transmis}), the pycno-nuclear reaction of  \citet{SalpeterVanHorn69}  are enhanced  by a large factor.

Using the  \citet{gasques2005}, and \citet{yakovlev2006} rates, in which the pycno-nuclear term is separate from the thermal one, the functions  $a_{ij}$ must be redefined. Finally, the reaction rate \cite{kitamura2000} is more complicated to treat,  because the  temperature dependence is intrinsic in the formulation and  cannot be singled out easily.

(i) {\em Gasques and Yakovlev rates}.  There are two approximations depending on the composition of the plasma.
If the plasma is made of pure $\rm ^{12}C$ with $\rm ^1H$ impurities, the  new $a_{ij}$ are

\begin{equation}
  a_{ij} = \lambda [\frac{3}{4\pi n}]^{1/3}
\label{alpha_aij}
\end{equation}
where $\lambda=0.56$  for the reaction $\rm ^{1}H + ^{12}C$, and $\lambda=0.73$ for the reaction $\rm ^{4}He + ^{12}C$. Other combinations of reactants (such as $\rm ^1H + ^{16}O $) can be easily obtained from the general formalism above, and  the effect of impurities is expected to be smaller. While using the rate by \citet{SalpeterVanHorn69},  the correction is directly applied to $R_0$, using   the \cite{gasques2005} and \citet{yakovlev2006} rates,  the old
$a_{ij}$ coefficients of eqn. (\ref{aij}) are replaced by  eqn. (\ref{alpha_aij}.

\begin{equation}\label{aij_1}
a_{ij} = {1\over 2} \big[ \big({3Z_i \over 4\i n_e}\big)^{1/3} + \big({3Z_j \over 4\pi n_e}\big)^{1/3} \big] \, .
\end{equation}
In such a case the effect is less evident and smaller than before.

In the case of a MCP (for instance $^{12}C + ^{16}O$), the new $a_{ij}$ coefficients are

\begin{equation}
                   a_{ij} = 0.5178 a_{CO} = 0.5178 \frac{(6^{1/3} + 8^{1/3})}{2}
                       \big[\frac{3 m_u}{4\pi 0.5 \rho} \big]^{1/3}
\end{equation}\label{aij_2}
where $a_{CO}$ is derived from eqn. (\ref{aij_1}) above.

(ii) {\em Kitamura Rate}. In this case,  at low densities the $a_{ij}$ are those of the classical formulation given by eqn. (\ref{aij_1}), whereas at high density they must coincide with those of eqn. (\ref{aij_2}).
In order to transit smoothly from the thermal to the pycno-nuclear regime and to take the correction of local densities due to impurities into account, we assume that the $a_{ij}$ coefficients can be linearly interpolated in $\log\rho$ according to

\begin{equation}
a_{ij}' = a_{ij}^{pyc} + (a_{ij}^{th} - a_{ij}^{pyc}) \frac{10-\log\rho}{10-6}
\end{equation}\label{aij_int}
where $a_{ij}^{pyc}$ is defined according to eqn. (\ref{aij_2}) at $\log\rho=10$, and $a_{ij}^{th}$ is defined according to eqn. (\ref{aij_1}) at $\log\rho=6$.

\subsection{Results from the Modified Harmonic Oscillator}
In  the MHO formalism derived above, the total number of reactions per second per $cm^3$ is given by
\begin{equation}
P_0=n_{el} W
\end{equation}
where $n_{el}$ is the number density of the reacting elements. This  is related to the abundance by number $f_{el}$ by  $n_{el} = f_{el} N/V$ where $N$ is the total number of ions in the star and $V$ is the volume. In the simple case of reactions among identical ions, converting number densities to mass densities,  mass abundances, and number abundances is straightforward, whereas in the case of reactions among different elements the procedure is more complicated.
 If $N_j$, $X_j$ and $A_j$ are the number of ions of species $j$ with mass abundance $X_j$ and mass number $A_j$,

\begin{equation}\label{frac1}
    X_j = { N_j A_j m_u \over \sum N_l A_l m_u }
\end{equation}
where $m_u$ is the mass unit. If  a WD, made of pure carbon, is polluted by traces of hydrogen and helium ($N_H << N$ and $N_{He} << N$) we   obtain
\begin{equation}
X_H = {f_H \over 12} \qquad  {\rm and} \qquad  X_{He} = f_{He}  {4 \over 12} = { f_{He} \over 3 }.
\end{equation}
\label{frac2}
For a WD  of carbon and oxygen, whose mass abundances are related as $\rm X_C /X_O = \alpha$,  we derive

\begin{equation}
    X_H = {f_H \over 12} { [\alpha + (3/4) ] \over (\alpha +1) } \quad \quad X_{He} =
                         f_{He} {4\over 12} { [\alpha + (3/4) ] \over (\alpha +1) }
\end{equation}
\label{frac3}
for typical values of $\alpha$, $\rm X_H$ is about a factor of 10 lower than  $f_H$, and $\rm X_{He}$ is a factor of 3 lower than $f_{He}$.

\noindent
Since the mass and structure of a WD  depend on its central density, it is useful to express the  reaction rates  as a function of this parameter.
In Fig. \ref{energy_sh.fig} we show the energy generation rates for the $\rm ^{1}H+^{12}C$, $\rm ^{4}He + ^{12}C$ and $ \rm ^{12}C + ^{12}C$  reactions. These are calculated with the MHO rates with and without the impurities and associated density enhancement.  We also show the case of the $\rm ^{12}C+^{12}C$ reaction calculated with the MHO. The astrophysical factors $S(0)$ and  $Q$-values of the reactions $\rm ^{1}H+^{12}C$,  $\rm ^{4}He+^{12}C$   are given in Table \ref{tab5.tab}, those for the $\rm ^{12}C +^{12}C$ are in Table \ref{reazioni}. The abundances by number and by mass of the elements are $f_{H}=10^{-20}$, $f_{He}=10^{-6}$ and $f_{C}=1$ (or equivalently $\rm X_H= 10^{-21}$, $\rm X_{He} = 3\times 10^{-7}$, and $\rm X_C=1$). For the sake of comparison we show also the mean luminosity of a typical CO-WD,
$\rm 10^{31}\, erg \, s^{-1}$. In order to express the energy generation rates and WD luminosity  in units of $\rm erg\,s^{-1}\,cm^{-3}$, the  WD luminosity is divided by the WD volume estimated assuming a mean radius of  5000 km.

The results of Fig  \ref{energy_sh.fig} are  of paramount importance: they clearly show that in the case of hydrogen impurities (with  $f_H=10^{-20}$) the energy produced by the pycno-nuclear reactions at zero temperature  may exceed the typical luminosity of a WD  at a density $\rho \simeq 3\times 10^7g/cm^3$, corresponding  to a mass of about 1.05 $M_\odot$, significantly smaller  than the Chandrasekhar mass, see  Table \ref{tabrho_mass.tab}.  This is a potentially explosive situation. In the case of helium, the intersection density and corresponding WD mass are higher but still of interest ($\rm \rho \simeq 4\times 10^8 g\,  cm^{-3}$ and  about $1.3\, M_\odot$, respectively). Finally, the intersection with the $\rm ^{12}C+^{12}C$ line is at about $\rm 9.8\, g\, cm^{-3}$ and the corresponding WD mass is nearly equal to the Chandrasekhar limit.

\begin{table}
\centering
\caption{Parameters for the nuclear reactions of hydrogen and helium with carbon. }	
\begin{tabular}{|l r r|}
\hline
& S(0)     &Q \\
& MeV barn & MeV \\
\hline
$\rm ^{1}H  + ^{12}C$  & $1.34\times 10^{-3}$   & 1.94  \\
$\rm ^{4}He + ^{12}C$  & $1.00\times 10^{-3}$   & 7.16  \\
\hline
\end{tabular}
\label{tab5.tab}
\end{table}

As already discussed in \citet{shapiro1983}, the MHO  expression for the reaction rate is not very different from the original one obtained by \citet{SalpeterVanHorn69} taking into account other effects, like  electron screening and anisotropy. As a test, we compare the rates for the  $^{12}C+^{12}C$ reaction  derived from the MHO above and \citet{SalpeterVanHorn69}.  The results are shown in Fig.\ref{mod1.fig}. The agreement is remarkable. Furthermore, these rates are also in good agreement  with those   by \citet{gasques2005} for the same reaction.

\begin{figure*}
\centering
{\includegraphics[width=0.4\textwidth]{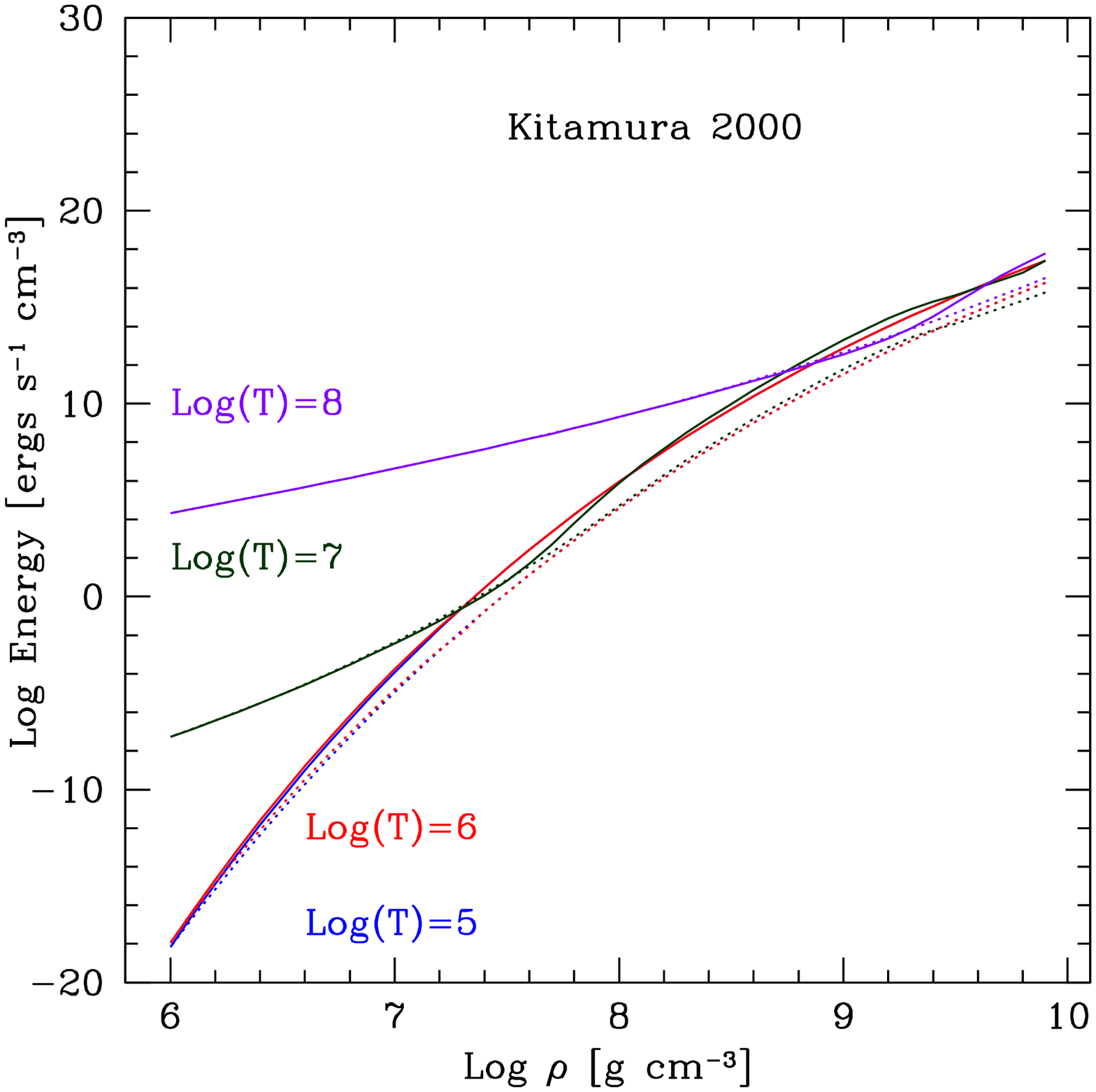}
 \includegraphics[width=0.4\textwidth]{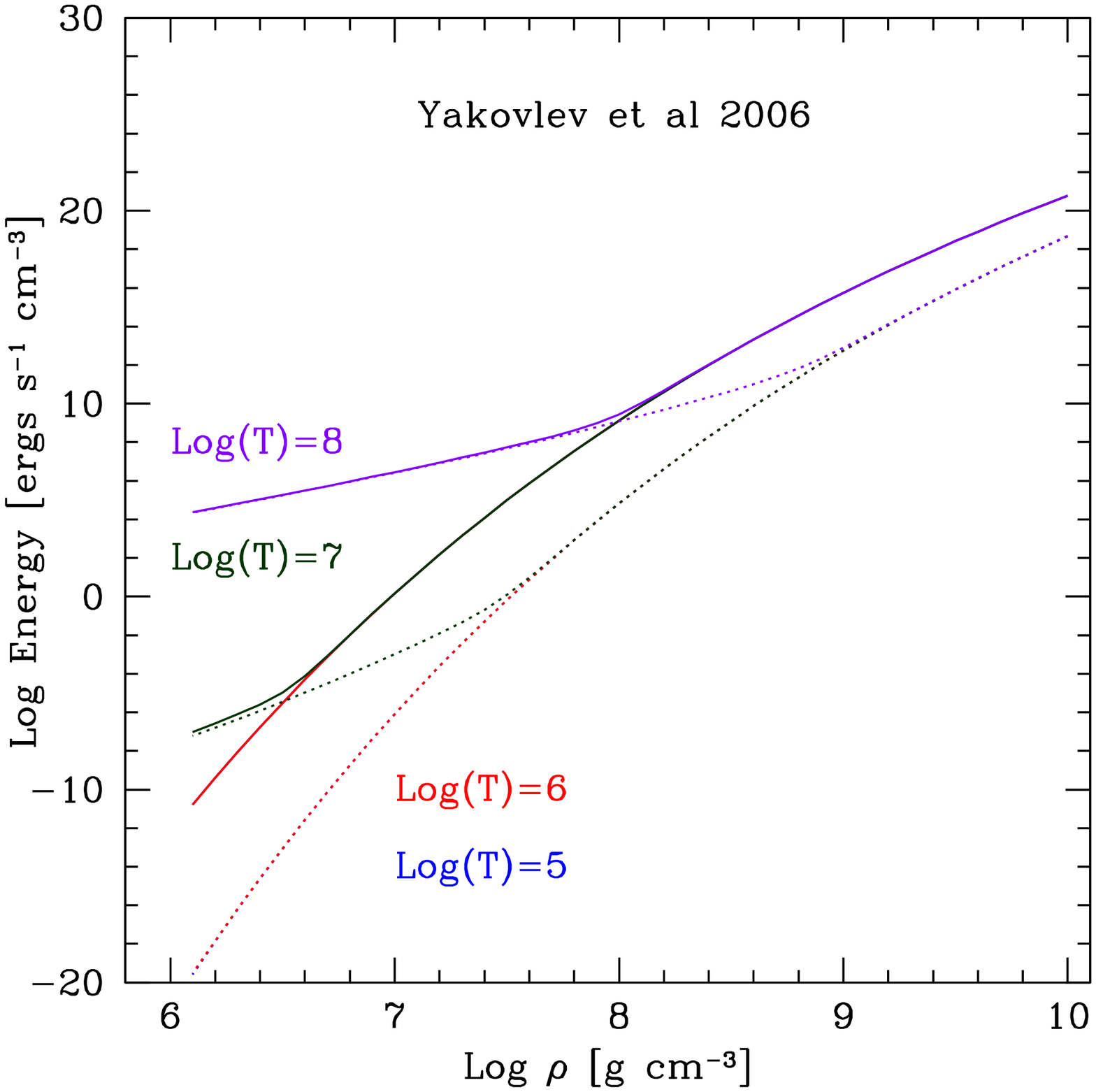}    }
\caption{\textbf{Left Panel}: energy rates per unit volume produced by  the reaction $\rm ^{12}C(^1H,\gamma)^{13}N$ with  hydrogen abundance $\rm X_H = 10^{-20}$ as a function of density and four different temperatures of the thermal branch. The nuclear  rates are from  \citet{kitamura2000}.The dotted lines represent the energy without local density correction while the continuous lines represent the model that considers local density correction. \textbf{Right panel}: the same as in the left panel but for reaction rates according to \citet{yakovlev2006}.The dotted lines represent the energy without local density correction while the continuous lines represent the model that considers local density correction.  }
\label{energy_ks_gs.fig}
\end{figure*}

\section{Can traces of H and He be present in the interior of a WD?}\label{evaluation}

The existence of traces of light elements in the core of CO-WDs is at the basis of our investigation.
Light elements  are the best candidates to consider, because the very high coulombian barriers of  high $Z$ elements would  quench the penetration probabilities to zero.

Undoubtedly, hydrogen and helium on the WD surfaces in small but significant  abundances are predicted theoretically
\citep[$\rm 10^{-9} \leq X_H \leq 10^{-5}$  according to ][]{millerbertolami2013,renedo2010} and are observed \citep{bergeron1990}.
How much hydrogen or helium can be present in the interior?

Following \citet{kippenhahn1990}, the rather high internal temperatures (from $10^8$ to $10^6$ K) of an aging   WD   set a limit to the possible hydrogen content in the interior. If hydrogen were present with a mass concentration $\rm X_H$, we would expect H-burning via the pp-chain. For average values $T = 5 \times 10^6$ K, $\rm \rho = 10^6 \, g\, cm^{-3}$,  the energy released  by the pp-chain $\epsilon_{pp} \simeq 5 \times 10^4 \, X_{H}^2$ $\rm erg\, g^{-1}\,s^{-1}$ to which for a $M=1 \, M_\odot$ the luminosity would be  $L/L_\odot \simeq M/M_\odot \times \epsilon_{pp} \simeq 2.5 \times 10^4 X_{H}^2$. The mean observed luminosity of WD $L/L_\odot \simeq 10^{-3}$ allows  $\rm X_H \leq 2\times 10^{-4}$, the upper limit for  $\rm X_H$ in WD interiors. Stability
considerations  indeed rule out that the luminosity of normal WD is generated by thermonuclear reactions, which was first pointed out by \citet{mestel_1952_a, mestel_1952_b}. Nuclear burning could only be expected in nearly cold configurations near $T=0$ by pycno-nuclear reactions.

According to \citet{kawaler1988} sedimentation and diffusion of elements will bring hydrogen and helium from the interior to surface and viceversa. Although this is a slow process, the initial content of hydrogen can be as high as $10^{-6} M_\odot$.
It   is  worth noting that diffusion of hydrogen inwards is inhibited by the electron degeneracy \citep[see for instance][]{Iben1985}. In brief,  while  the tail   of  the  hydrogen
distribution chemically diffuses inward  as the WD cools down, actually it may reach a maximum depth, and with
further cooling, begins to retreat outwards. This is because the increasing
electron degeneracy halts the inward diffusion of hydrogen. Hence, surface hydrogen never reaches very deep layers.

We argue and try to demonstrate here that  traces of  light elements (below the upper limits given above)   left over by core  nuclear burnings during  the previous phases can still be present when the WD is cooling. This would be the analog of the situation in neutron stars, where traces of electrons, protons and even heavy nuclei are expected to exist  in the medium   of free neutrons \citep{shapiro1983}. We show that hydrogen abundances in the range  $\rm 10^{-22} \leq X_H \leq  10^{-14}$ may be present and cause  ignition of pycno-nuclear reactions. Analogous considerations can be made for helium.

\subsection{Hydrogen and helium contents expected in stellar interiors at the end of the AGB phase}

The correct answer to the above question would be detailed numerical models of stellar interiors in which the abundances of all elements undergoing nuclear processing, convective mixing, diffusion and sedimentation (whenever appropriate) are followed in a great detail to even  very low values. In current stellar models, the abundances of the chemical elements are calculated by solving complicated networks of differential equations governing the local creation / destruction of the elemental abundances according to the rates of the  nuclear reactions that are involved, and the  efficiency of the diffusive/convective mixing.

A widely adopted short-cut to the above complex integration technique and quite heavy computational burden is to set equal to zero the abundances of  elements  when they fall below some reasonably small values, unless otherwise required. This is particularly true for those elements that initially are very abundant, such as  hydrogen and  helium. They indeed are customarily  set to zero when $\rm X_H \simeq 10^{-8}$ and $\rm X_{He} \simeq 10^{-8}$. In the case of the central H- and He-burnings this is also taken to mark  the end of the evolutionary phase  \citep[see for instance ][and references therein]{Bertelli1994,Bressan2012}  when describing the physical ingredients and technical   details of the Padova stellar evolutionary code.
Furthermore, requesting that all elements that are gradually transformed  into heavier species (and therefore decrease their abundance) are followed to extremely low values is hardly viable and time consuming when  extended grids of stellar models are considered. Therefore,  proving or disproving that traces of hydrogen and helium
(at the abundances that will be used in the present study) can survive previous burnings and be present in WDs cannot be easily assessed with current evolutionary codes to disposal.

However, the elementary theory of thermo-nuclear reactions suggests that elements at very low abundances can survive even the nuclear phase in which they are burned into heavier species.  In brief,  looking at the case of hydrogen, the typical lifetime of a proton against a proton in the pp-chain or against CNO nuclei in the CNO-cycle  is expected to  decrease with increasing the temperature and to increase with decreasing $\rm X_H$ \citep[see][]{Iben2013a, Iben2013b}. In other words,  as the abundance of hydrogen   decreases, it is more and more difficult to  destroy  it further, i.e. the abundance of hydrogen never becomes zero.  Similar considerations can be made for helium (and any other element in general).

We try here  two simple experiments designed to estimate the abundances  $\rm X_H$ and $\rm X_{He}$ left from the nuclear burnings along the whole evolutionary history from the zero age main sequence to the beginning  of the TP-AGB phase that terminates with generation of  CO WD (the TP-AGB phase  can be neglected because of its short lifetime compared to that of previous phases). Hydrogen and helium expected in WDs have survived all previous nuclear burnings.

To this aims we consider the evolutionary sequences of the
3, 5 and 6 $M_\odot$ stars with solar-like composition [X=0.723, Y=0.260, Z=0.017] that are the progenitors of massive CO WDs  e.g. the entries of  Table \ref{mi_mco_mwd.tab}. The sequences are taken
from the Padova library of stellar models by \citet{Bertelli2009}.  For each star  we know  all relevant physical quantities as a function of time both in the centre (temperature, density, abundances of elements, size of the convective and H-exhausted core, etc.) and at the surface (total and partial luminosities, effective temperature, etc.).

Since the central H- and He-burning (CNO and 3$\alpha$, respectively) in these stars are point-like sources at the centre, we may use  simple energy conservation arguments and write

\begin{equation}
\epsilon_{i}(X_i, T, \rho) = {L_i \over M}
\label{epsi_lum}
\end{equation}

\noindent
where the index $i$ stands for H- or He-burning, and $L_i$ are the partial luminosities (the total luminosity is  $L=L_H + L_{He} + L_G$, where  $L_G$ is the contribution from gravitational contraction, usually negligible). This equation simplifies the complexity of energy production (nuclear burning and gravitational release in convective/radiative conditions, the latter  depending on the stellar mass)  to an ideal situation in which a gram of matter produces the amount of energy   radiated  by the surface   per unit mass and unit time. Using the above equation we implicitly neglect the contribution of the gravitational contraction/expansion to the luminosity (a reasonable approximation for most of the stellar lifetime).
Furthermore, for the sake of simplicity, we consider only the nuclear burning occurring in the core and  ignore the nuclear burning  in the shell. In   the 3, 5, 6 $M_\odot$ stars,   hydrogen burns  via the CNO-cycle with a small contribution from the pp-chain, and  helium burns via the 3$\alpha$ process. Finally, in first approximation we consider nuclear burning only in radiative conditions, neglecting continuous refueling by convection which may occur in stars of 3, 5 and 6 $\, M_\odot$.

For the energy release by H- and He-burning we adopt  analytical expressions  from classical textbooks
\citep{kippenhahn1990,CoxGiuli2004}:

\noindent (i)  pp-chain,
\begin{eqnarray}
\epsilon_{pp}&=&2.06\times 10^6 f_{pp} g_{pp} \rho X_H^2 T_6^{-0.66} e^{(-33.81 T_6^{-0.33})} \\
g_{pp} &=& 1+0.0012 T_6^{0.33} + 0.0078 T_6^{0.66}+0.0006 T_6  \nonumber \\
f_{pp} &=& 1+0.25 \rho^{0.5}  T_6^{-1.5}  \nonumber
\label{pp_chain}
\end{eqnarray}

\noindent (ii)  CNO-cycle
\begin{eqnarray}
\epsilon_{CNO}&=& 8.7\times 10^{27}g_{14,1}X_{CNO} X_H \rho T_6^{-2/3}e^{-152.28/T_6^{1/3}}   \\
g_{14,1}&=&  1+0.003 T_6^{1/3}-0.0078T_6^{2/3}-0.00015 T_6  \nonumber
\label{cno_cycle}
\end{eqnarray}

\noindent
(iii) and  3$\alpha$
\begin{eqnarray}
\epsilon_{3\alpha}&=&5.09\times 10^{11}f_{3\alpha} X_{He}^3 \rho^2 T_8^{-3}e^{(-44.027/T_8)}\\
f_{3\alpha}&=& exp{(2.4\times 10^{-3} \rho^{1/2} / T_8^{3/2})}                  \nonumber
\label{helium3a}
\end{eqnarray}

\noindent
where $T_6$ and $T_8$ are the temperatures in units of $10^6$ and $10^8$ K, respectively,  the functions $g_{14,1}$ and $ g_{3\alpha}$ are the screening factors, $\rm X_H$, $\rm X_{He}$, $\rm X_{CNO}$ are the abundances by mass of hydrogen, helium and CNO group, respectively;   $\epsilon$, and $\rho$ are in cgs units, $\rm erg\,g^{-1}\,s^{-1}$ and $\rm g\,cm^{-3}$, respectively; finally, the above expressions refer to situations in which all intermediate steps are at equilibrium.

   \begin{figure}
   \centering
   \resizebox{\hsize}{!}{\includegraphics{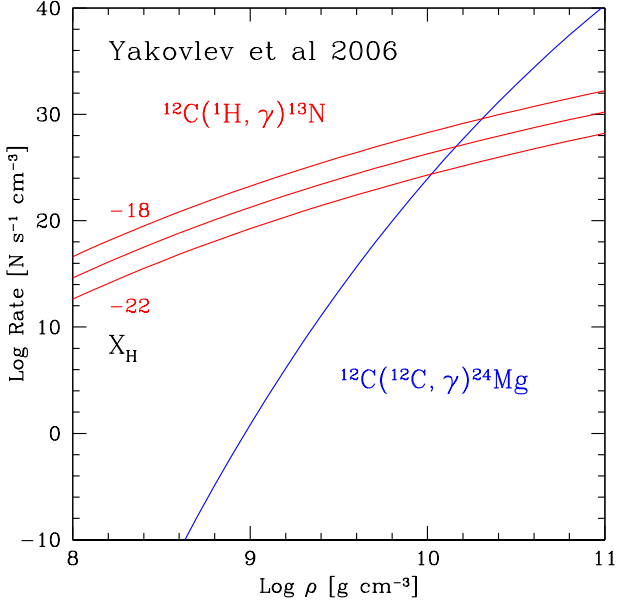}}
    \caption{We compare the rates of the pycno-nuclear reactions $\rm ^{12}C(^1H,\gamma)^{13}N$ (dashed red lines) and $\rm ^{12}C(^{12}C, \gamma)^{24}Mg$ (solid blue line) according to   the rates by
     \citet{yakovlev2006} and different values  of $\rm X_H$ as indicated. From top  to bottom $\log X_H =-18$, -20 and -22.  }
   \label{rate_xh.fig}
    \end{figure}

\begin{figure*}
\centering
{\includegraphics[width=0.4\textwidth]{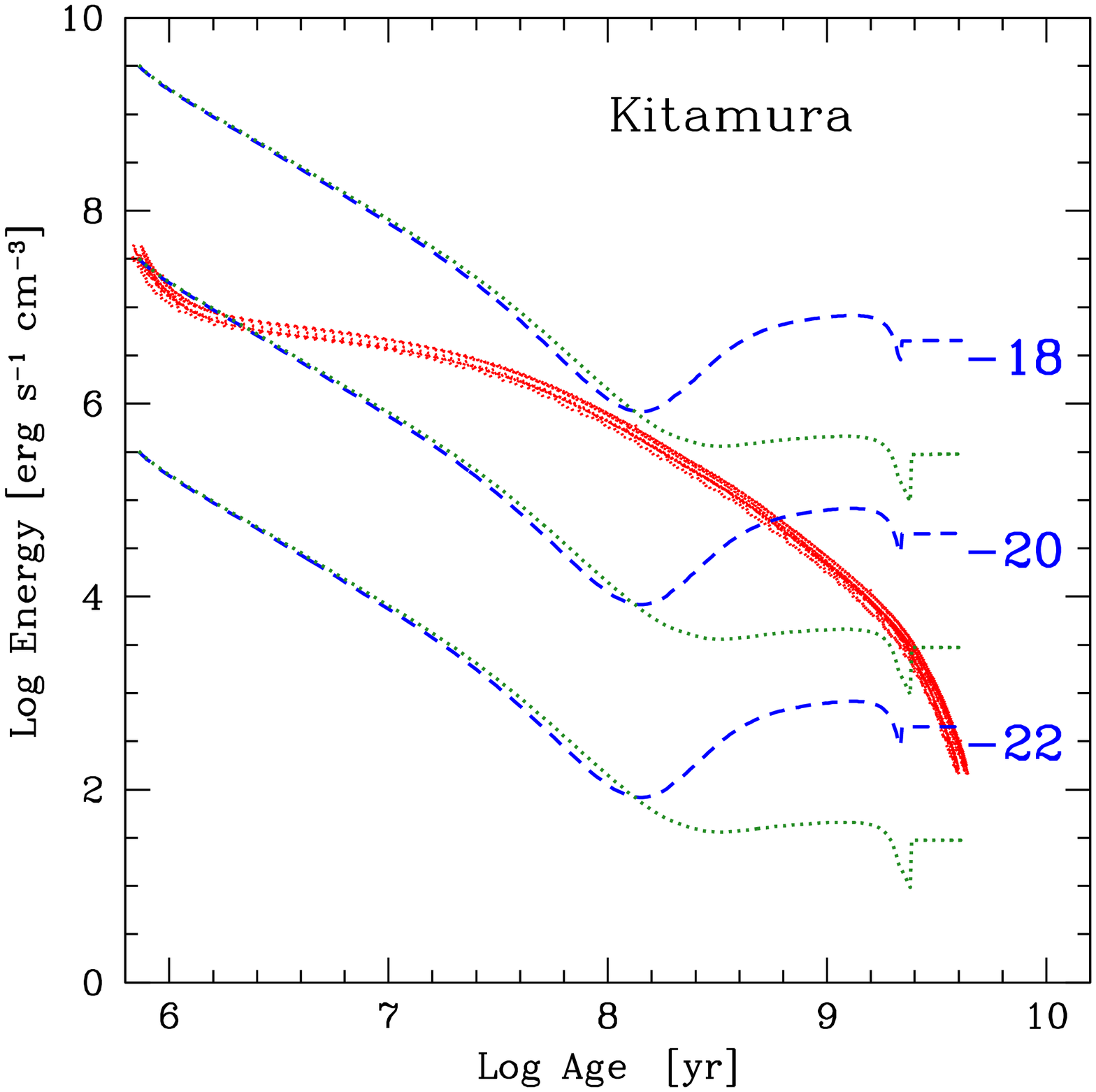}
 \includegraphics[width=0.4\textwidth]{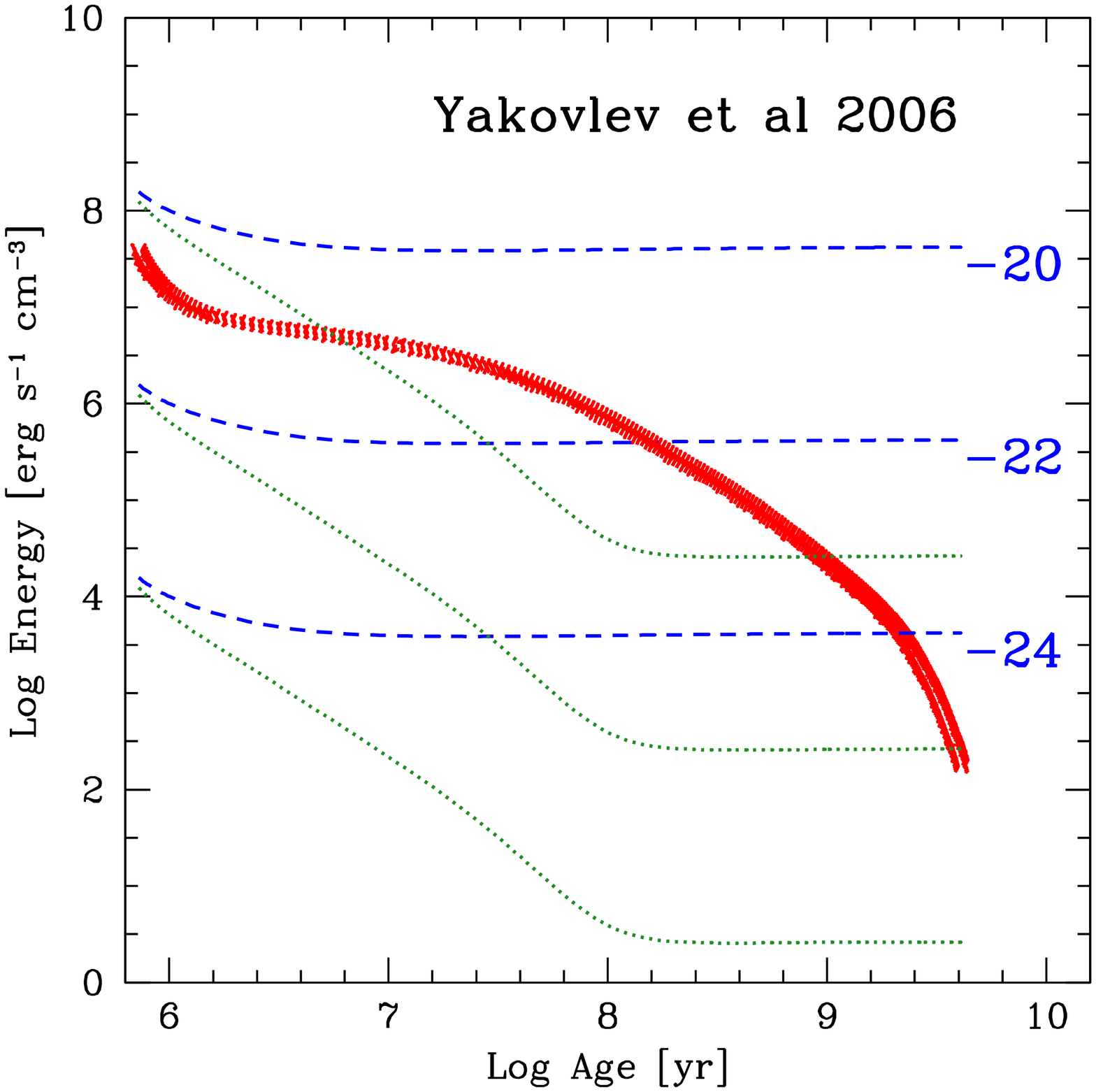}
\caption{\textbf{Left Panel}:  We plot the pycno-nuclear energy generation rates based on the \citet{kitamura2000} model  for the reaction $\rm ^{12}C(^{1}H,\gamma)^{13}N$ and three different values of hydrogen  (from $\rm X_H=10^{-18}$ to $\rm X_H=10^{-22}$) as shown by the blue continuous lines and compare these with the luminosity of the WD (red dotted line)  according to the models of \citet{althaus1997,althaus1998} for a 1.2 $M_\odot$ WD. The released nuclear energy and WD luminosity are per unit volume in units of $\rm erg\,s^{-1}\,cm^{-3}$. To do so, the luminosity of the WD is divided by the WD volume using the data provided by \citet{althaus1997,althaus1998}. The nuclear rates are calculated using the central values of temperature and luminosity of the WD model we have adopted.  The green lines represent the energy without local density correction while the blue lines represent the model that considers local density correction.
\textbf{Right Panel}:  the same as in the left panel but for the nuclear energy rates based on the  \citet{yakovlev2006} and  different values of $\rm X_H$ from  $\rm X_H=10^{-20}$ to $\rm X_H=10^{-24}$. }
 \label{evol_all.fig}       }
 \end{figure*}

(i)\textbf{Analytical}.  An estimate of the hydrogen abundance left from the nuclear burning after the main sequence
and the helium burning phase can be given analytically in the following way considering the relative weigh of pp-chain and CNO-cycle in the H-burning process.
Let $X_{H,r}$ be the hydrogen abundance at a certain time $t$.  At proceeding nuclear burning over a time interval
$\Delta t$, the fraction of consumed hydrogen is $X_{H,c} = X_{H,i}-X_{H,r}$, where $X_{H,i}$ is the initial abundance of hydrogen. Therefore $\Delta X_{H,c} = - \Delta X_{H,r}$. The fraction $X_{H,c}$ is in turn governed by the equation
 \begin{equation}
\frac{\Delta X_{H,c} \rho}{m_p}=\frac{\epsilon \rho \Delta t}{Q}
\end{equation}
where $\epsilon$ is  a function of $X_{H,r}$. In particular
for the case of the pp-chain, $\epsilon$  it is proportional to $X_{H,r}^2$. Therefore we can write in differential form
\begin{equation}
 -\frac{dx}{x^2}=A_{pp} dt
\end{equation}
where $x$ stands for the current value of $X_{H,r}$.  Upon  integration we obtain
\begin{equation}
 x_f=\frac{1}{\frac{1}{x_i}+A_{pp} \Delta t}
\end{equation}
where $x_i$ and $x_f$ are the values of $X_{H,r}$ at the beginning and at the end of the generic time step.

In the case of the CNO-cycle $\epsilon$ is a linear function in $X_{H,r}$. Therefore we have
\begin{equation}
 -\frac{dx}{x}=A_{cno} dt
\end{equation}
that gives
\begin{equation}
 x_f=x_i e^{-A_{cno} \Delta t}
\end{equation}
Taking the weighted mean value of the two channels and extending the integration over the whole time interval fro the zero age main sequence to the end of he-burning, we obtain
\begin{equation}
 \langle X \rangle =\frac{A_{pp} X_H^{pp}+A_{cno} X_H^{cno} }{A_{pp}+A_{cno}}\simeq 10^{-12}-10^{-13}
\end{equation}
for a progenitor  from 3 to $6\, M_{\odot}$. The evolutionary history of the stars comes in via the central temperature and density as functions of the time. The results for the central hydrogen content along the evolutionary sequences are shown in Table \ref{hydrogen_content}.

\begin{table}
\centering
\caption{Hydrogen abundances in the innermost regions at the end of the Main Sequence and core He-burning,  and beginning of the TP-AGB phases according to two simple models.}
\begin{tabular}{|l l l l|}
\hline
Mass & Phase& $X_H$       & $X_H$\\
     &      &  Analytical & Numerical\\
\hline
3 $M_\odot$& Hb  & $2.9\times 10^{-4 }  $  & $ 1.7\times 10^{-7 } $ \\
           & Heb & $3.6\times 10^{-9 }  $  & $ 1.3\times 10^{-14 } $ \\
           & AGB & $2.4\times 10^{-13}  $  & $ 3.3\times 10^{-18} $ \\
5 $M_\odot$& Hb  & $8.8\times 10^{-5 }  $  & $ 2.4\times 10^{-7  } $ \\
           & Heb & $2.7\times 10^{-9 }  $  & $ 4.2\times 10^{-14 } $ \\
           & AGB & $1.2\times 10^{-12}  $  & $ 7.8\times 10^{-18 } $ \\
6 $M_\odot$& Hb  & $3.7\times 10^{-5 }  $  & $ 2.8\times 10^{-7  } $ \\
           & Heb & $2.2\times 10^{-9 }  $  & $ 7.3\times 10^{-14 } $ \\
           & AGB & $1.9\times 10^{-12}  $  & $ 1.2\times 10^{-17 } $ \\
\hline
\end{tabular}
\label{hydrogen_content}
\end{table}

(ii) \textbf{Numerical}. The above result can be refined making use of eqn. (\ref{epsi_lum}),  considering that in principle the pp-chain  and CNO-cycle  can occur simultaneously, however with  significantly different efficiency, and taking into consideration the structure of the stars during their evolutionary history. In principle if one or more nuclear burning regions are present in a star, eqn. (\ref{epsi_lum}) implies
\begin{equation}
\sum_i \int_0^M \sum_j \epsilon_{i,j} dm  = \sum_i {L_i }
\label{sum_eps}
\end{equation}
where $i$ indicates the phase, and $j$ the type of burning (pp, CNO, and 3$\alpha$ as appropriate).
If only central H-burning is present (main sequence), eqn. (\ref{sum_eps}) reduces to $\epsilon_H \times M  = L_H $; when central and shell H-burnings are present it becomes $\epsilon_{H,core} \times M_{core} + \epsilon_{H,shell} \times M_{shell} = L_{H,core} + L_{H,shell} = L_H$; with more complicated nuclear stratifications and phases the generalization of eqn. (\ref{sum_eps}) is obvious.
The first case is simple to treat and $L_H$ is the true luminosity generated by core H-burning. The second case is more complicated and cannot be solved analytically for obvious reasons. Fortunately, some approximation are possible: first compared to the main sequence lifetime it is short lived and can be neglected. To this aim we have remove the contribution to the nuclear energy generation by the the shell and suitably rescaled the luminosity $L_H$ (this is possible because all details of the stellar models are to our disposal).

In the case of central core H-burning, isolating  the dependence on $\rm X_H$ in eqns. (\ref{pp_chain}) and (\ref{cno_cycle}),  we can write
\begin{equation}
\epsilon_{pp} + \epsilon_{cno} \simeq \epsilon_{pp}^o X_H^2 + \epsilon_{cno}^o X_H   = L_{H,core}/M
\label{eps_H}
\end{equation}
where the quantities with superscript $o$  are of obvious definition and are immediately known from eqns. (\ref{pp_chain}) and (\ref{cno_cycle}), and $L_{H,core}$ is the contribution to the total luminosity by the sole core H-burning. This quantity is known from the  stellar models we are using. Eqn. (\ref{eps_H}) is a second order algebraic equation to be solved for $\rm X_H$ as a function of the  temperature and density of the mass element (typically a small volume at the center) along the evolutionary sequence of a star.  The results at the end of the core H-burning, core He-burning, and beginning of the TP-AGB phase are listed in Table \ref{hydrogen_content}.

Similar reasoning can be followed to derive the helium abundance. We recast eqn. (\ref{eps_H}) as
\begin{equation}
\epsilon_{3\alpha}  = \epsilon_{3\alpha}^o X_{He}^3 = L_{He, core}/M
\end{equation}
and solve it for $X_{He}$. With this procedure we  estimate that the abundance of helium in the core at the start of  TP-AGB phase is   about $\rm X_{He}=10^{-6}$ in the three stellar models.

No estimate is made of the amounts of central hydrogen and helium consumed during the TP-AGB phase because of the complexity of the stellar evolution during this phase. However, this is less of a problem because  the TP-AGB is short-lived compared to the previous evolutionary history so that it can be neglected.

\begin{figure}
   \centering
{\includegraphics[width=0.4\textwidth]{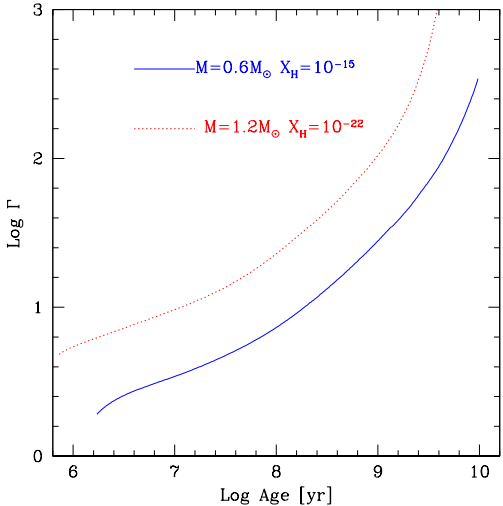} }
\caption{The Coulomb coupling parameter $\Gamma_{ij} = {Z_1 Z_2 e^2 \over a_{ij} k_B T}$ as function of age for the two extreme values of the WD masses under consideration. The underlying  nuclear reaction  is $\rm ^{1}H + ^{12}C$. The enhancement in the local density induced by the contaminant hydrogen is included (see eqn. 48 and the solution $R'_0= 0.5178 R_0$). The blue lines show the $0.6 M_\odot$ WD with $\rm X_H = 10^{-15}$, whereas the red lines are the same but for the $1.2 M_\odot$ WD with $\rm X_H = 10^{-22}$. All other combinations of mass and $\rm X_H$ abundances fall in between. The corresponding Coulomb coupling parameter with no enhancement in the local density are  obtained by scaling the values on display by the factor 1/0.5178. }
\label{gamma}
\end{figure}

\subsection{Decrease of $\rm X_H$ and $\rm X_{He}$ during the WD cooling sequence}

At the beginning of the WD cooling sequence the luminosity is sustained by CNO burning on the surface.  We want to assess whether part of the luminosity may still be due to minor nuclear burning in the interior (if it can occur at all), thus possibly further lowering  the inner content of hydrogen.
In order to follow the hydrogen and  helium consumption by nuclear burnings  during the early stages of WD cooling, we used the WD evolutionary sequences of \citet{renedo2010} in which the surface burnings (in the H and He-shells) are followed in great detail so that they can be taken  as a reference models. Using the same  procedure described above,  we estimate that the  central hydrogen and helium abundances  in our test models of 3, 5 and 6 $\, M_\odot$ during the time interval in which the surface shells are active, approximately  $\sim 4\times 10^6 \, \rm yr$,   decrease  only by a modest amount (roughly less than a factor of 10), i.e. the new abundances are $\rm X_H=10^{-19} - 10^{-18}$ and $\rm X_{He}=10^{-7}$.

\subsection{General Remarks}

To conclude, the   experiment and arguments presented above suggest that traces of hydrogen and helium  can be left over in the interiors of a WD at the end of the whole nuclear history via the thermal channels of nuclear reactions.
However it is worth emphasizing that major drawbacks and  uncertainties are present. In brief

(i) There is a large difference between the results obtained with the two methods. We prefer to consider those from the numerical approach to be more realistic because they are tightly related to detailed stellar models (temperatures, densities and luminosities).

(ii) We have used the classical reactions rates for the pp-chain, CNO-cycle and 3$\alpha$ group at the equilibrium. This prevents us from following the temporal history of the  abundances of the intermediate elements and reactions. To highlight the issue, if  some hydrogen survives the stage of core H-burning, we  would  expect that  any trace of  hydrogen   should  be completely burnt during  the helium  core burning, as  a result of  the reactions $\rm ^{12}C+H {\rightarrow}  ^{13}C$ and $\rm ^{13}C+H {\rightarrow} ^{14}N$.  Eventually, all $\rm ^{14}N$ should be converted into  $\rm ^{22}Ne$ via  $\alpha$-captures.   This  sequence of  events cannot  be described correctly  by our  CNO-cycle equilibrium rates. This is perhaps the point of major uncertainty.

(iii) Detailed evolutionary models would be the right way of assessing whether traces of light elements could be present in the interiors of WDs at the beginning of their cooling sequence and, if so, whether there exists any trend and/or lower limit at varying the initial mass of the progenitor star.

(iv) An alternative to  calculating complete stellar models could be the ideal case of an elementary cell of matter in which extended networks of nuclear reactions among  a  number of elementary species are let occur as a function of temperature,  density and patter of initial chemical parameters. To simulate the centre of a real star, temperature and density as a function of time could be taken from detailed stellar models.

\noindent
Work is in progress along the lines of items (iii) and (iv).

Despite the above limitations of the present approach and waiting for careful numerical investigations, we consider the above estimates of   $\rm X_H$ and $\rm X_{He}$  as  \textsl{free parameters} within the  ranges we derived above, and proceed to investigate the effects they would induce on the evolution of WDs of different mass.

\begin{figure}
 \centering
{\includegraphics[width=0.4\textwidth]{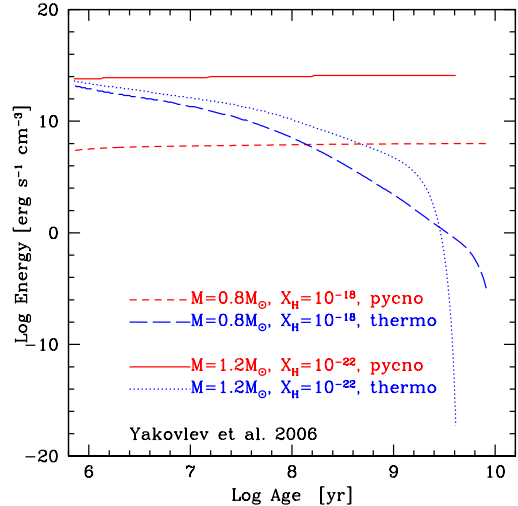} }
\caption{The nuclear energy rates $R_{ij}(\rho,T)=R_{ij}^{pyc}(\rho)+\Delta R_{ij}(\rho,T)$ of  \citet{yakovlev2006}. The first term is the pycno contribution (short dashed and solid red lines) and the second one the thermal component (long dashed and dotted blue lines).
Two values of the WD mass and two values of $\rm X_H$ are displayed, i.e. 0.8$M_\odot$ with $\rm X_H=10^{-18}$ and 1.2 $M_\odot$ with $\rm X_H=10^{-22}$.     The energy generation rates  are in $\rm erg\, cm^{-3}\, s^{-1}$.}
   \label{therm_pycn}
      \end{figure}

\section{Results for the  Kitamura and Yakovlev reaction rates}\label{results_Kita_Yako}

The \citet{kitamura2000}, \citet{gasques2005} and \citet{yakovlev2006} formalisms \citep[see also][]{Beard2010} were  originally tailored for reactions  like $ \rm ^{12}C+^{12}C$,
$\rm ^{12}C+^{16}O$, $\rm ^{16}O+^{16}O$. However, they can be extended to   reactions involving impurities  of light elements, and to incorporate the enhancement in the local density. We have used the reaction rates of \citet{kitamura2000}  to study the reactions $\rm ^{1}H + ^{12}C$ and/or $\rm ^{4}He + ^{12}C$. We will not examine here the similar reactions occurring with  oxygen  because of the higher atomic number $Z$. The relative $S$ factors and $Q$ values of the two reactions are listed in Table \ref{tab5.tab}. The abundance of carbon in these test calculations is $\rm X_C \simeq 1$.  The results are more general than those of the \citet{SalpeterVanHorn69} rates because they take also into account  the effects of the temperature, extending from   the thermal  to the pycno-nuclear regime.  However, our results match those  obtained with the \citet{SalpeterVanHorn69} rates
in the low temperatures and high densities regime. Results for the  $\rm ^{12}C(^{1}H, \gamma)^{13}N$ reaction with $\rm X_H=10^{-20}$ are presented in  left panel of Fig. \ref{energy_ks_gs.fig}, in which the generated energy  is plotted as a function of central density.
The same procedure has been applied to the \citet{yakovlev2006} formalism. The results for the same reaction and hydrogen abundance are shown in the right panel of Fig. \ref{energy_ks_gs.fig}. In both cases, four different temperatures  have been considered.

In both panels the thermal branches are visible only for the $10^7$ and $10^8$ K temperatures. All curves merge together beyond a certain value of the density that depends on the temperature and source of the rates.  The two groups of rates significantly differ both in the thermal and pycno-nuclear regime: the \citet{kitamura2000} rates are higher in the thermal and lower in the pycno-nuclear regime compared to those \citet{yakovlev2006}. The rates are enhanced by the increase in the local density. This dependence  is shown  by the results  in Fig. \ref{rate_xh.fig} which displays the $\rm ^{1}H +^{12}C$ and $\rm ^{12}C + ^{12}C$ reaction rates, limited to the case of \citet{yakovlev2006}, for three values of $\rm X_H$ as indicated.  These results can be compared to those
of Fig. \ref{energy_sh.fig}, scaled to the same energy units and  multiplied by the $Q$-value. The energy generation rate exceeds the  mean luminosity of a WD at relatively low  densities.

\begin{figure}
   \centering
{\includegraphics[width=0.4\textwidth]{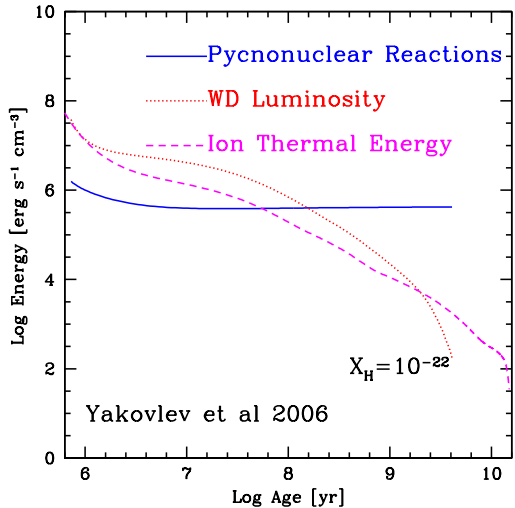} }
    \caption{The pycno-nuclear energy rates based on the \citet{yakovlev2006} model for an abundance concentration     of hydrogen $\rm X_H=10^{-22}$ (blue continuous line) compared to the luminosity of the WD
    (red dotted line) according to the models of \citet{althaus1997,althaus1998} for a 1.2 $M_\odot$ WD. The dashed magenta line represents the energy of the ions. The luminosity is in the same units as the energy generation rates ($\rm erg\, cm^{-3}\, s^{-1}$).}
   \label{cool_lum_triangle.fig}
      \end{figure}

\section{Including the real evolution of a WD}\label{evolution}
WDs of given masses follow cooling sequences along which the luminosity varies with time. Therefore we relax now the simplifying assumption of a constant mean luminosity  by using the evolutionary sequences of WDs of different mass and compare them with the energy released by nuclear burning.
The  cooling sequence in the luminosity vs effective temperature plane is determined by the central density and mean molecular weight of the electrons (i.e. the chemical composition of the WD). Therefore the really meaningful comparison is between the energy production of the WD governed by its temperature and density (the central values) and its current luminosity.

\textbf{Taking the WD structure into account}.  In our analysis we need to compare the total luminosity with the total nuclear energy generation inside ($L \equiv \int \epsilon dM $) along the cooling sequence. To this aim we need the internal structure of  the WD. The cooling sequence provides the age from the beginning of the WD phase, the luminosity, effective temperature,  and the  central temperature and density. At each stage we integrate the structure equations of a WD of given mass, chemical composition, and central density in the so-called zero-temperature approximation. This provides the stratification of mass and density as functions of the radius. Owing to the high thermal conductivity of degenerate electrons (see below), the temperature tends to become more and more uniform from the center to the surface at increasing age of the WD (lowering of the mean temperature by cooling). In order to evaluate the reaction rate as a function of the position it is sufficient to assume that the temperature throughout the star is equal to its central value. This may not be correct toward the surface, but  the reaction rates are nearly zero over there  because of the low density (the nuclear energy production, if any takes place, is mostly limited to the central regions).  Therefore we assume that the rate of energy production along the radial distance is $\epsilon[T_c,\rho(r) ]$, and proceed to calculate the  mean energy production per unit volume,    $\langle {\epsilon  }\rangle$, and  unit mass,   $\langle {\epsilon \over \rho} \rangle$. The luminosity per unit volume of a WD is given by
\begin{equation}
 {L \over V} = {4 \pi R^2 \sigma T_{eff}^4 \over V}
 \end{equation}
where $V$ is the total volume of the WD. This luminosity must be compared with

\begin{equation}
\langle {\epsilon  }\rangle = { \int_0^R \epsilon[T_c,\rho(r) ] \times 4 \pi r^2 dr \over
                           \int_0^R 4\pi r^2 dr }
\end{equation}
in $\rm erg\, cm^{-3}\, s^{-1}$.

In addition to this we will make use of the energy generated per unit mass and time given by
\begin{equation}
\langle {\epsilon \over \rho} \rangle = { \int_0^R \epsilon [T_c, \rho(r)]  \rho^{-1} \times  4 \pi r^2 dr \over
                           \int_0^R 4\pi r^2 dr }
\end{equation}
in $\rm erg\, g^{-1}\, s^{-1}$.

These quantities can be expressed by those  evaluated at the center $[\epsilon]_c$  and $[{\epsilon\over \rho}]_c$
by means of the relations $\langle \epsilon \rangle = [\epsilon]_c (V_c/V)$ and $\langle {\epsilon \over \rho} \rangle = [\epsilon]_c \times (V_c/V)$,
where $V_c$ is the volume of a small sphere of  radius $R_c$ around the center. The radius $R_c$ is found to range from $6.5 \times 10^7$ to $2.5 \times 10^7$ cm when the WD increases from 0.6 to 1.2 $M_\odot$ mass. This radius  is about a factor 1/20 smaller than the real radius  of WDs of the same mass. This means that the nuclear energy is mainly generated in a small sphere around the center, thus
justifying the approximation $\langle \epsilon \rangle \simeq \epsilon_c \times (V_c/V)$ and  $\langle {\epsilon / \rho } \rangle \simeq (\epsilon_c/ \rho_c) \times (V_c/V)$.

\textbf{Results}. For the sake of illustration, let us consider  the evolutionary sequence of a CO-WD  of 1.2 $M_\odot$ (the border value above which the mean density is high enough to enter the General Relativity domain) calculated by  \citet{althaus1997,althaus1998}. All necessary information  about central temperature, central density and luminosity, etc is available. The composition of the WD is $\rm X_C \simeq 0.5$ and $\rm X_O \simeq 0.5$.  We are particularly interested in determining  when the energy rates given by   \citet{kitamura2000} or by \citet{yakovlev2006},  modified for the effects of contaminants on the local densities,  exceed the luminosity.  Fig. \ref{evol_all.fig}  shows the variation of the WD luminosity (the red thick dotted curve) as a function of time and the energy generated by the reaction $\rm ^{12}C(^{1}H,\gamma)^{13}N$ for different abundances $\rm X_H$ and keeping   $\rm X_C$ constant (the thin dashed blue and and the thin dotted green curves). The dotted green lines are for nuclear rates that neglect the local density enhancement caused by impurities, whereas the dashed blue lines are for nuclear rates that take this into account.

There are a number of important points to note:

(i) When the enhancement in local density is not considered, the nuclear rates show a similar trend,  they initially decrease,  but at given WD age  they become flat (see the green curves in both panels of Fig. \ref{evol_all.fig}). The little kick in the \citet{kitamura2000} rates is a numerical artefact  in his relations passing from fluid to solid state. The initial decrease of the nuclear rate is the signature of the thermal regime and that the main source of energy is the internal energy of the ions, whereas the levelling of  the rate is when the pycno-nuclear regime takes over.

(ii) When the enhancement in local density is taken into account, this  trend is  only typical of  the \citet{kitamura2000} rates. This is partially because we used a linear interpolation to calculate the  interionic distance in transiting  from the low-density, thermal regime to the high-density pycno-nuclear regime (see eqn. \ref{aij_int}). The nuclear rates of \citet{yakovlev2006}  are nearly flat all over  the age range. The pycno-nuclear regime dominates the rate from the very beginning. This is partially due to the mass we have  chosen in which the density is very high ($\rm \rho_c \simeq 1.8 \times 10^8\, g \, cm^{-3}$ for the $1.2\, M_\odot$ WD). For lower masses the previous trend is recovered.

(iii) For the \citet{kitamura2000} model the pycno-nuclear energy exceeds the luminosity of the WD for  $\rm X_H=10^{-21}$ after 6 Gyr (see the left panel of Fig. \ref{evol_all.fig}). Assuming instead the \citet{yakovlev2006} rates, we reach the critical situation  for  $\rm X_H=10^{-22}$ even before 1 Gyr (see the right panel of Fig. \ref{evol_all.fig}).

To clarify the above issues first we examine the variation of the Coulomb coupling parameter $\Gamma_{ij}$ along the cooling sequences of the WD models. $\Gamma_{ij}$ for the $\rm ^1H + ^{12}C$ reaction is shown in Fig.\ref{gamma}. It is soon evident that the reaction occurs in the third and fourth regimes   of the group of five that have been discussed in Section \ref{fiveregimes} as long as the age is younger than 1 Gyr for the $1.2\, M_\odot$ and 8 Gyr for the $0.6\, M_\odot$, respectively. More precisely, initially the nuclei are bound to the lattice sites, so that the reaction occurs between highly thermally excited nuclei, which oscillate with frequencies higher than the plasma frequency and have energies greater than the zero point energy of the plasma; later they enter the so-called thermally enhanced pycno-nuclear regime but the melting temperature is not reached yet. Only WDs  older than the above limits  enter the pure pycno-nuclear regime.

Second we look at the    separate contributions to the total rate by the thermally enhanced  and pure pycno-nuclear regimes  according to the \citet{yakovlev2006} formalism.
In  eqn. (26), the total rate is made by the sum of two terms
$R_{ij}(\rho,T)=R_{ij}^{pyc}(\rho)+\Delta R_{ij}(\rho,T)$, where the first term is the pycno contribution and the second one the thermal component. The evaluation is made for  two values of the WD mass and two values of $\rm X_H$, i.e. 0.8$M_\odot$ with $\rm X_H=10^{18}$ and 1.2 $M_\odot$ with $\rm X_H=10^{-26}$. The results are shown in  Fig. \ref{therm_pycn}, where the blue long dashed and dotted curves are for the thermally enhanced component,  and the red short dashed and solid curves are for the pure pycno component, respectively.

As already anticipated, the contribution from the thermally enhanced channel is significant for the $0.8 M_\odot$ WD, it gradually decreases as the WD mass (density) increases, and is fully masked by the pycno-nuclear one for the 1.2 $M_\odot$ star and beyond.

\subsection{Adding up all sources of energy:  critical $\rm X_H$ abundances and ages}\label{sources_energy}
We  now include all energy sources, i.e.  the generation by any of the five possible regimes for the pycno-nuclear reactions (including also the so-called thermally enhanced terms in the early stages of WD cooling sequences) and  the thermal energy of the ions.

\textbf{The proto-type WD of $1.2 M_\odot$}. Following \citet{yakovlev2006} we consider only the  $\rm ^{12}C(^{1}H,\gamma)^{13}N$  reaction,  activated by the traces of hydrogen left over by the previous phases and still present in the WD.   This is one of the three starting reactions that ignite the CNO-cycle, which however cannot be completed, owing to the extremely low abundances of the intermediate elements. This reaction has a $Q$-value of 1.94 MeV compared to the 25.02 MeV when the cycle is completed, releasing a factor of 12.5 less energy.    The nuclear energy release, the ion internal energy, and  the comparison luminosity  are all expressed per unit volume of the WD. The volume is calculated from the luminosity and effective temperature of the WD along its cooling sequence. All quantities are provided by \citet{althaus1997,althaus1998} and \citet{renedo2010}.

\begin{figure*}
\centering
{\includegraphics[width=5.5truecm,height=7.0truecm]{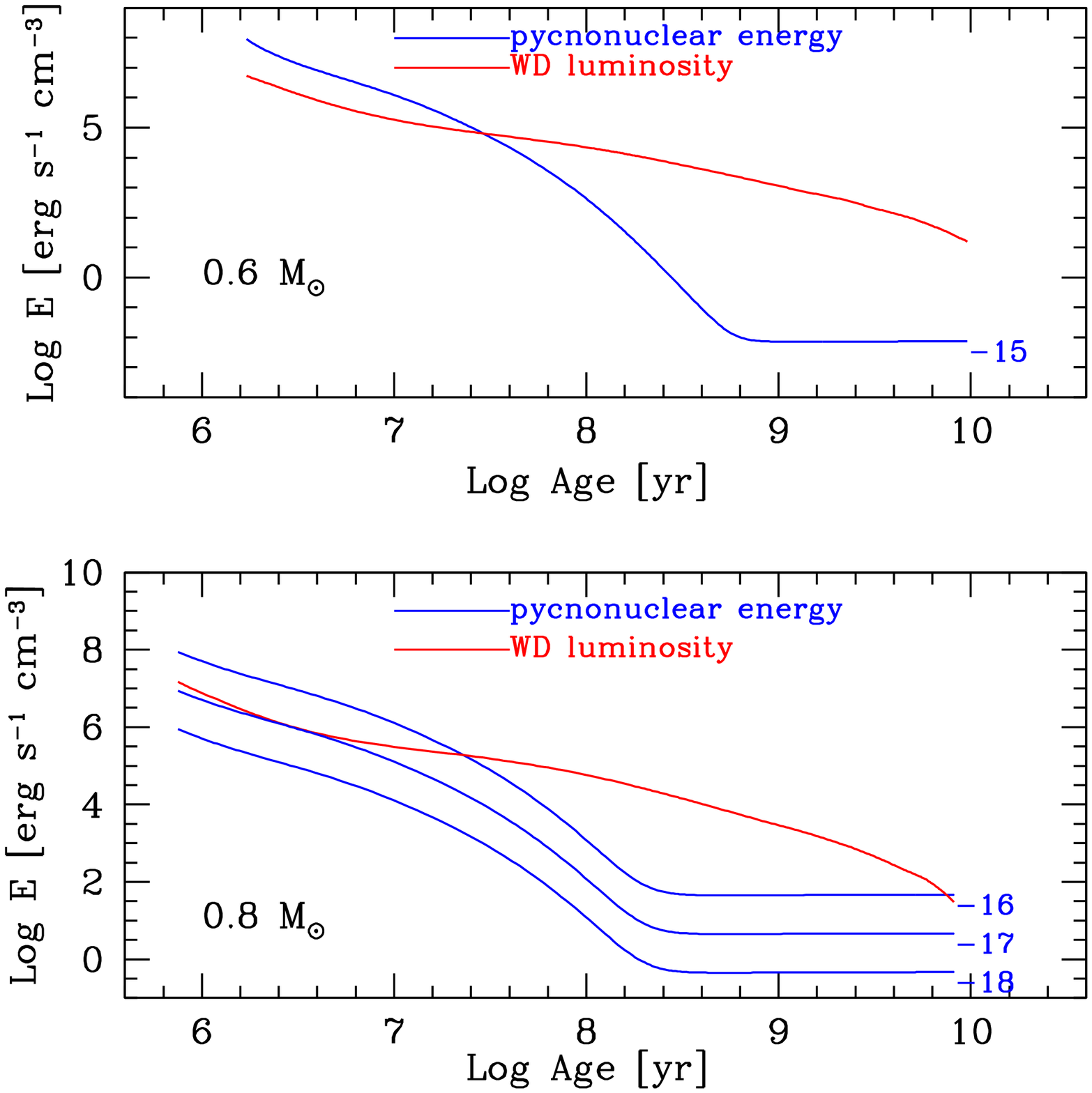}
 \includegraphics[width=5.5truecm,height=7.0truecm]{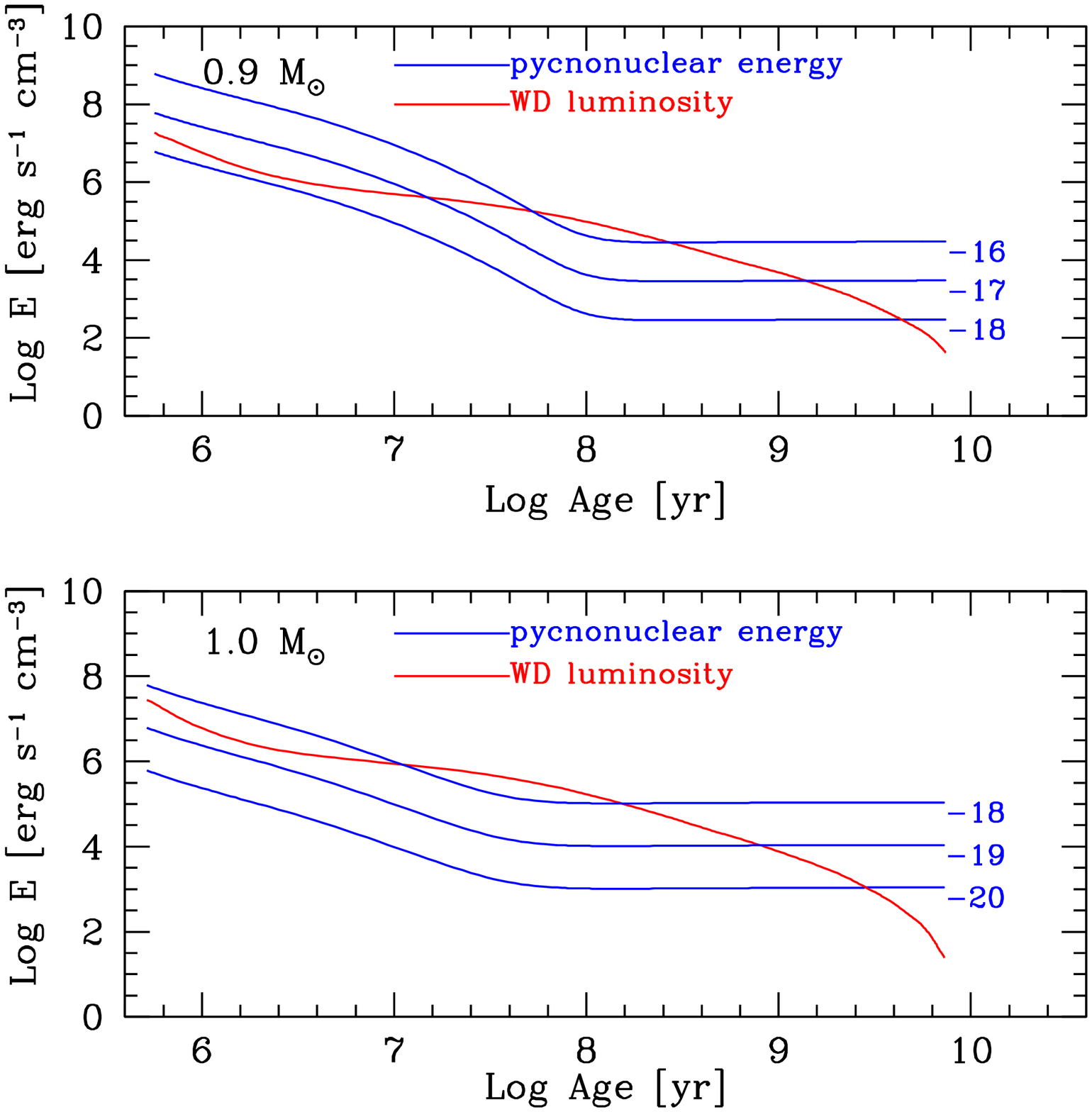}
 \includegraphics[width=5.5truecm,height=7.0truecm]{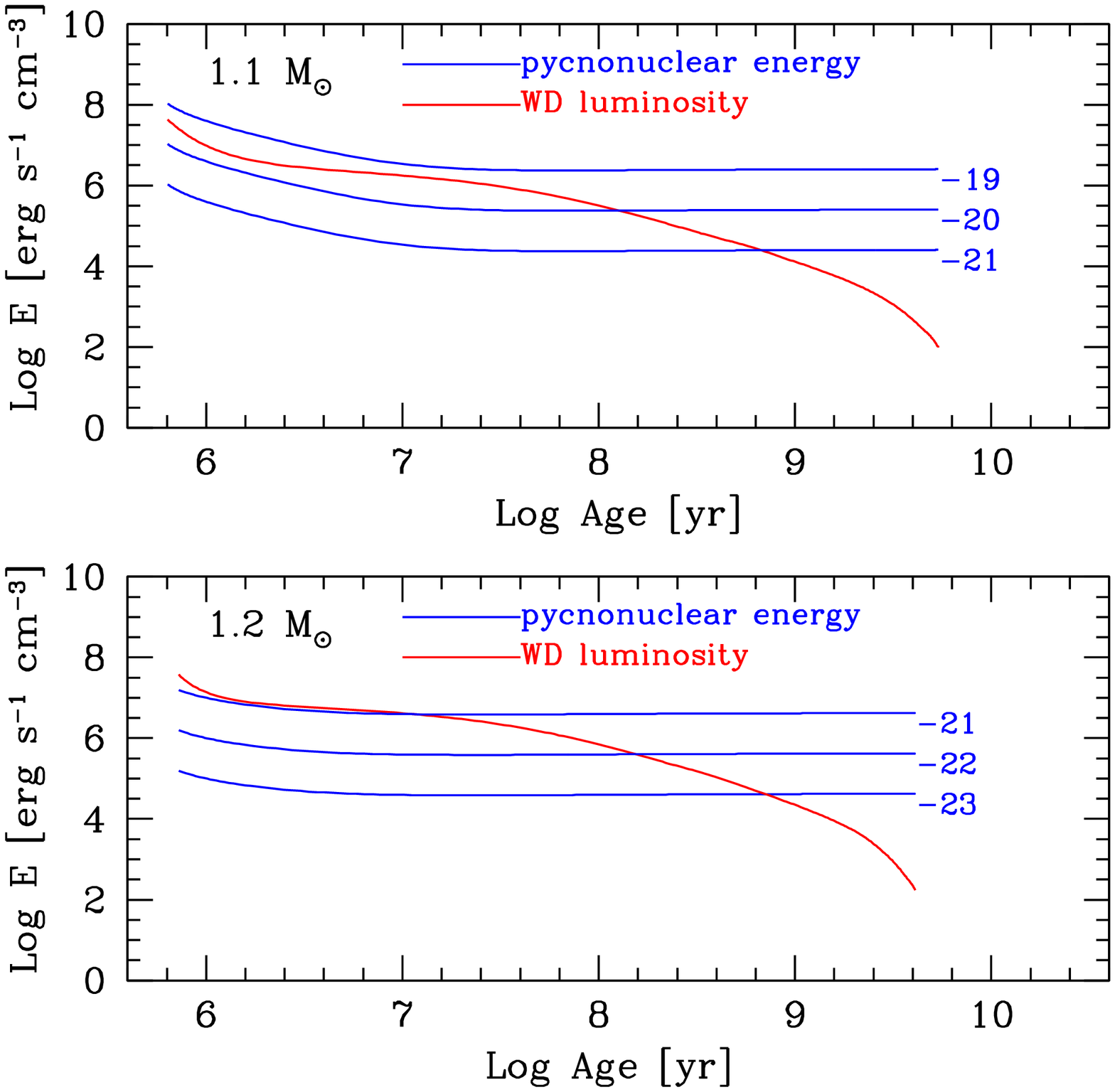}   }
\caption{Energy rates per unit volume produced by   hydrogen impurities with different  abundances of $\rm X_H$ as indicated for WD of different mass. The reaction on display is the  $\rm ^{12}C(^1H,\gamma)^{13}N$ according to  \citet{yakovlev2006}. The luminosity is in the same units as the energy generation.  }
\label{allmass123.fig}
\end{figure*}

\begin{table*}
\centering
\caption{Logarithm of the age in yr at which the energy generation $\langle \epsilon / V \rangle  $ by the nuclear reaction $\rm ^{12}C(^1H, \gamma)^{13}N$ in $\rm erg\,cm^{-3}\,s^{-1}$ crosses the WD luminosity  in $\rm erg\,cm^{-3}\,s^{-1}$ for different $M_{WD}$ in $M_\odot$ and hydrogen abundances $\rm X_H$. Ages in italic corresponds to very early intersections of the two curves when the temperature is very high and neutrino cooling is very efficient. In these cases, the nuclear energy is most likely carried away and radiated by the star. The other ages in roman correspond to cases in which  the energy produced by nuclear reactions could be trapped inside the WD because the energy production by nuclear reactions overwhelms the luminosity and neutrinos have become much less efficient in removing the energy.   The luminosities and radii (volumes) of the WDs are from the cooling sequences of \citet{althaus1997,althaus1998}. }	
\begin{tabular*}{104.5mm}{|l| c   c  c c  c c  c c     c|}
\hline
\multicolumn {1}{|l|}{} &
\multicolumn{8}{c}{$\log X_H$}&
\multicolumn{1}{c|}{}      \\
$M_{WD}$&    -15   & -16      & -17         & -18       & -19  & -20 & -21 & -22 & -23  \\
\hline
0.6     &    {\it 7.5}   &          &             &           &      &     &     &     &      \\
0.8     &          & {\it 7.4}; 9.8 & {\it 6.5}; $>$10  & $>$10     &      &     &     &     &      \\
0.9     &          & {\it 7.6}; 8.4 & {\it 7.2}; 9.2    & 9.8       &      &     &     &     &      \\
1.0     &          &          &             & {\it 7.0}; 8.2  & 8.8  & 9.6 &     &     &      \\
1.1     &          &          &             &           &      & 8.2 & 8.8 &     &      \\
1.2     &          &          &             &           &      &     & 7.0 & 8.2 & 8.9  \\
\hline
\end{tabular*}
\label{tab_age_xh.tab}
\end{table*}

The temporal evolution of the WD luminosity (red dotted line), nuclear energy release (blue solid line) and thermal energy of the ions (magenta dashed line) are shown in Fig. \ref{cool_lum_triangle.fig}.   Since the ion internal energy is derived here from simple expressions, neglecting the effect of degenerate electrons and the variation of the degrees of freedom in the crystal lattice at very low temperatures, we do not extend it to very high densities or low temperatures, characteristic of  very old ages.  First of all, we notice that the abundance $\rm X_H$ of the residual hydrogen cannot exceed $10^{-21}$, otherwise the nuclear energy release by the sole $\rm ^{12}C(^{1}H,\gamma)^{13}N$ reaction during  the initial stages of WD cooling is comparable to or even exceeds the total luminosity, thus bringing the WD to a risky regime at which a nuclear runaway may start.  See also  \citep[][]{renedo2010} for the energy production by the surface CNO during the same phase. We will see below that the upper limit to $\rm X_H$ to avoid early nuclear runaway changes  with the WD mass, it increases at decreasing mass. Incidentally, this constraint may be a way of estimating and calibrating the maximum content of residual hydrogen in a CO-WD.

As cooling proceeds, the energy generation by nuclear reactions in the thermally-enhanced regime decreases  and the  light contaminants do not have any effect until  the pure pycno-nuclear channel begins.
In the case of the $1.2\, M_\odot$ star with the hydrogen abundance of  $\rm X_H=10^{-22}$, the  nuclear energy production   exceeds the luminosity   at the age of about $2.5\times 10^8$ as shown in Fig. \ref{cool_lum_triangle.fig}.   $\rm X_H$ greater than  the above value would lead the WD to intersect  the luminosity curve during the thermally enhanced phase, i.e. soon after (or shortly later) the formation of the WD itself (a possibility that cannot be firmly excluded and is temporarily left aside, see below).

\textbf{Exploring the solution space}. Since the abundances $\rm X_H$ cannot be assessed a priori, it is safe to consider them as free parameters whose values fall within a plausible range  bounded by (a) the maximum value above which the energy production exceeds the WD luminosity during the early stages of the thermally enhanced regime, (b) the lower value below which the energy production equals  the WD luminosity at ages older than the age of the Universe. This grid of possible solutions is shown in the various panels of   Fig. \ref{allmass123.fig} and the corresponding ages of intersection are listed in Table \ref{tab_age_xh.tab} as function of the WD mass and $\rm X_H$. It is worth noting here that intersections occurring during the earliest stages of the cooling sequences are unlikely because cooling by neutrinos is very efficient so that a release of nuclear energy per unit volume and time exceeding the luminosity per unit volume does not trigger any potential instability but simply the energy excess is carried out by radiation and neutrinos. In contrast, the intersections occurring at later stages when the pycno-nuclear regime is already in place and neutrino cooling is over are much more interesting because they can potentially lead to unstable situations.

As expected, at given $M_{WD}$ the intersection occurs later with decreasing $\rm X_H$. For values lower than those indicated, the intersection   occur at ages much older than the current age of the Universe (and thus of  little interest here).
We note that the intersection occurs  at lower and lower abundances and younger ages with increasing  WD mass.
In fact,  the luminosity per unit volume (evaluated at the typical age of $10^9$ yr) increases by a factor of ten from  the 0.8 to the 1.2 $M_\odot$ WD.  Also the central and the mean density of a WD  increase with the mass, and so does the efficiency of the pycno-nuclear reactions. Consequently, the values of $\rm X_H$  for which the intersection may occur within the age of the Universe systematically decrease with increasing WD mass:  $\rm X_H \simeq 10^{-18}$ for the
$0.8 M_\odot$ star and $\rm X_H \simeq  10^{-23}$ for the $1.2 M_\odot$.

Finally, we note that the systematic decrease of  the "permitted" values of $\rm X_H$ at increasing WD mass is compatible with the past thermal history of the progenitor star. In fact,  the mean temperature in a low mass progenitor is lower than the mean temperature in a more massive one, clearly having effects on   nuclear burning.  For this reason   it is tempting to suggest that  $\rm X_H$ decreases with  increasing  WD mass.

Concluding this section, we remind the reader that the rates of energy generation as function of  age have been calculated at constant $\rm X_H$  and $\rm X_C$. In reality, $\rm X_H$ (and $\rm X_C$) should decrease with time.  Therefore a real WD should move along a path in the panel corresponding to its mass gradually shifting to lines of decreasing $\rm X_H$. Consequently, the age at which the liberated energy would equal  the luminosity cannot be exactly determined without the aid of real models of WD stars including the energy generated by the reactions between contaminants and the carbon (oxygen) nuclei. The ages reported in Table \ref{tab_age_xh.tab}  are merely indicative of the expected trend.

\begin{table*}
\centering
\caption{
Columns (1) through (11) are:  (1) $M_{WD}$ the WD mass in $M_\odot$; (2)  $\log X_H$; (3) $\log Age$ in yrs; (4) the central density  $\rho_c$  in $g\, cm^{-3}$; (5) the central temperature $T_c$ in K; (6) the reaction rate $R_{H,C}$ in $n\, s^{-1}\, cm^{-3}$; (7) the energy generation rate  $\epsilon$ in $\rm erg\,  cm^{-3}\, s^{-1}$; (8)  the energy generation rate ${\epsilon_\lambda \over \rho }$ per gram in the elementary volume $\lambda^3$  in $erg\, g^{-1}\, s^{-1}$; (9) the energy difference $\Delta E=$  WD luminosity per unit volume - nuclear energy generation per unit volume in $erg\, cm^{-3}\, s^{-1}$. The negative sign in brackets and  in front of $\log |\Delta E|$ means that $\Delta E$ is negative; (10) the time in sec required to process all [H+C]-reactions in a gram of matter; (11) the total opacity $\kappa \simeq \kappa_c$ in $cm^2\, g^{-1}$. The mean free path of thermal conduction is $\lambda= 1/(\rho\kappa)$. The reaction  and energy generation rates per unit mass and time are $R_{H,C}/\rho$ and $\epsilon/\rho$, respectively. The Q-value of the [H+C]-reaction is  $3.108 \times 10^{-6}$ erg/reaction. The threshold value for C-burning $\epsilon_{CC} = 10^2 erg\,  g^{-1}\, s^{-1}$.  Finally, $\epsilon_\lambda \over \rho $ is expected to be equal to $\epsilon_{CC}$.
}	
\begin{tabular*}{164.2mm}{ |l|  c c c  c  r r  c r c  c| }
\hline
          &         &          &         &     &         &          &         &     &     &      \\
$M_{WD}$ &
         $ \log X_H$ &
                    $\log Age$ &
                               $\log \rho_c$  &
                                         $\log T_c$ &
                                               $\log R_{H,C}$ &
                                                         $\log \epsilon$ &
                                                                    $\log {\epsilon_\lambda \over \rho }$ &
                                                                             $\log |\Delta E|$ &
                                                                                        $\log Time$  &
                                                                                               $\log \kappa$  \\
          &    &    &         &    &    &            &   &   &    &      \\
\hline
0.6&    -14&   8.35177&   6.55997&   7.19447&   7.10141&   1.59388&   1.92175&   (-)3.94112&   7.95889&  -4.33077\\
0.6&    -15&   8.24095&   6.55871&   7.25346&   6.79210&   1.28457&   1.99387&   (-)4.07929&   7.25004&  -4.20843\\
0.6&    -16&   8.13146&   6.55737&   7.30867&   6.43479&   0.92725&   1.99437&   (-)4.20756&   6.59058&  -4.09341\\
0.6&    -17&   8.02755&   6.55603&   7.35761&   5.99829&   0.49075&   1.87563&   (-)4.31829&   6.01263&  -3.99095\\
0.6&    -18&   7.89861&   6.55429&   7.41307&   5.62932&   0.12179&   1.86722&   (-)4.44405&   5.36593&  -3.87426\\
0.6&    -19&   7.75130&   6.55223&   7.47019&   5.26896&  -0.23858&   1.87873&   (-)4.57408&   4.71112&  -3.75329\\
0.6&    -20&   7.56597&   6.54952&   7.53408&   4.96728&  -0.54026&   1.99383&   (-)4.72370&   3.99713&  -3.61688\\
0.6&    -21&   7.38734&   6.54662&   7.58990&   4.56002&  -0.94752&   1.95181&   (-)4.86916&   3.39188&  -3.49641\\
0.6&    -22&   7.16977&   6.54240&   7.65074&   4.18331&  -1.32423&   1.97544&   (-)5.06910&   2.75617&  -3.36288\\
\hline
0.8&    -14&   8.52201&   7.02594&   7.09378&   9.17690&   3.66936&   1.97868&   (-)3.93600&   6.45311&  -5.27969\\
0.8&    -15&   8.31731&   7.02427&   7.21050&   8.39026&   2.88273&   1.95233&   (-)4.37791&   6.19206&  -5.04049\\
0.8&    -16&   8.18340&   7.02304&   7.28099&   7.81197&   2.30443&   1.86132&   (-)4.55444&   5.71492&  -4.89531\\
0.8&    -17&   8.03733&   7.02158&   7.35189&   7.42634&   1.91880&   1.95147&   (-)4.72227&   5.06044&  -4.74862\\
0.8&    -18&   7.89409&   7.02008&   7.41467&   7.02794&   1.52040&   1.96527&   (-)4.86833&   4.43377&  -4.61807\\
0.8&    -19&   7.74364&   7.01846&   7.47349&   6.60320&   1.09566&   1.92477&   (-)5.00203&   3.83823&  -4.49514\\
0.8&    -20&   7.56920&   7.01658&   7.53337&   6.18817&   0.68063&   1.90075&   (-)5.13687&   3.23409&  -4.36931\\
0.8&    -21&   7.35847&   7.01436&   7.59534&   5.78730&   0.27976&   1.90494&   (-)5.27617&   2.61641&  -4.23826\\
0.8&    -22&   7.11746&   7.01181&   7.65611&   5.36457&  -0.14296&   1.87995&   (-)5.41917&   2.02231&  -4.10874\\
\hline
0.9&    -14&   9.07445&   7.26987&   6.74175&  11.97203&   6.46449&   1.96448&  (+)6.46393&   3.94219&  -6.36532\\
0.9&    -15&   8.83613&   7.26875&   6.90389&  10.96806&   5.46052&   1.94121&  (+)5.44860&   3.94616&  -6.03731\\
0.9&    -16&   8.54620&   7.26701&   7.07901&   9.96173&   4.45419&   1.99767&  (+)3.94001&   3.95236&  -5.68135\\
0.9&    -17&   8.27756&   7.26491&   7.23405&   8.96598&   3.45844&   1.94997&   (-)4.62444&   3.94478&  -5.36434\\
0.9&    -18&   8.01843&   7.26247&   7.36559&   8.10553&   2.59800&   1.93500&   (-)4.95998&   3.76529&  -5.09338\\
0.9&    -19&   7.81745&   7.26038&   7.45365&   7.49842&   1.99088&   1.94260&   (-)5.16469&   3.29955&  -4.91068\\
0.9&    -20&   7.63707&   7.25849&   7.52097&   6.98654&   1.47901&   1.89308&   (-)5.31693&   2.76507&  -4.77012\\
0.9&    -21&   7.42745&   7.25636&   7.58690&   6.53473&   1.02719&   1.88136&   (-)5.46365&   2.18577&  -4.63166\\
0.9&    -22&   7.17789&   7.25393&   7.65287&   6.10282&   0.59528&   1.88488&   (-)5.60437&   1.59335&  -4.49220\\
\hline
1.0&    -14&   9.39692&   7.53170&   6.41961&  14.54245&   9.03491&   1.92506&  (+)9.03491&   1.65015&  -7.40431\\
1.0&    -15&   9.24047&   7.53132&   6.58673&  13.54068&   8.03315&   1.92948&  (+)8.03313&   1.65191&  -7.06854\\
1.0&    -16&   9.04733&   7.53067&   6.75247&  12.53808&   7.03054&   1.92630&  (+)7.03028&   1.65451&  -6.73474\\
1.0&    -17&   8.80496&   7.52952&   6.92807&  11.53392&   6.02639&   1.98335&  (+)6.02026&   1.65867&  -6.37987\\
1.0&    -18&   8.52880&   7.52782&   7.09582&  10.52770&   5.02016&   1.99405&  (+)4.84002&   1.66488&  -6.03919\\
1.0&    -19&   8.26222&   7.52570&   7.25066&   9.52116&   4.01363&   1.92983&   (-)4.85314&   1.67121&  -5.72304\\
1.0&    -20&   7.93496&   7.52246&   7.41753&   8.52990&   3.02236&   1.96429&   (-)5.29860&   1.65732&  -5.37960\\
1.0&    -21&   7.62916&   7.51914&   7.54204&   7.64720&   2.13966&   1.88934&   (-)5.58367&   1.49859&  -5.12084\\
1.0&    -22&   7.31205&   7.51584&   7.64026&   7.01341&   1.50588&   1.94940&   (-)5.79626&   1.04672&  -4.91482\\
\hline
1.1&    -14&   9.54505&   7.82986&   6.14979&  16.91043&  11.40289&   1.98658& (+)11.40289&  -0.44007&  -8.37871\\
1.1&    -15&   9.44801&   7.82976&   6.31254&  15.90966&  10.40213&   1.96387& (+)10.40213&  -0.43931&  -8.05259\\
1.1&    -16&   9.32710&   7.82956&   6.47463&  14.90856&   9.40103&   1.93754&  (+)9.40103&  -0.43821&  -7.72746\\
1.1&    -17&   9.16675&   7.82915&   6.64704&  13.90679&   8.39926&   1.97362&  (+)8.39925&  -0.43644&  -7.38110\\
1.1&    -18&   8.97695&   7.82848&   6.80741&  12.90427&   7.39673&   1.93788&  (+)7.39649&  -0.43391&  -7.05817\\
1.1&    -19&   8.74844&   7.82733&   6.97449&  11.90026&   6.39273&   1.94316&  (+)6.38691&  -0.42991&  -6.72059\\
1.1&    -20&   8.46042&   7.82544&   7.14373&  10.89416&   5.38663&   1.96240&  (+)5.21465&  -0.42381&  -6.37693\\
1.1&    -21&   8.17092&   7.82287&   7.31018&   9.88766&   4.38012&   1.96812&   (-)5.23897&  -0.41736&  -6.03697\\
1.1&    -22&   7.83702&   7.81916&   7.47749&   8.88169&   3.37415&   1.98576&   (-)5.67932&  -0.41220&  -5.69232\\
\hline
1.2&    -14&   9.54748&   8.19896&   5.90131&  19.12970&  13.62216&   1.99528& (+)13.62216&  -2.38222&  -9.39229\\
1.2&    -15&   9.48817&   8.19894&   6.06513&  18.12936&  12.62183&   1.97860& (+)12.62183&  -2.38189&  -9.06438\\
1.2&    -16&   9.41359&   8.19889&   6.22851&  17.12885&  11.62132&   1.95941& (+)11.62132&  -2.38138&  -8.73723\\
1.2&    -17&   9.32198&   8.19879&   6.39158&  16.12812&  10.62058&   1.93851& (+)10.62058&  -2.38064&  -8.41051\\
1.2&    -18&   9.19360&   8.19855&   6.56513&  15.12693&   9.61939&   1.98084&  (+)9.61939&  -2.37945&  -8.06243\\
1.2&    -19&   9.03909&   8.19810&   6.72674&  14.12516&   8.61762&   1.95172&  (+)8.61760&  -2.37768&  -7.73777\\
1.2&    -20&   8.83129&   8.19725&   6.89917&  13.12217&   7.61464&   1.98801&  (+)7.61417&  -2.37470&  -7.39049\\
1.2&    -21&   8.60565&   8.19587&   7.06372&  12.11783&   6.61030&   1.97738&  (+)6.59886&  -2.37036&  -7.05788\\
1.2&    -22&   8.34100&   8.19411&   7.22991&  11.11314&   5.60561&   1.97863&  (+)5.19378&  -2.36567&  -6.72080\\
\hline
\end{tabular*}
\label{tab_energies.tab}
\end{table*}

\begin{figure*}
\centering
{\includegraphics[width=7.0truecm,height=7.0truecm]{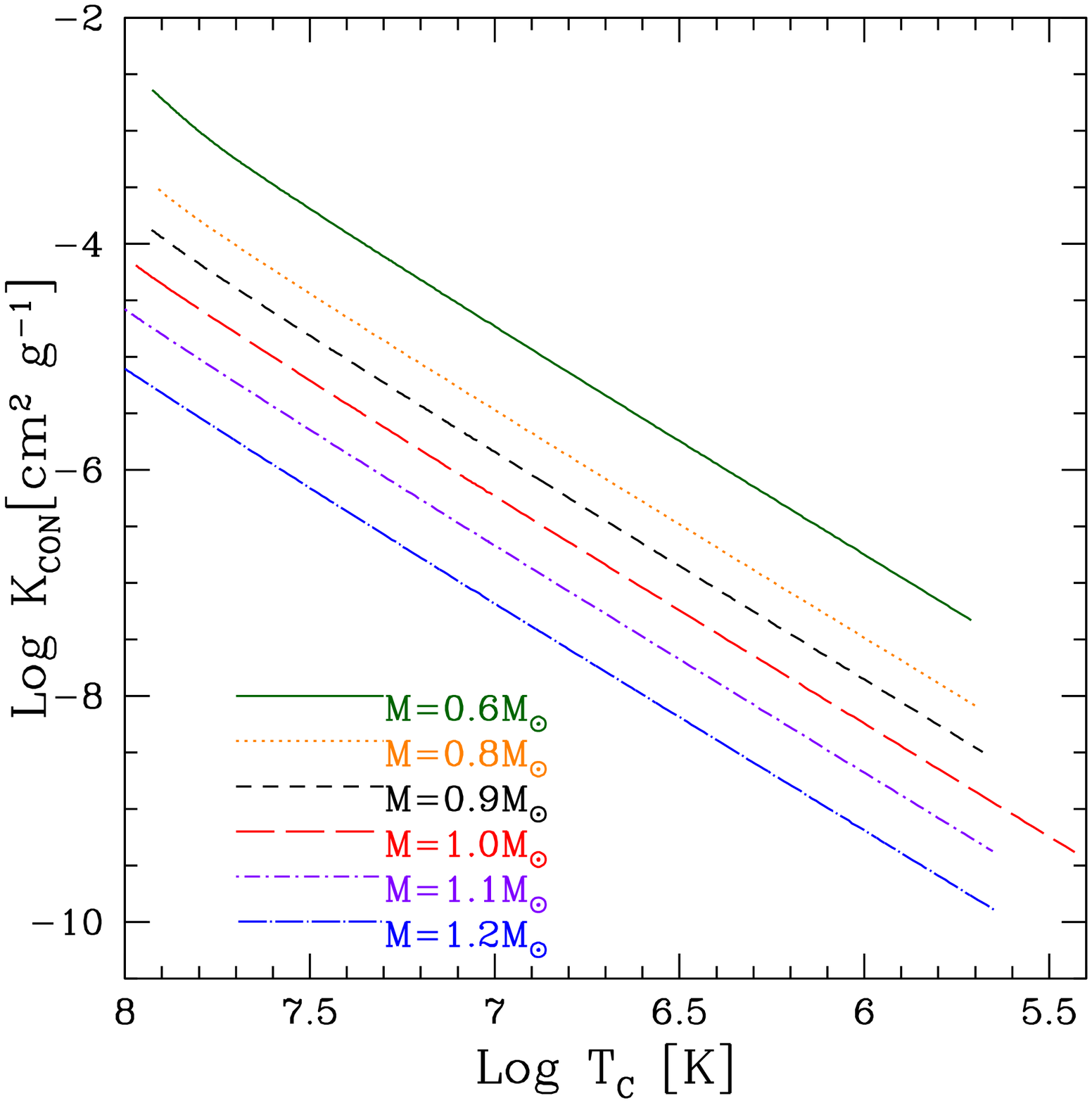}
 \includegraphics[width=7.0truecm,height=7.0truecm]{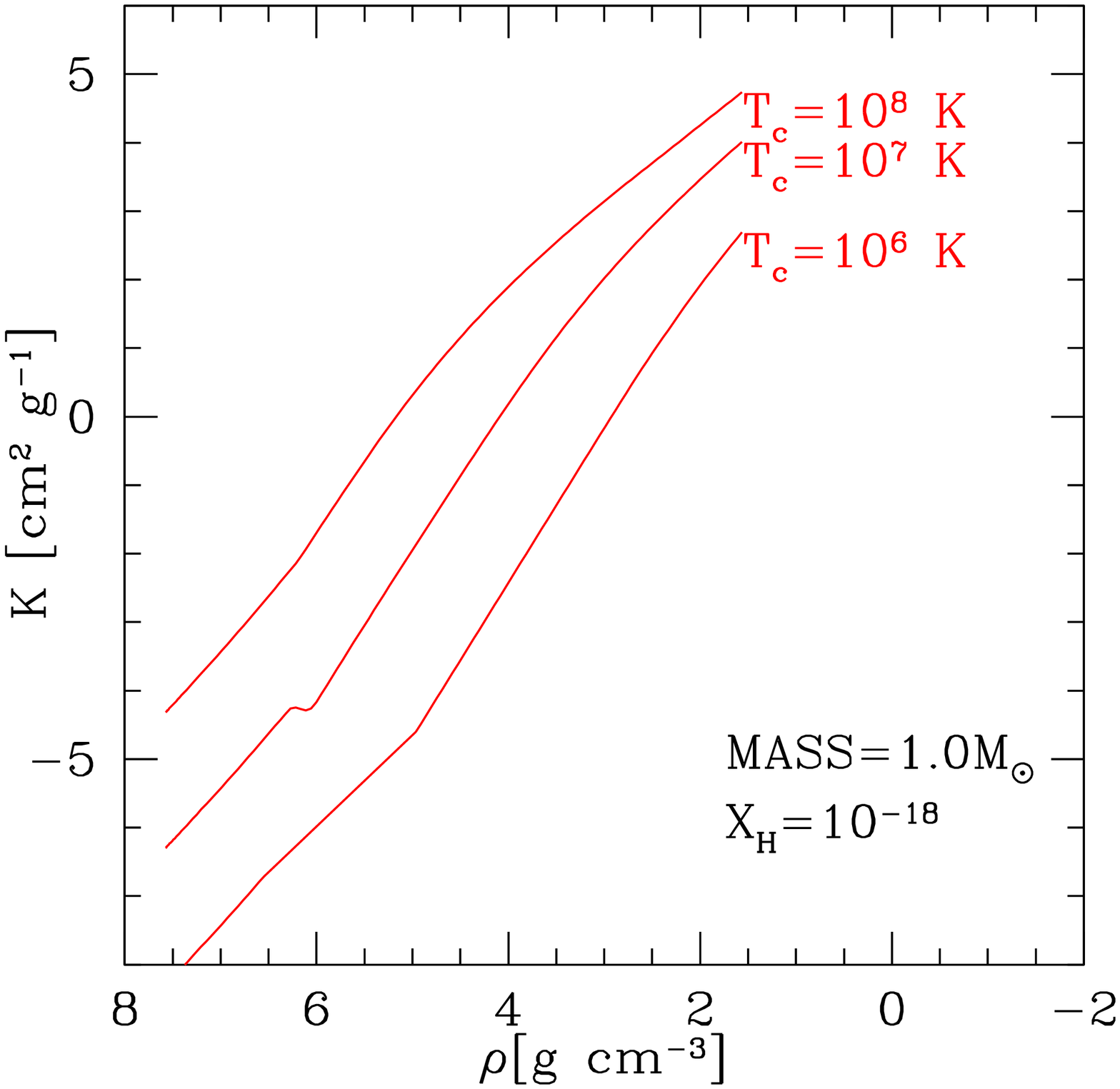}   }
\caption{\textbf{Left Panel}: The run of the central conductive opacity $\kappa_c$ (in $cm^{2}\, g^{-1}$) as a function of the central temperature $T_c$ along the cooling sequence in WDs of different mass as indicated. \textbf{Right Panel}: the run of the conductive opacity from the center to the surface in the 1.0 $M_\odot$ WD for three different values of the central temperature  $T_c$. The other masses have a similar trend. The cooling sequences and models of internal structure are from \citet{althaus1997,althaus1998}.  }
\label{kappa_path}
\end{figure*}

\subsection{To be, or not to be: that is the question (William Shakespeare, Hamlet 3/1)}

Adapting Hamlet's famous soliloquy to our contest, the question is whether or not WDs containing traces of light elements (hydrogen in the case we have considered) during  cooling   produce enough nuclear energy via the $\rm ^1H + ^{12}C$ reaction to balance and exceed the WD luminosity. A fraction of the energy would be kept inside the WD creating the physical conditions for  C+C-burning, and likely initiating a nuclear runaway followed by explosion.

\subsubsection{A Fuse for C-ignition}

The condition $\langle \epsilon  \rangle \geq L_{WD}/V$ does not necessarily imply that C-burning is started by the $\rm [H+C] $ reaction and, by proceeding toward C-deflagration (or C-detonation), cause the explosion of the WD.  When the energy-luminosity condition occurs in  late  cooling stages, i.e.  when  neutrino production does not occur, it only implies  that part of the energy can be trapped in the star. For the nuclear runaway  a mechanism must provide  the threshold energy for carbon-carbon ignition.

According to \citet{Nomoto1982a, Nomoto1982b} and \citet{kitamura2000}, the condition   to ignite
the Carbon-Carbon  reaction over a wide range of temperatures ($10^7 - 10^9$ K) and densities ($10^8 - 10^{10} \rm \, g\, cm^{-3}$) in dense matter is $R_{CC}\times Q_{CC} /\rho = 10^{-5} \rm W \, g^{-1} \equiv 10^2 \rm \, erg\, g^{-1} \, s^{-1}$ where $R_{CC}$ is the reaction rate (see Sect.\ref{nuclear} above) and $Q_{CC}$ is the energy released per reaction ($Q_{CC}=13.931\rm \, MeV$). This condition is hardly
reached via the sole [H+C]-reaction in presence of very low  abundances of hydrogen. \textit{We suggest that in the physical situation of the WD interiors, in which   $\Gamma > 1$ is soon reached by cooling and  the nuclear burning occurs in regimes from thermally enhanced to pure pycno-nuclear, the threshold limit for C+C-ignition is reached in a different way}.

\textbf{Elementary burning cells}. Let us start  considering a single $\rm ^1H + ^{12}C$ reaction. It occurs in a medium in which the carbon and oxygen nuclei are already in  liquid, partially crystallized or even fully crystallized conditions, and therefore have  reduced their mobility. Consequently,  the energy   deposited by this reaction will preferentially be given to the neighbouring nuclei, and then  shared with  a wider environment by conduction and radiation with opacities $\kappa_c$ and $\kappa_r$, respectively,  and total opacity  $ \kappa = {\kappa_c \times \kappa_r \over \kappa_c + \kappa_r}$. In the physical conditions of WDs,
thermal conduction dominates over radiative transport
\citep[see the results of detailed calculations described by][and references therein]{Iben2013a, Iben2013b}, therefore $\kappa \simeq \kappa_c$. For the purposes of this study we  adopt the analytic fits of \citet{Iben1968, Iben1975} of the numerical conductive opacities by
\citet{HubbardLampe1969} and \citet{Canuto1970}. The associated mean free path of thermal conduction is
\begin{equation}
 \lambda=\frac{1}{\kappa \rho}
\end{equation}
Both $\kappa_c$ and $\lambda$ vary significantly  with the temperature and density and in turn  with the evolutionary stage and mass of the WD.  Looking at the case of the $0.9\, M_\odot$ WD, the conductive opacity in the centre varies from $4 \times 10^{-3}\, \rm cm^2\, g^{-1}$ at the beginning
($T_c \simeq 8 \times 10^7\rm K, \rho_c \simeq 1.6\times 10^7 \, g \, cm^{-3} $) to about  $3 \times 10^{-8}$ at  the end
($T_c \simeq 5\times 10^5 \rm K, \rho_c \simeq 1.8\times 10^7 g\, cm^{-3}$) of the cooling sequence. The mean free path of the energy transportation is of
the order of
$\lambda \simeq 4.5 \times 10^{-4}$ cm and 16.7 cm, respectively. Furthermore, within a WD of given mass,
the conductive opacity $\kappa_c$ increases from the centre to the surface and the mean free path does the opposite.
In Fig. \ref{kappa_path} we show the run of the central value of the conductive opacity  along the cooling sequences for all the WD models to our disposal (left panel) and limited to the case of the $1 \, M_{\odot}$ the variation of $\kappa_c$  from the centre to the surface for three selected models characterized by their central temperature (right panel). However, since the nuclear energy by the [H+C]-reaction is expected to occur within  a central sphere of rather small dimensions, in the analysis below we may approximate the mean effective conductive opacity of a WD to its central value. In any case, the mean free path of conduction is much longer than the interionic distance $a_{ij}$ which varies from $8\times 10^{-9}$ to $2 \times 10^{-9}$ cm  from a WD of $0.6\, M_\odot$ to a WD of  $1.2\, M_\odot$.

In this scenario, it is plausible to conceive that most of the energy emitted by a single [H+C]-reaction is  acquired by the nuclei contained in a small  volume, which is  approximated to a cube of edge $\lambda$  centered on the H-nucleus. In the volume $\lambda^3$ there are   $N_C$ nuclei of carbon and $N_O$ nuclei of oxygen. This volume is named the \textit{elementary burning cell}. The ratio $(\lambda/a_{ij})^3$   indicates of the total number of nuclei inside the cell,
$N_C+N_O \equiv 2\times N_C = (\lambda/a_{ij})^3$. Depending on the local values of total opacity and density, we estimate  $N_C \simeq 10^{16} -  10^{19}$.
The $Q_{HC} \simeq 1.95$ Mev of a single [H+C]-reaction is the source of energy  for the  nuclei of carbon and oxygen in the cell. The  amount of  energy to be shared per unit mass in the cell $\lambda^3$ is
\begin{equation}
 \frac{Q}{\rho \lambda^3}
\end{equation}
where $Q_{HC}$ is expressed in erg and $\rho \lambda^3$ is the mass of the cell. The above ratio  for typical values of density ($10^8\rm\, g\, cm^{-3}$) and  mean free path ($4 \times 10^{-4}$ cm)  turns to be in the range 0.01 to 0.001.
We multiply this quantity by the total rate $R$ (in $\rm N\, cm^{-3}\, s^{-1}$) of the [H+C]-reactions to obtain the total energy to be shared.   This energy  is mostly used to heat  the carbon and oxygen nuclei in the elementary cell.

The above limit for C-ignition of about $10^2 \rm \, erg\, g^{-1} \, s^{-1}$ has to be applied to the quantity of carbon in the elementary burning cell (the dimensions of which already take conduction into account).

In this model in which the WD medium is populated by many elementary cells around the hydrogen nuclei, the
condition for C-ignition must be suitably scaled and applied to the environment of the cells. We define the quantities
\begin{equation}
 \epsilon_{\lambda} = \epsilon  \times  { 1\, cm^3 \over \lambda^3}
\end{equation}

\begin{equation}
{ \epsilon_{g,\lambda}}= \epsilon_{g} \times {1\, cm^3 \over  \lambda^3} = \frac{\epsilon_\lambda }{\rho }
\end{equation}

\noindent
where $\epsilon$ is the energy per $\rm cm^3$, and $\epsilon_\lambda$  the energy per $\rm cm^3$ in the volume $\lambda^3$, and $\epsilon_{g,\lambda}$ the energy per gram in the same volume.  We show that the energy deposited by a single [H+C]-reaction, when trapped into the elementary cell, is sufficient to heat  and finally ignite the C-nuclei inside the cell (see the entries in Table \ref{tab_energies.tab}).

Because of the above considerations, along the evolutionary sequence of a WD of assigned mass we calculate the reaction  rate and associated energy production rate of the [H+C]-reaction both per unit volume and unit mass  (expressed in  $\rm N\, cm^{-3} \, s^{-1}$,  $\rm erg \, cm^{-3} \, s^{-1}$,   $\rm N\, g^{-1} \, s^{-1}$ and $\rm erg \, g^{-1} \, s^{-1}$, as appropriate) and   also the quantity $\epsilon_\lambda / \rho$ (the energy generated in the elementary volume $\lambda^{3}$ sampled  by the mean free path of conductive energy transport).  We remind the reader that  the above reaction rate depends also on  $\rm X_H$ (here considered as a free parameter).

\textbf{Criterion for initiating a nuclear runaway}. Given these premises, along the cooling sequence of a WD of a given mass we identify  the time at which the  energy input by the [H+C]-reaction to the elementary cell, $[\epsilon_\lambda / \rho]$ falls below the threshold value for C-ignition  $\epsilon_{CC} = 100 \rm \, erg \, g^{-1} s^{-1}$. Since we are moving along a cooling sequence from high to low values of $T_c$ (the density remains nearly constant), during all previous stages, the energy $\epsilon_\lambda/ \rho$ is always above the threshold value. Along the same cooling sequence we search the time at which the condition $\langle \epsilon  \rangle \geq L_{WD}/V$ is verified, thus initiating the energy trapping. \textsl{The two conditions must be simultaneously satisfied to start C-burning in a situation of energy trapping}.

The results of these calculations are summarized in  Table \ref{tab_energies.tab} which  for each value of  $\rm X_H$ lists a few key quantities at the stage  at which the nuclear energy generation by the $\rm ^1H + ^{12}C$ reaction satisfies the conditions $\langle \epsilon  \rangle \geq L_{WD}/V$ and  $\epsilon_\lambda / \rho \leq 10^2 \rm erg\,  g^{-1}\, s^{-1}$. The physical quantities listed in Table \ref{tab_energies.tab} are: (1) the WD mass  $M_{WD}$ in solar units; (2) the hydrogen abundance   $\log X_H$; (3) the logarithm of the age in years  at which for the first time  the  elementary cell does not reach the conditions for  C+C-ignition; (4)-(5) the corresponding central density and temperature $\rho_c$ in  $\rm g\, cm^{-3}$ and $T_c$ in $K$; (6) the reaction rate per unit volume and time  $R_{H,C} $ in $\rmn\,  cm^{-3} \, s^{-1}$; (7) the energy generation rate per unit volume and time  $\epsilon$  in $\rm erg\,  cm^{-3}\, s^{-1}$;  (8) the energy generation rate per unit mass and time within the elementary cell of volume  $\lambda^3$   ${\epsilon_{\lambda} \over \rho }$ in $erg\,s^{-1}\,g^{-1}$; (9) the difference $\Delta E = \langle \epsilon  \rangle - L_{WD}/V $ in  $\rm erg\, cm^{-3} \, s^{-1}$. To display $\Delta E$ on a logarithmic scale, we take the absolute value adding  a negative (positive) sign  when the difference is negative (positive).  The positive sign is for the nuclear energy  generation exceeding the luminosity;  (10) logarithm of the time in s required to process all [H+C]-reactions in a gram of matter;  and finally (11) the total opacity $\kappa \simeq \kappa_c$ in $\rm cm^2\, g^{-1}$.

At this stage of the analysis,  we use three parameters: the abundance $\rm X_H$, the age ($\tau_1$) at which the nuclear energy generation rate per unit volume exceeds the WD luminosity per unit volume (see the intersections shown in Fig \ref{allmass123.fig} and the entries of Table \ref{tab_age_xh.tab}), and the age $\tau_2$ at which $\epsilon_\lambda / \rho$ falls below $\epsilon_{CC}$ in the elemental burning cell for the first time.  The abundance $\rm X_H$ can be further constrained by imposing that the nuclear rate per unit volume during the first cooling stages remains  smaller than the WD luminosity per unit volume. This avoids very early ignition, which is unlikely for the majority of WDs   since we do observe most of them to live and cool for a long time.
This means that for each WD mass there is an upper limit of $\rm X_H$  which decreases with increasing WD mass as already suggested by the intersections in Fig. \ref{allmass123.fig}.

\begin{figure*}
\centering
{
\includegraphics[width=5.3truecm,height=5.0truecm]{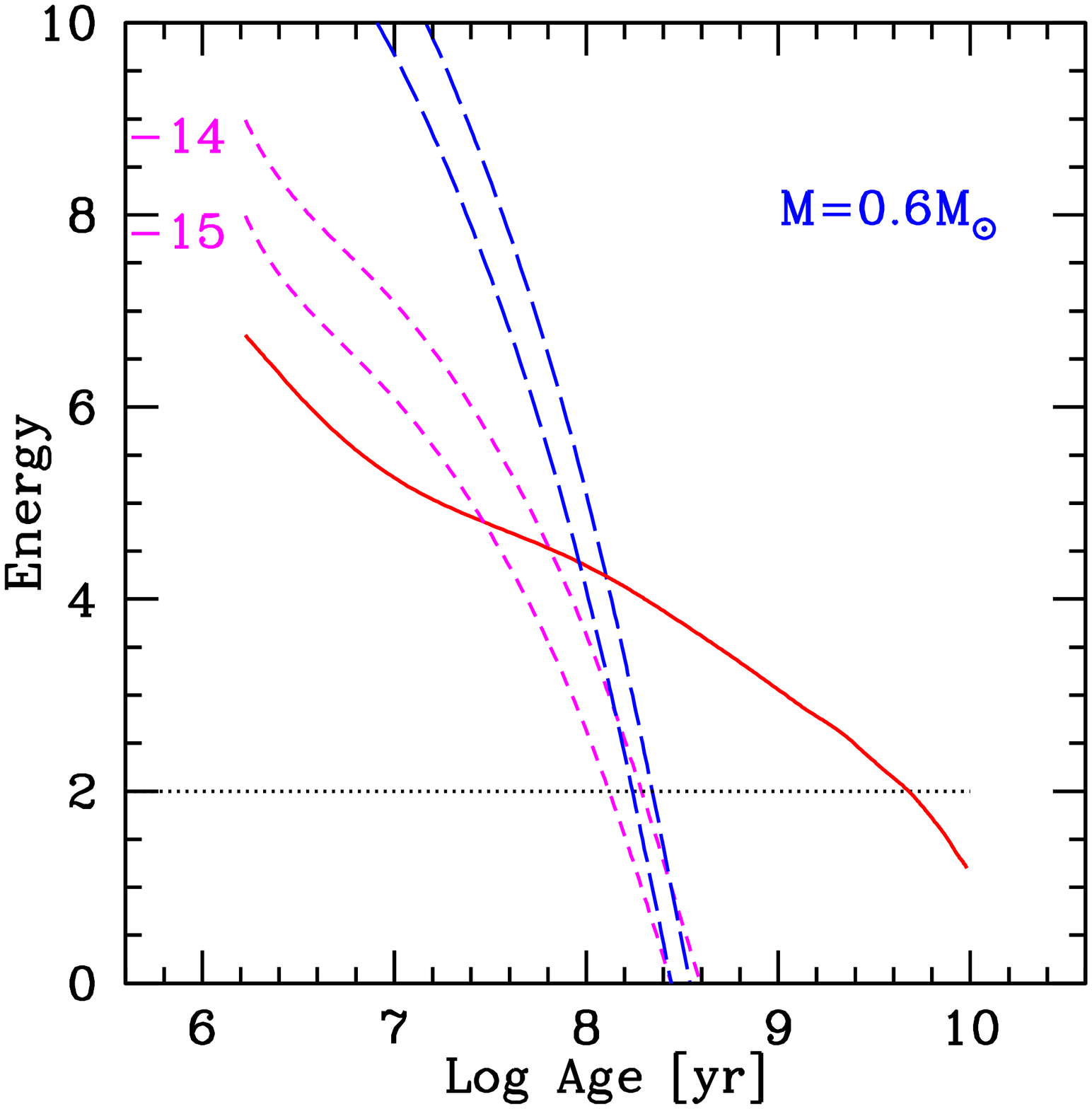}
\includegraphics[width=5.3truecm,height=5.0truecm]{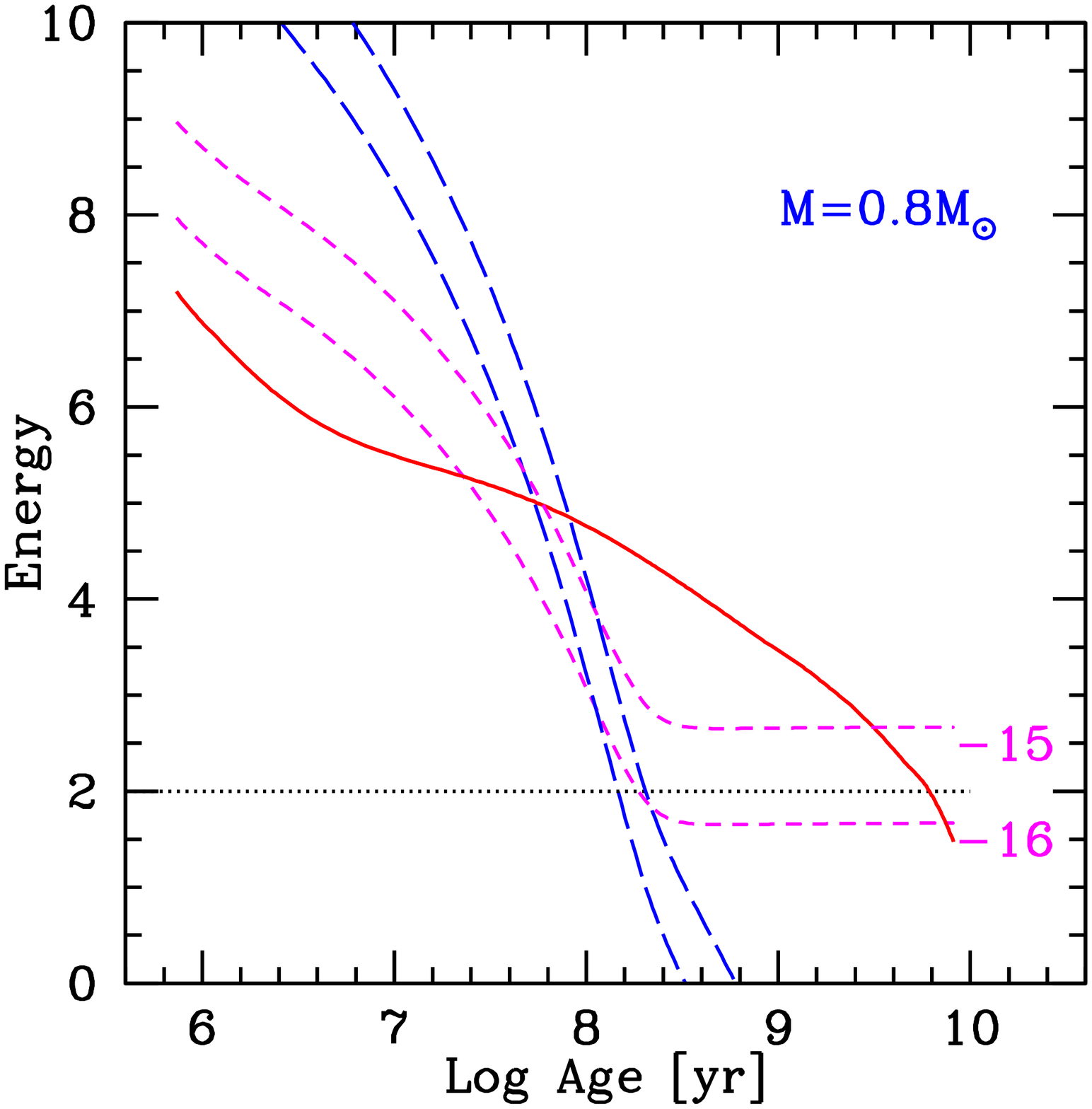}
 \includegraphics[width=5.3truecm,height=5.0truecm]{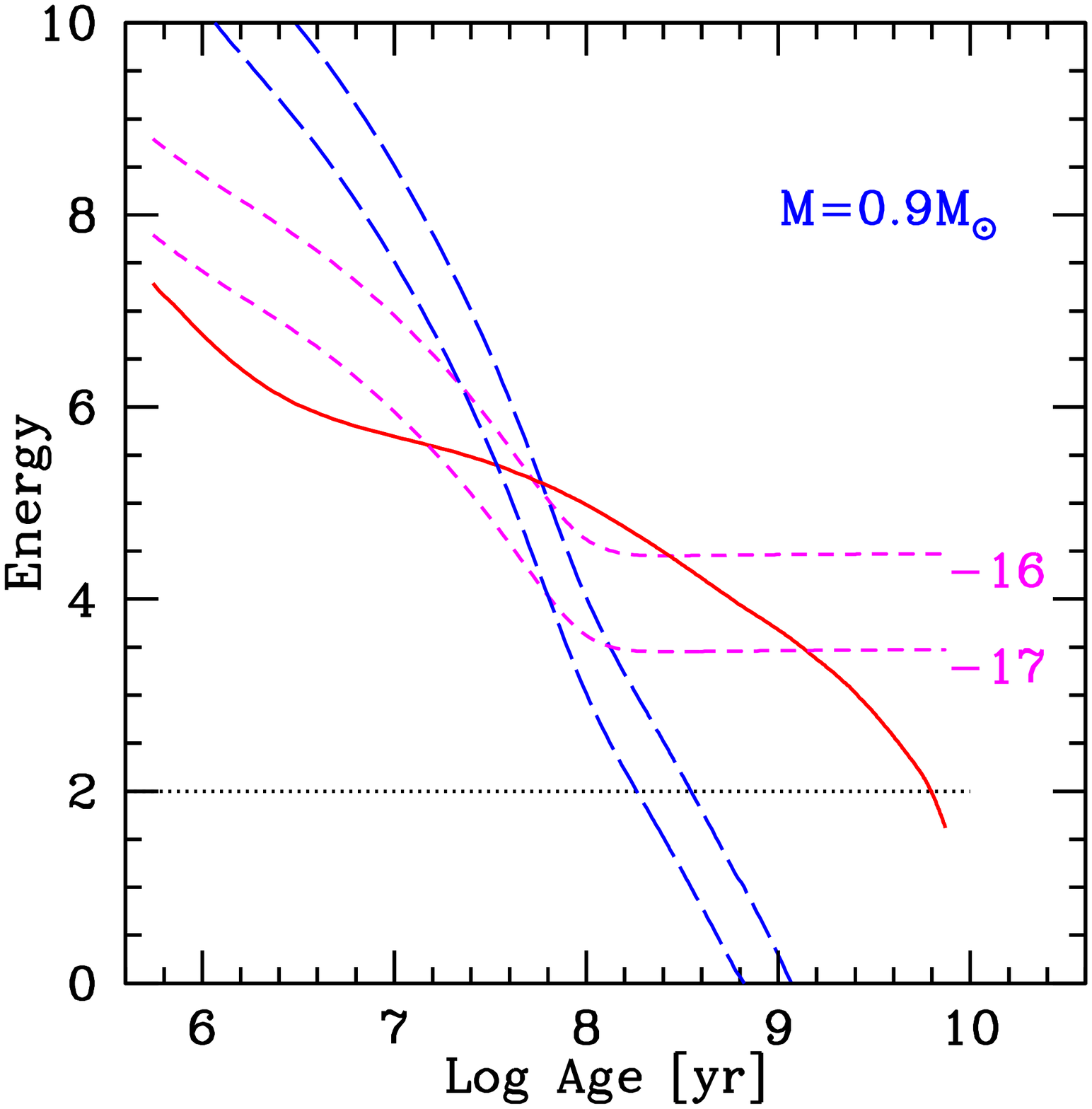}
 \includegraphics[width=5.3truecm,height=5.0truecm]{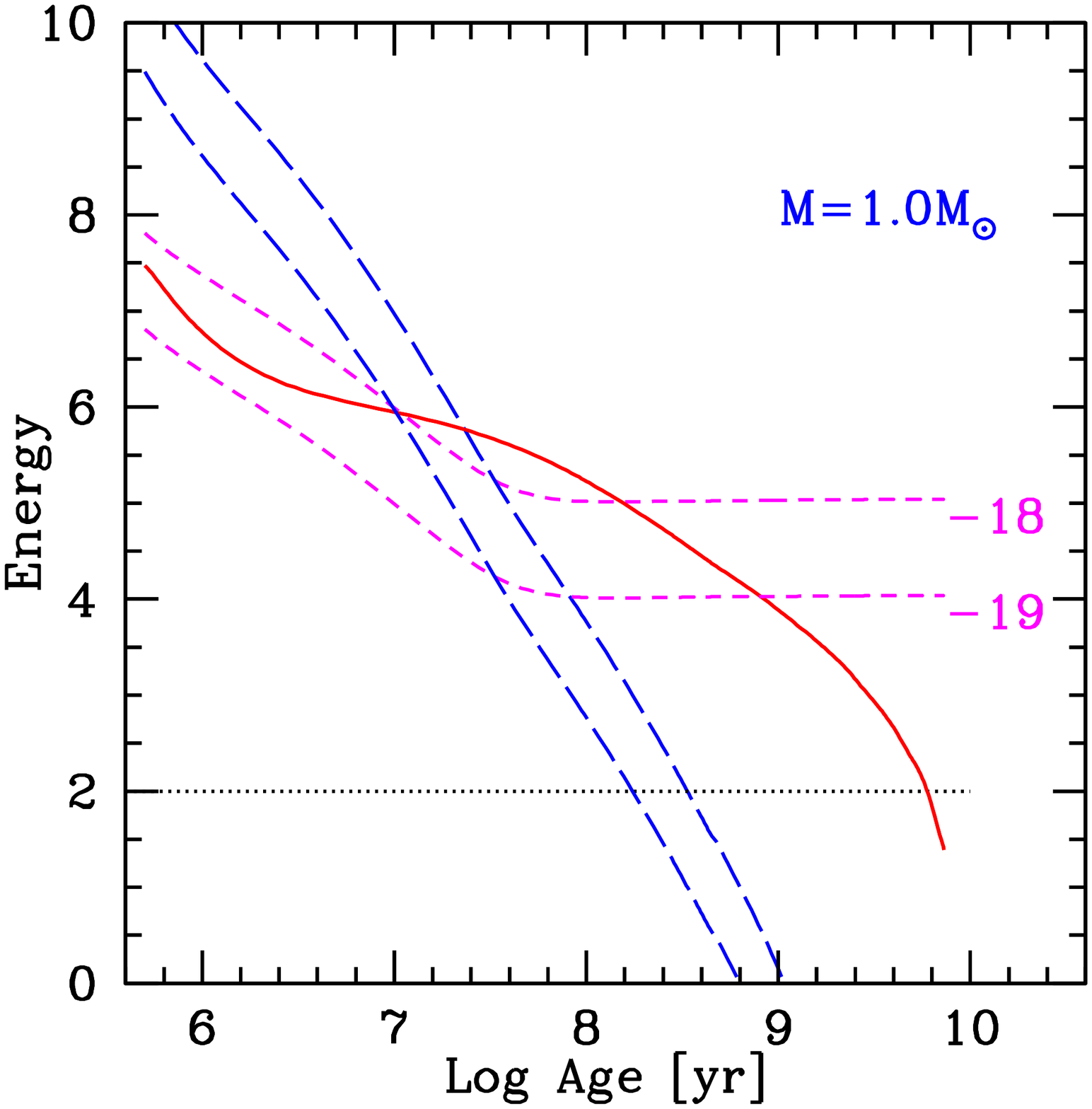}
 \includegraphics[width=5.3truecm,height=5.0truecm]{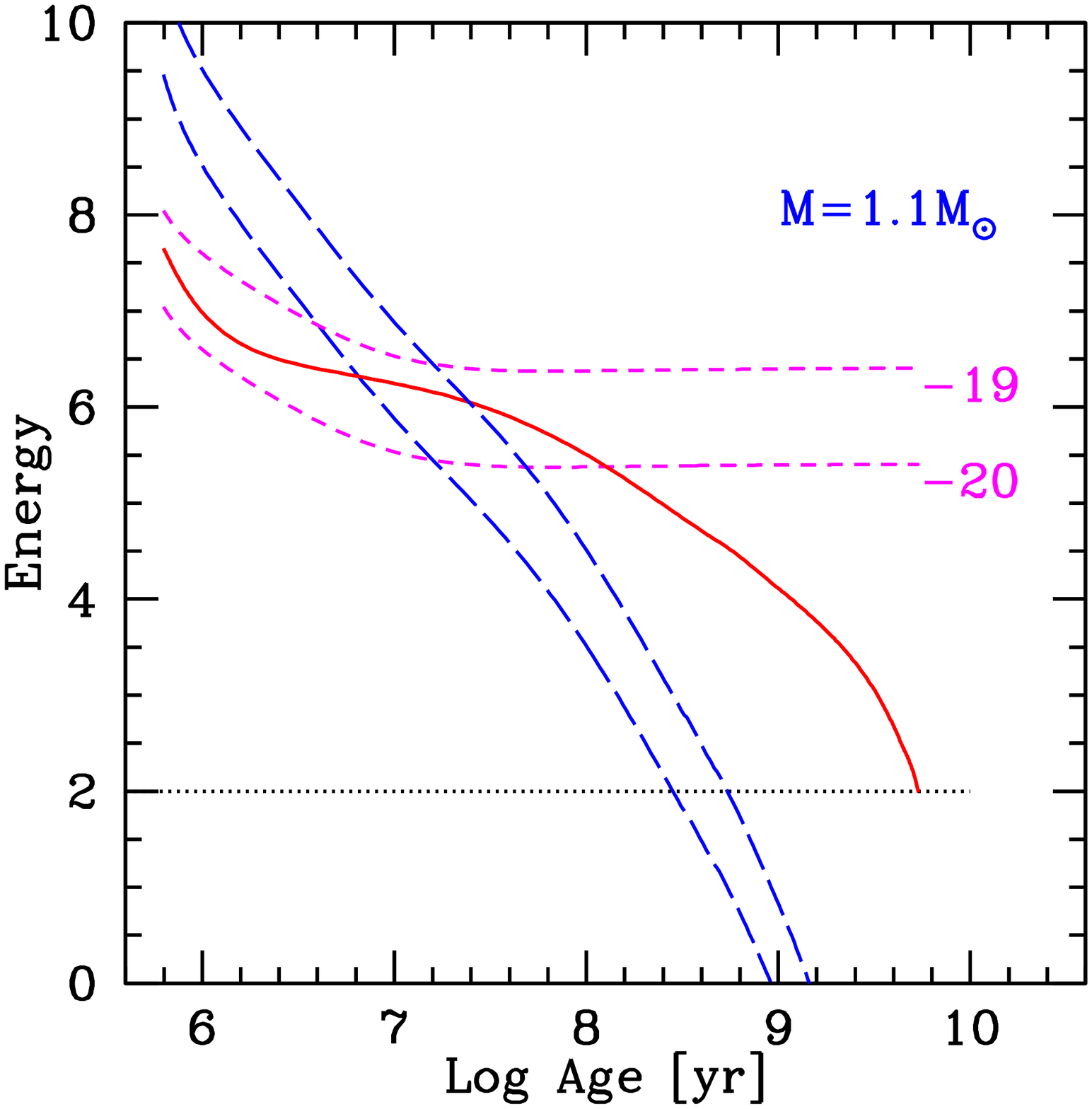}
 \includegraphics[width=5.3truecm,height=5.0truecm]{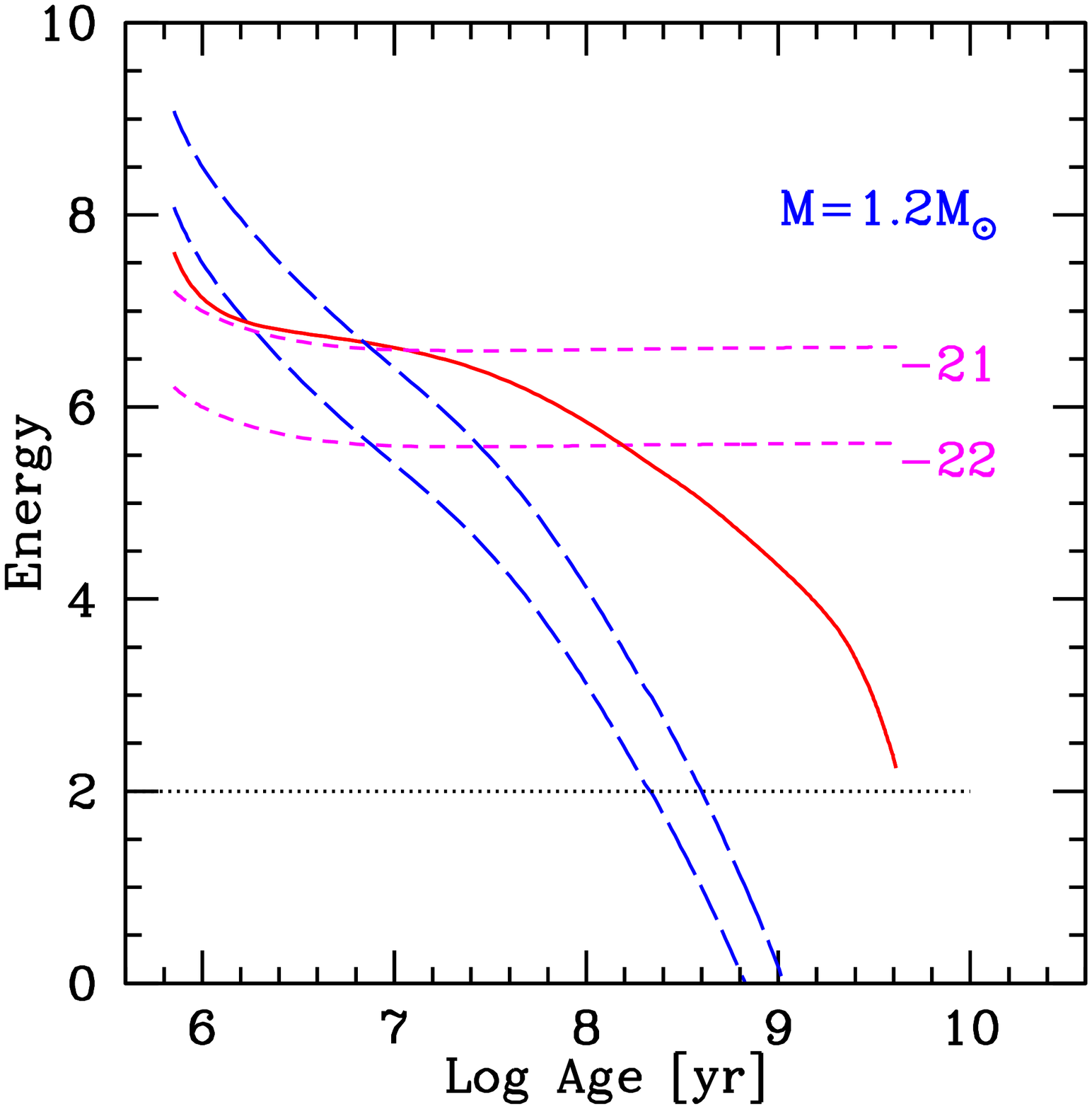}
 }
\caption{ The  WD luminosity per unit volume (thick red solid line), the energy per unit volume produced by the [H+C]-reaction for two reasonable values of  $\rm X_H$ (the thin, short-dashed magenta lines) that change with the WD mass,  the energy $\epsilon_\lambda/ \rho$ acquired by a typical burning cell for the same values of $\rm X_H$ (the thin, long-dashed green lines),   and finally the threshold value $\epsilon_{CC}$ (the horizontal, dotted black line), all of which are  plotted as a function of the logarithm of the age in years. The various energies are plotted on the same logarithmic scale. From left to right, WDs with different masses are displayed, i.e. 0.6, 0.8, 0.9, 1.0, 1.1, and 1.2 $M_\odot$ respectively. }
\label{erg_lambda}
\end{figure*}

An explosion may only occur with $\tau_1 \leq \tau_2$, as illustrated in Fig. \ref{erg_lambda} where   the WD luminosity per unit volume, the energy per unit volume produced by the [H+C]-reaction for two reasonable values of $\rm X_H$,  the energy $\epsilon_\lambda/ \rho$ for the same values of $\rm X_H$,   and finally the threshold value $\epsilon_{CC}$, all of them as function of time and on the same logarithmic scale.    Looking at the panels of  Fig. \ref{erg_lambda}, the condition can be met by WDs with mass 0.9, 1.0, 1.1 and 1.2 $M_\odot$ whereas it is missed by WDs with mass smaller that 0.8 $M_\odot$.  In the latter stars, the  energy production falls below $\epsilon_{CC}$ at ages $\tau_2 << \tau_1$.  The condition  could be met, however, by a 0.85 $M_\odot$ WD, and finally the 0.8 $M_\odot$ WD is border line.

\textbf{Global estimates of energy production}.
In order  to better highlight the above numerical results, we present some analytical estimates of (i) the total energy released by the [H+C]-reactions, (ii) the  fraction of this energy released  inside  an elementary cell,  to assess whether it is sufficient to increase the  temperature of C-nuclei contained in the cell to the threshold value for C-burning, and (iii)  the total amount of energy produced  by C-burning in the whole WD:

(i) [H+C]-burning. In a WD  there are $\rm X_H \times M_{WD} \times M_\odot$  (the $M_{WD}$ mass is in solar units) grams of hydrogen or equivalently a total number of hydrogen nuclei

$$N_H = {X_H \times M_{WD} \times M_\odot \over  A_H \times m_u } $$
\noindent
approximating $M_{WD} = 1\, M_\odot$, $M_\odot = 1.989 \times 10^{33}$ g and using $A_H = 1$ we obtain

$$N_H = 1.19 \times 10^{57} \times X_H$$

\noindent
Adopting for $\rm X_H$ a typical value of $10^{-20}$ and considering the energy generated by each [H+C]-reaction, i.e.
$3.12 \times 10^{-6}$ erg, we obtain a total energy release of $E_{[H+C]}=10^{31}$ erg.  For a typical WD volume of $\simeq 5.2 \times 10^{26}\, cm^3$ the nuclear energy per unit volume is about $2 \times 10^5$,  much higher than the WD luminosity per unit volume $\simeq 8\times 10^3$.

(ii) [C+C]-ignition. In order to verify  whether the energy deposited by the [H+C]-reactions  in the elementary cell may increase the temperature of the latter to that of C-ignition ($\simeq 0.6-0.7\times 10^9$ K), we start from   \citet{Nomoto1982a,Nomoto1982b,kitamura2000} condition for C-ignition,
the cell acquires  the energy ${\epsilon_\lambda \over \rho} \times ({\rho \over {\langle  R_{ij} \rangle} \times \lambda^3} )$. This energy   goes into  the total energy of the particles in the elemental cell, i.e.  ${3 \over 2} k_B T \times N_V  $, where $N_V$ is the total number of particles in the cell, i.e. the equation

\begin{equation}
  {\epsilon_\lambda \over \rho} \times \left({\rho \over {\langle R_{ij} \rangle} \times \lambda^3 }\right)  = {3\over 2} k_B\, T \, N_V
\end{equation}
is verified.
According to its definition, ${\epsilon_\lambda / \rho} \equiv \epsilon_{CC}  \simeq 10^2 \rm \,
erg \, g^{-1}\, s^{-1}$ and $\langle R_{ij}\rangle = \langle \epsilon_{ij} \rangle / Q_{HC} $. In addition,  each elementary cell contains only one nucleus of hydrogen, so  in a cell the [H+C]-reaction occurs only once and therefore the cell acquires  only this specific amount of energy.  Consequently,

\begin{equation}
 T = {2 \over 3}\, {\epsilon_{CC}} \times \left({ A_C  \times m_u  \over  k_B  \times \lambda^3}\right)
    \times \left( {1 \over \langle R_{ij} \rangle  \times \lambda^3  } \right)   \times \eta
\end{equation}
\noindent
where $N_V$ has been replaced by $(\rho \times \lambda^3)/( A_C \times m_u)$,   $A_C=12$,  and all other  symbols have their usual meaning. For typical values of  $ \lambda \simeq 10^{-4} - 10^{-5}$ and $\langle R_{ij} \rangle \simeq 10^{10} - 10^{11}$, we estimate a temperature in the range  $10^8 - 10^{14} $ K.  In the above equation, we  have also introduced the  $\eta $ parameter  to take into account  that part of the energy may escape
from the cell. While the lower limit temperature is below the threshold for C-burning, the temperature at upper limit would require $\eta \simeq 10^{-4}$. This means that a small adjustment of the parameters and/or the effective  energy acquired by the cell would yield the right temperature.   The numerical calculations show that the ignition temperature is reached all cases we have considered.

(iii) [C+C]-burning. Once  C-burning is ignited  in the elementary cells, how much energy is released by burning all the carbon nuclei contained in the cells?  The C+C reaction has a $Q_{CC}$-value of 13.93 Mev ($2.23\times 10^{-5}$ erg). The number of C pairs in a cell is about $N_C/2$ (for solid medium). With the aid of the reaction rates of \citet{yakovlev2006} (see Sect.\ref{nuclear} above), we evaluate the rate ( $\rm N\, cm^{-3} \, s^{-1}$)  at the temperature of
$6\times 10^8$ K and density of $10^8 \, \rm g \, cm^{-3}$. The rate is about $2.15\times 10^{22} \rm \,  cm^{-3}\, s^{-1}$ and the  energy generation rate is $4.8 \times 10^{18} \rm \, erg\, g^{-1} \, s^{-1}$. With the above rates, all carbon nuclei in a cell are destroyed in a tiny fraction of a second.

Since the total number of burning cells is  equal to the total number of hydrogen nuclei and the total number
of [C+C]-reactions per cell is $N_C/2$, the total energy release  $E_{C+C}$  is
\begin{eqnarray}
E_{C+C} &=& 2.23 \times 10^{-5} \times M_{\odot} \times  M_{WD}  \times N_0 \times X_H \times {N_C\over 2} \nonumber \\
          &=& 1.33\times 10^{52} \times M_{WD} \times X_H \times N_C
\end{eqnarray}
\noindent

\noindent
where $M_{WD}$ is  in  solar units, and $N_0=6.025\times10^{23}$  the Avogadro number.  For $\rm X_H= 10^{-20}$,  $N_C\simeq 10^{18}-10^{19}$, and $M_{WD}= 1$ the above equation yields  $E_{C+C} \simeq   10^{50} - 10^{51} \, \rm erg$.  This energy is  comparable to or larger than   the total gravitational energy of a WD, varying betwen     $|\Omega|=5.5 \times 10^{50}$ erg for the $0.6\times M_\odot$ to $|\Omega| = 5.8 \times 10^{51}$ erg for the $1.2 \times M_\odot$. Once started, [C+C]-burning will likely cause a thermal runaway followed by the explosion of the star.

\textbf{The fuse for C-ignition}. Since in any elemental volume, thanks to the energy released by the [H+C]-reaction the condition for local, mild C-ignition is met, we name this two steps process the "\textbf{\it fuse for C-ignition}". Most likely, once the fuse is activated, the additional release of energy by C-burning  makes it even stronger,  propagating  it to neighbouring regions, and activating complete C-burning that gradually moves into the thermally driven regime.

\begin{figure*}
\centering
{\includegraphics[width=5.5truecm,height=5.5truecm]{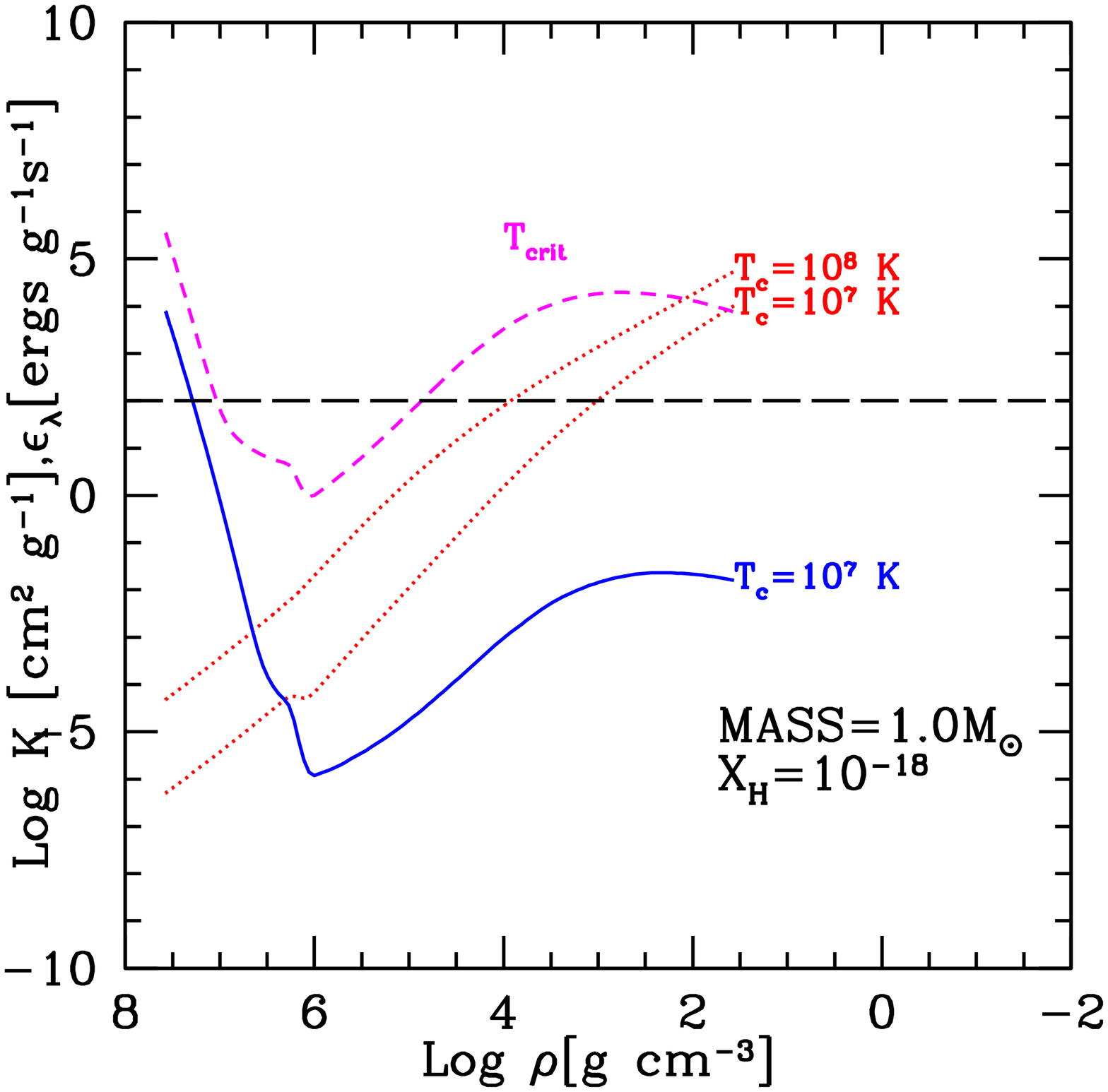}
 \includegraphics[width=5.5truecm,height=5.5truecm]{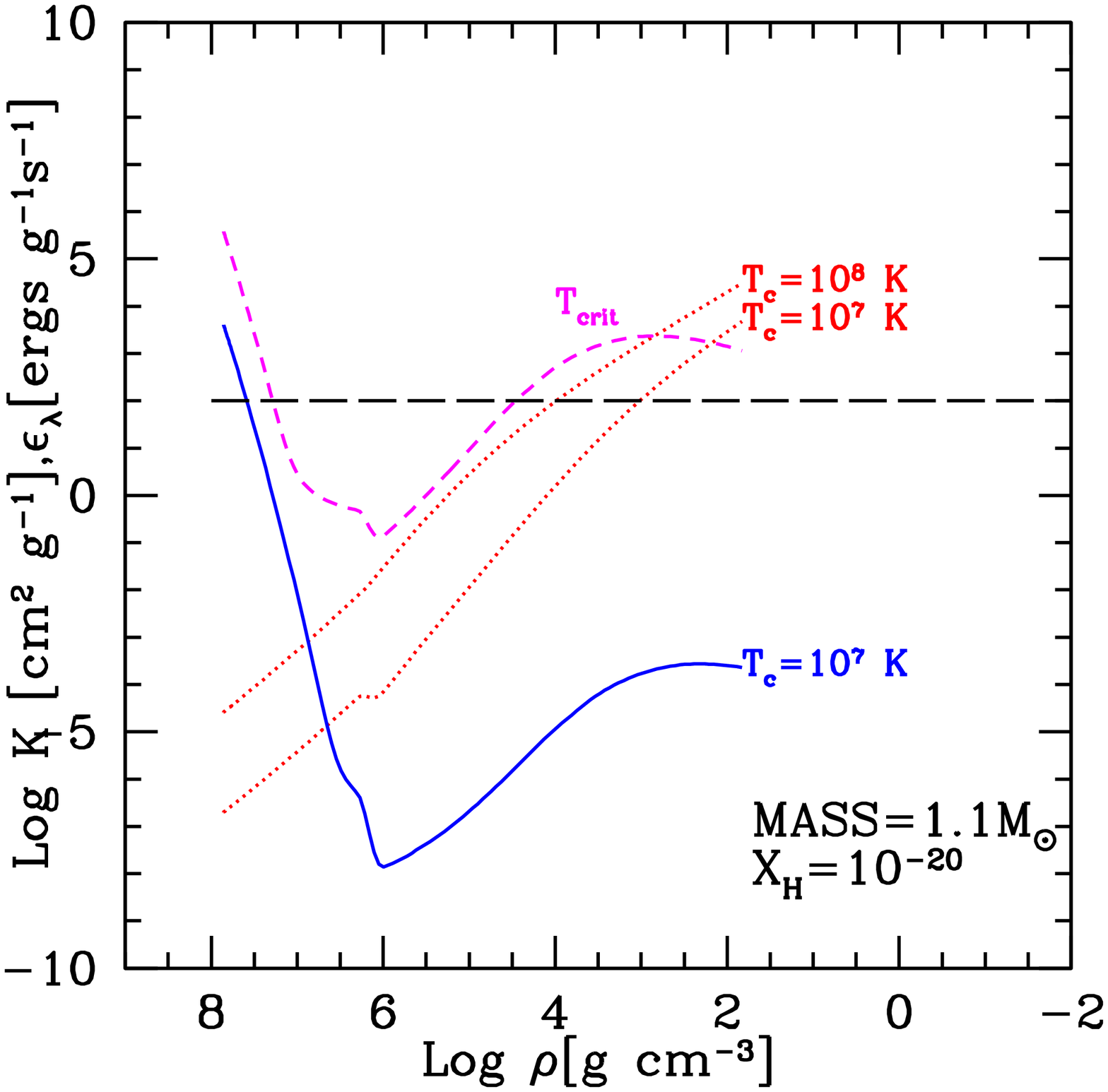}
 \includegraphics[width=5.5truecm,height=5.5truecm]{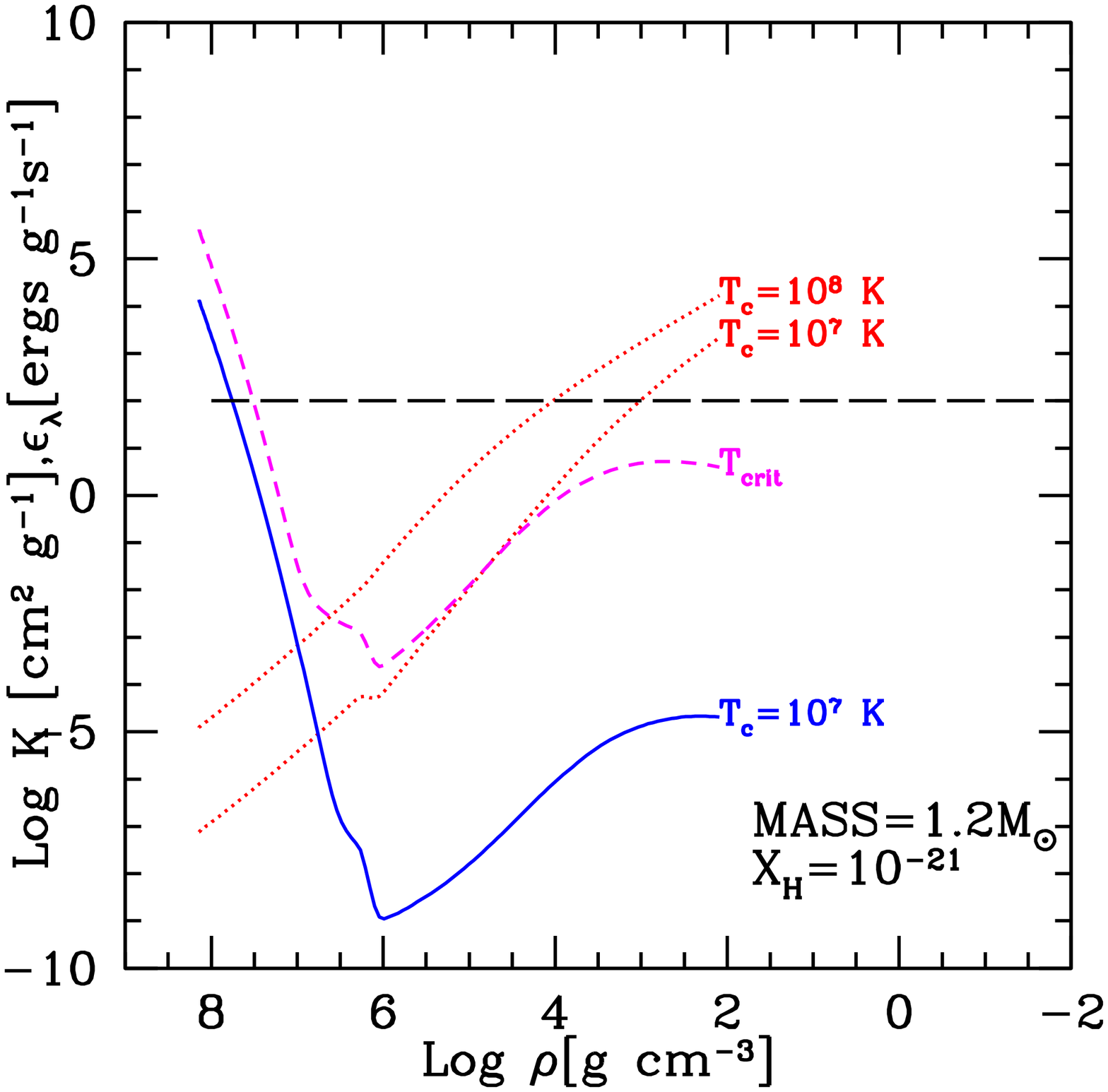}
 }
\caption{ The run of the energy $\epsilon_\lambda / \rho$ in $\rm erg\, g^{-1} \, s^{-1}$ produced inside  an elementary  cell by the [H+C]-reaction as a function of radial distance from the centre. The distance from the centre is described by the  density decreasing from  the center toward the surface. The energy production is presented for two values of the central temperature $T_c$ (assumed also to be  the temperature throughout the WD):  $T_{crit}$ the value at the stage when $\epsilon_\lambda / \rho$ becomes for the first time $\leq  \epsilon_{CC} = 10^2 \rm \, erg\, g^{-1}\, s^{-1}$, and a lower value $T_c=10^{7} \rm \, K$ for comparison. The dotted lines show the run of the conductive opacity for two values of $T_c$ as indicated. All the quantities are plotted on the same logarithmic scale. Two intersections are possible for the curve labelled $T_{crit}$ with the horizontal line  corresponding to  $\log \epsilon_{CC} =2$ in the case of the 1.0 and 1.1 $M_\odot$ WDs. The first one is at the center and the second one  occurs at $\log \rho \simeq 5$ and 5.5, for the 1.0 and 1.1 $M_\odot$ stars, respectively. Off-center C-ignition is possible. Only the intersection at the center occurs for the 1.2 $M_\odot$. No off-center C-ignition is  expected. }
\label{kappa_energy}
\end{figure*}

\textbf{Off-center C-ignition}.
The analysis we have made so far relies  on the model of elementary cells,  based on the central values of conductive opacity, temperature, and density. We have already shown that the conductive opacity and mean free path of thermal transport vary along the radius of a WD (see the right panel of Fig. \ref{kappa_path}). Therefore it is worth investigating if the energy condition for C-ignition in the elementary cells can be met also in more external regions (off-center ignition).
To this aim we show in Fig. \ref{kappa_energy} the energy production by the  [H+C]-reaction as a function of radial distance measured here by the value of local density $\rho(r)$.  The energy is evaluated adopting the temperature of the models listed in Table \ref{tab_energies.tab}, shortly indicated here by $T_{crit}$, and also the value of $10^7$ for comparison, and for $\rm X_H$ the value indicated by the results of Table \ref{tab_energies.tab} for each WD mass.  This energy production is indicated in Fig. \ref{kappa_energy} by the heavy solid red line for $T_{crit}$  and the dashed blue line for $T=10^7\rm\, K$. Together with this we plot the conductive opacity as a function of $\rho(r)$ and the two temperatures in question. Finally, we show the threshold energy for C-ignition (horizontal line). All quantities are shown on the same logarithmic scale. The three panels correspond to three values of the WD mass as indicated. The intersection of the threshold value $\epsilon_{CC}$ and the $T_{crit}$ curve is possible for the 1.0 and 1.1 $M_\odot$ WDs whereas it is missing for the 1.2 $M_\odot$. The intersection occurs at the centre (situations that we have already considered)  and at  some intermediate region ($\rho \simeq 10^5 \, \rm g\, cm^{-3}$). Off-center ignition is therefore possible.

\subsubsection{General remarks}
 To summarize, the main results of our analysis are:

(i) The above reaction always  occurs in physical conditions with $\Gamma_{ij} \geq 1$, transiting from the third  (thermally-enhanced) to the fifth regime (pure pycno) of burning (see Section \ref{fiveregimes}). This is well evident looking at the variation of the rate of energy generation per unit time and volume by the [H+C]-reaction as a function of the age (along the cooling sequence). The energy generation rate is according to \citet{yakovlev2006}, eqn. (26), i.e. made of two terms $R_{ij}^{pyc}(\rho) + \Delta_{ij}(\rho,T)$ whose meaning is straightforward. At high temperatures (early stages of the cooling sequence) the second term dominates and the rate steadily decreases, however past a certain temperature, the first term drives the rate and, since the density in a WD is essentially fixed by the mass and remains nearly constant as long as the mass does not change, the rate levels off. We may picture this by saying that the density driven  nuclear regime acts as a wedge to stabilize the otherwise ever decreasing temperature dependent rates.

(ii) In low mass WDs, the third regime is initially more efficient than the fifth regime and the latter overwhelms the former only past a certain age, whereas in high mass WDs the fifth regime always prevails. This trend is due to the increase of central and mean density with the WD mass.

(iii) Depending on $\rm X_H$ and $M_{WD}$, the sum of nuclear energy and ion internal energy can equal the luminosity at any age along the cooling sequence.

(iv) Massive  WDs are more sensitive to the instability threshold that can be reached even with very small hydrogen abundances. Therefore  the reaction $\rm ^{1}H+^{12}C$ is easily ignited and a nuclear runaway can be started as soon as massive WDs are born. For a typical value of $\rm X_H \simeq 10^{-20}$,  the above situation should occur for $ M_{WD} \geq 1.0$.  Although WDs of this mass are rare, the possibility  cannot be discarded and it may be possible to identify such progenitors.

(v) WDs of lower mass, about 0.9 $M_\odot$, are more stable and encounter the threshold condition only after a certain amount of time has elapsed since their formation (in the present analysis about 1 Gyr). WDs of 0.8 $M_\odot$ are at the borderline, in the sense that the two conditions could be met for $\rm -14 < X_H \leq -13$. WDs of about 0.85 $M_\odot$ could be  the transition stars. Finally WDs of even lower mass cool down to lw temperature without reaching the threshold energy for C-ignition.

(vi) We suggest that a future investigation should address the possible correlation between the WD  mass (and its progenitor) and  $\rm X_H$ (or $\rm X_{He}$),  which   translates into an additional dependence of the explosion time  on the WD mass.

(vii) Finally,  the onset of nuclear runaway should be a consequence of the positive gravo-thermal specific heat of a  mixture of nuclei and electrons whose equation of state is basically driven by the highly degenerate electrons, see the discussion by \citet{kippenhahn1990} and  \citet{mestel_1952_a, mestel_1952_b}.

To conclude,  there is an ample range of possibilities for  WDs contaminated by the presence of traces of light elements in their interiors to reach the critical stage at  which  nuclear burning is activated,  followed  by nuclear runaway and consequent disruption of the star.

This semi-analytical, orders-of-magnitude analysis does not yield more information. Only detailed numerical models  can give the correct answer to the Hamlet question.

\section{Conclusions}\label{conclusions}
In this paper we  investigated the possible effects of  impurities on the pycno-nuclear reaction rates in CO-WDs. We reviewed  the present-day  pycno-nuclear reaction rates, highlighting  the peculiarities of each model and the use that has been done until now. We introduced two important modifications in the existing expressions:

i) We extended the \citet{SalpeterVanHorn69} and  \citet{shapiro1983} reaction rate, to calculate the variation of the coulombian potential induced by  the presence   of a lighter ion of hydrogen  or helium in an arbitrary node of  the C-O ion lattice. Thus we extended  the original formulation for  pycno-nuclear reaction rates   conceived for OCPs  to  BIMs, including the reactions $\rm ^{1}H+^{12}C$ and/or $\rm ^{4}He+^{12}C$.

ii) We evaluated the displacement of nearby nuclei produced by the presence of impurities of different charge and estimated the change in local density, and applied this to  the MHO,  \citet{kitamura2000},
 and \citet{yakovlev2006} expressions.

Using the above revision of the  theoretical rates, we made a preliminary analysis  using the MHO formalism to study the pycno-nuclear reactions between light nuclei (hydrogen or helium) present in extremely low abundances  and heavier nuclei like carbon and oxygen, the basic constituents of WDs.

Encouraged by the results obtained with the MHO rates, we  used the more sophisticated descriptions of nuclear reactions in high density environments by \citet{kitamura2000} and \citet{yakovlev2006} to explore the effects  of impurities of light elements in triggering the above reactions.

The main results of the present study suggest that:
\begin{itemize}

\item[-] The presence of hydrogen even in extremely low concentrations (from $10^{-16}$ to $10^{-21}$) can raise
the pycno-nuclear reaction rates  in density intervals from   $\rm 10^7$ to $\rm 10^8\, g \, cm^{-3}$. The same is true for helium  at somewhat higher threshold densities.

\item[-] In the case of hydrogen, the above density interval corresponds to WD  masses from $\simeq 0.85\, {\rm to} \, 1.2 M_\odot$, well below the known limit of the Chandrasekhar mass.

\item[-] In WDs in this mass range, the energy released by pycno-nuclear reactions like $\rm ^1H + ^{12}C$ may trigger the ignition of CC-burning in a two steps process that we have named \textbf{\textit{ the fuse of C-ignition}}. The age at which this is expected to occur depends on the WD mass and abundance of residual hydrogen. The fuse-induced C-ignition is likely followed by thermal runaway according to the classical mechanisms.

\item[-] Even WDs with masses as low as   0.85 $M_\odot$ may experience nuclear runaway.

\end{itemize}

Our results could in principle radically change not only the current understanding of the structure and evolution of WDs but also imply that single WDs may be progenitors of type Ia SNe.  We may have discovered an alternative channel to  SNa Ia explosions. In this way we may be able (i) to  explain  the star formation rate dependence of the SNa Ia rate
\citep[e.g.,][]{Mannucci2006}; (ii) to provide some clues to interpreting the observational data
on the  ejected mass distribution of type Ia SNe showing a significant rate of non-Chandrasekhar-mass progenitors of mass as low as  $0.8\, M_\odot$ \citep{Scalzo2014}; and (iii) to account for the SNe exploding inside Planetary Nebulae (shortly named SNIPs)  in alternative to the core-degenerate scenario in which a WD merges with the hot core of an AGB star on a time interval $ \leq 3\times 10^8$ yr  since the WD formation \citep[see][for more details ]{Tsebrenko2014a,Tsebrenko2014b}. With our models, a single CO-WD may reach the explosion stage soon after the formation if sufficiently massive ($\geq 1.0 \, M_\odot$) and sufficiently rich in residual hydrogen ($\rm X_H \simeq 10^{-19} - 10^{-20}$). The expected time delay after formation can be as low as about a few ten of thousand years.

Therefore, before proceeding further it is mandatory to remind the reader of the limitations of the present approach.

\begin{itemize}

\item[-] Prior to any other consideration, it is worth recalling that the success of the proposed model for the evolution of CO WDs relies on the existence of traces of light elements like hydrogen and helium  (the former in particular) that survived  previous nuclear burnings.  Complete stellar models from the main sequence to the WD stages in which the  abundances of elements are followed throughout the various nuclear burnings to the values used in this study (i.e. $\rm X_H$ in the range $10^{-18}$ to $10^{-20}$) are not presently available. To cope with this, we have presented some plausible arguments to sustain  this possibility that eventually has been adopted as a working hypothesis. Null abundances for the light elements in WD interiors cannot be firmly excluded. Stellar models  checking this major issue are mandatory.

    \item[-] Even if  our computations rely  on  state-of-the-art WD models, the correct approach would be to calculate and follow in time  stellar models responding to the new source of energy   in the course of evolution.  The present models, although acceptable for very low nuclear rates,     fail to represent the real physical structure of a WD in  presence of  large energy production. The WD structure  may be deeply altered  and follow a different evolutionary     history that is not easy to foresee at present.

    \item[-] Our calculations do not take  into
    account yet the additional energy release due to elements stratification, solid state transition (latent heat)  and     gravitational contraction.

    \item[-] Only complete, self-consistent models would allow us to correctly determine  the amount of energy generated by nuclear reactions, and to  deal with the energy transport problem, rigorously comparing  production versus transport of energy.

\end{itemize}

Future work should be  the computation of a self-consistent model able to respond to  changing physical conditions, indicating the exact age at which the under-barrier reactions become important and  the  structural response of a WD  to the novel energy input.  After the WD cools  to the  temperature for the activation of the under-barrier channel in the liquid/solid phase, we foresee three possible scenarios: (a) reheating with consequent increase of the cooling lifetime;  (b)  rejuvenation to another type of object;  (c)  explosion as type Ia SNa  (or a different type?).

We  urge  reconsidering  the whole subject of
nuclear reactions in these extreme conditions, in particular in presence of impurities that could deeply change our current understanding of the energy sources in WDs.

\textit{If the results of our exploratory project are confirmed by further investigation,  important implications for the currently accepted scenario for  type Ia SNe will follow. The binary origin of type Ia SNa explosion would be no longer strictly necessary. Isolated WDs with masses well below the Chandrasekhar limit  may reach the threshold for pycno-nuclear burning and consequent SNa explosion due to the survival of traces of light elements. These impurities  may remain inactive for long periods of time and be activated only when the WDs  reach the liquid-solid  regime. Owing to the  large range of WD masses that could be affected by the presence of impurities and undergo  thermal runaway and consequent  SNa explosion, the nature of standard candle so far attributed to type Ia SNe may not be true. Because of the far reaching implications, the whole subject deserves careful future investigation.}

\section*{Acknowledgements}
We would like to deeply thank Dr. Maurizio Salaris  for his  friendship, for the many illuminating discussions, and for providing us his cooling sequences and WD models. We like to thank Drs.  S. Ichimaru and H. Kitamura
for  patiently replying to the numberless emails sent by PT and the many very useful explanations and comments.
We acknowledge  the critical discussion with Dr. K. Shen and finally, the very helpful comments of the unknown referee.

\bibliographystyle{mn2e}               
\bibliography{mnemonic,MN_14_2494_MJ}    

\begin{thebibliography}{}

\bibitem[\protect\citeauthoryear{{Althaus} \& {Benvenuto}}{{Althaus} \&
  {Benvenuto}}{1997}]{althaus1997}
{Althaus} L.~G.,  {Benvenuto} O.~G.,  1997, ApJ, 477, 313

\bibitem[\protect\citeauthoryear{{Althaus} \& {Benvenuto}}{{Althaus} \&
  {Benvenuto}}{1998}]{althaus1998}
{Althaus} L.~G.,  {Benvenuto} O.~G.,  1998, MNRAS, 296, 206

\bibitem[\protect\citeauthoryear{{Althaus}, {Garc{\'{\i}}a-Berro}, {Isern},
  {C{\'o}rsico} \& {Miller Bertolami}}{{Althaus} et~al.}{2012}]{althaus2012}
{Althaus} L.~G.,  {Garc{\'{\i}}a-Berro} E.,  {Isern} J.,  {C{\'o}rsico} A.~H.,
    {Miller Bertolami} M.~M.,  2012, A\&A, 537, A33

\bibitem[\protect\citeauthoryear{{Althaus}, {Miller Bertolami} \&
  {C{\'o}rsico}}{{Althaus} et~al.}{2013}]{althaus2013}
{Althaus} L.~G.,  {Miller Bertolami} M.~M.,    {C{\'o}rsico} A.~H.,  2013,
  A\&A, 557, A19

\bibitem[\protect\citeauthoryear{{Althaus}, {Panei}, {Miller Bertolami},
  {Garc{\'{\i}}a-Berro}, {C{\'o}rsico}, {Romero}, {Kepler} \&
  {Rohrmann}}{{Althaus} et~al.}{2009}]{althaus2009}
{Althaus} L.~G.,  {Panei} J.~A.,  {Miller Bertolami} M.~M.,
  {Garc{\'{\i}}a-Berro} E.,  {C{\'o}rsico} A.~H.,  {Romero} A.~D.,  {Kepler}
  S.~O.,    {Rohrmann} R.~D.,  2009, ApJ, 704, 1605

\bibitem[\protect\citeauthoryear{{Beard}}{{Beard}}{2010}]{Beard2010}
{Beard} M.~L.,  2010, PhD thesis, University of Notre Dame

\bibitem[\protect\citeauthoryear{{Bergeron}, {Wesemael}, {Fontaine} \&
  {Liebert}}{{Bergeron} et~al.}{1990}]{bergeron1990}
{Bergeron} P.,  {Wesemael} F.,  {Fontaine} G.,    {Liebert} J.,  1990, ApJL,
  351, L21

\bibitem[\protect\citeauthoryear{{Bertelli}, {Bressan}, {Chiosi}, {Fagotto} \&
  {Nasi}}{{Bertelli} et~al.}{1994}]{Bertelli1994}
{Bertelli} G.,  {Bressan} A.,  {Chiosi} C.,  {Fagotto} F.,    {Nasi} E.,  1994,
  106, 275

\bibitem[\protect\citeauthoryear{{Bertelli}, {Girardi}, {Marigo} \&
  {Nasi}}{{Bertelli} et~al.}{2008}]{Bertelli2008}
{Bertelli} G.,  {Girardi} L.,  {Marigo} P.,    {Nasi} E.,  2008, A\&A, 484, 815

\bibitem[\protect\citeauthoryear{{Bertelli}, {Nasi}, {Girardi} \&
  {Marigo}}{{Bertelli} et~al.}{2009}]{Bertelli2009}
{Bertelli} G.,  {Nasi} E.,  {Girardi} L.,    {Marigo} P.,  2009, A\&A, 508, 355

\bibitem[\protect\citeauthoryear{{Bressan}, {Marigo}, {Girardi}, {Salasnich},
  {Dal Cero}, {Rubele} \& {Nanni}}{{Bressan} et~al.}{2012}]{Bressan2012}
{Bressan} A.,  {Marigo} P.,  {Girardi} L.,  {Salasnich} B.,  {Dal Cero} C.,
  {Rubele} S.,    {Nanni} A.,  2012, MNRAS, 427, 127

\bibitem[\protect\citeauthoryear{{Brown} \& {Sawyer}}{{Brown} \&
  {Sawyer}}{1997}]{brown1997}
{Brown} L.~S.,  {Sawyer} R.~F.,  1997, Reviews of Modern Physics, 69, 411

\bibitem[\protect\citeauthoryear{{Canuto}}{{Canuto}}{1970}]{Canuto1970}
{Canuto} V.,  1970, ApJ, 159, 641

\bibitem[\protect\citeauthoryear{{Catal{\'a}n}, {Isern}, {Garc{\'{\i}}a-Berro}
  \& {Ribas}}{{Catal{\'a}n} et~al.}{2008}]{catalan2008}
{Catal{\'a}n} S.,  {Isern} J.,  {Garc{\'{\i}}a-Berro} E.,    {Ribas} I.,  2008,
  MNRAS, 387, 1693

\bibitem[\protect\citeauthoryear{{Chandrasekhar}}{{Chandrasekhar}}{1939}]{chandrasekhar1939}
{Chandrasekhar} S.,  1939, {An introduction to the study of stellar structure}

\bibitem[\protect\citeauthoryear{{Chiosi}, {Bertelli} \& {Bressan}}{{Chiosi}
  et~al.}{1992}]{ChiBerBre1992}
{Chiosi} C.,  {Bertelli} G.,    {Bressan} A.,  1992, ARA\&A, 30, 235

\bibitem[\protect\citeauthoryear{{C{\'o}rsico} \& {Althaus}}{{C{\'o}rsico} \&
  {Althaus}}{2014}]{Corsico2014}
{C{\'o}rsico} A.~H.,  {Althaus} L.~G.,  2014, ApJL, 793, L17

\bibitem[\protect\citeauthoryear{{Dewitt}, {Slattery}, {Baiko} \&
  {Yakovlev}}{{Dewitt} et~al.}{2001}]{DeWitt2001}
{Dewitt} H.,  {Slattery} W.,  {Baiko} D.,    {Yakovlev} D.,  2001,
  Contributions to Plasma Physics, 41, 251

\bibitem[\protect\citeauthoryear{{DeWitt}, {Slattery} \& {Yang}}{{DeWitt}
  et~al.}{1992}]{DeWitt1992}
{DeWitt} H.~E.,  {Slattery} W.~L.,    {Yang} J.,  1992, in International
  Conference on the Physics of Strongly Coupled Plasmas, Rochester, NY, 17-21
  Aug. 1992 {Monte Carlo simulation of the OCP freezing transition}.
pp 17--21

\bibitem[\protect\citeauthoryear{{Eisenstein}, {Liebert}, {Harris}, {Kleinman},
  {Nitta}, {Silvestri}, {Anderson}, {Barentine}, {Brewington}, {Brinkmann},
  {Harvanek}, {Krzesi{\'n}ski}, {Neilsen} Jr., {Long}, {Schneider} \&
  {Snedden}}{{Eisenstein} et~al.}{2006}]{eisenstein2006}
{Eisenstein} D.~J.,  {Liebert} J.,  {Harris} H.~C.,  {Kleinman} S.~J.,  {Nitta}
  A.,  {Silvestri} N.,  {Anderson} S.~A.,  {Barentine} J.~C.,  {Brewington}
  H.~J.,  {Brinkmann} J.,  {Harvanek} M.,  {Krzesi{\'n}ski} J.,  {Neilsen} Jr.
  E.~H.,  {Long} D.,  {Schneider} D.~P.,    {Snedden} S.~A.,  2006, ApJS, 167,
  40

\bibitem[\protect\citeauthoryear{{Fowler}, {Caughlan} \& {Zimmerman}}{{Fowler}
  et~al.}{1975}]{Fowler75}
{Fowler} W.~A.,  {Caughlan} G.~R.,    {Zimmerman} B.~A.,  1975, ARAA, 13, 69

\bibitem[\protect\citeauthoryear{{Fujimoto}}{{Fujimoto}}{1982a}]{Fujimoto1982b}
{Fujimoto} M.~Y.,  1982a, ApJ, 257, 767

\bibitem[\protect\citeauthoryear{{Fujimoto}}{{Fujimoto}}{1982b}]{Fujimoto1982a}
{Fujimoto} M.~Y.,  1982b, ApJ, 257, 752

\bibitem[\protect\citeauthoryear{{Gasques}, {Afanasjev}, {Aguilera}, {Beard},
  {Chamon}, {Ring}, {Wiescher} \& {Yakovlev}}{{Gasques}
  et~al.}{2005}]{gasques2005}
{Gasques} L.~R.,  {Afanasjev} A.~V.,  {Aguilera} E.~F.,  {Beard} M.,  {Chamon}
  L.~C.,  {Ring} P.,  {Wiescher} M.,    {Yakovlev} D.~G.,  2005, PhRvC, 72,
  025806

\bibitem[\protect\citeauthoryear{{Hubbard} \& {Lampe}}{{Hubbard} \&
  {Lampe}}{1969}]{HubbardLampe1969}
{Hubbard} W.~B.,  {Lampe} M.,  1969, ApJS, 18, 297

\bibitem[\protect\citeauthoryear{{Iben} Jr.}{{Iben}}{1968}]{Iben1968}
{Iben} Jr. I.,  1968, ApJ, 154, 557

\bibitem[\protect\citeauthoryear{{Iben} Jr.}{{Iben}}{1975}]{Iben1975}
{Iben} Jr. I.,  1975, ApJ, 196, 549

\bibitem[\protect\citeauthoryear{{Iben} Jr.}{{Iben}}{2013a}]{Iben2013a}
{Iben} Jr. I.,  2013a, {Stellar Evolution Physics, Volume 1: Physical Processes
  in Stellar Interiors}

\bibitem[\protect\citeauthoryear{{Iben} Jr.}{{Iben}}{2013b}]{Iben2013b}
{Iben} Jr. I.,  2013b, {Stellar Evolution Physics, Volume 2: Advanced Evolution
  of Single Stars}

\bibitem[\protect\citeauthoryear{{Iben} Jr. \& {MacDonald}}{{Iben} \&
  {MacDonald}}{1985}]{Iben1985}
{Iben} Jr. I.,  {MacDonald} J.,  1985, ApJ, 296, 540

\bibitem[\protect\citeauthoryear{{Iben} Jr. \& {Renzini}}{{Iben} \&
  {Renzini}}{1983}]{IbenRenzini1983}
{Iben} Jr. I.,  {Renzini} A.,  1983, ARA\&A, 21, 271

\bibitem[\protect\citeauthoryear{{Ichimaru}}{{Ichimaru}}{1982}]{Ichimaru1982}
{Ichimaru} S.,  1982, Reviews of Modern Physics, 54, 1017

\bibitem[\protect\citeauthoryear{{Ichimaru} \& {Kitamura}}{{Ichimaru} \&
  {Kitamura}}{1999a}]{ichimaru1999}
{Ichimaru} S.,  {Kitamura} H.,  1999a, Physics of Plasmas, 6, 2649

\bibitem[\protect\citeauthoryear{{Ichimaru} \& {Kitamura}}{{Ichimaru} \&
  {Kitamura}}{1999b}]{IchimaruKitamura99}
{Ichimaru} S.,  {Kitamura} H.,  1999b, Physics of Plasmas, 6, 2649

\bibitem[\protect\citeauthoryear{{Ichimaru}, {Ogata} \& {van Horn}}{{Ichimaru}
  et~al.}{1992}]{ichimaru1992}
{Ichimaru} S.,  {Ogata} S.,    {van Horn} H.~M.,  1992, ApJL, 401, L35

\bibitem[\protect\citeauthoryear{{Kawaler}}{{Kawaler}}{1988}]{kawaler1988}
{Kawaler} S.~D.,  1988, ApJ, 334, 220

\bibitem[\protect\citeauthoryear{{Kepler}, {Kleinman}, {Nitta}, {Koester},
  {Castanheira}, {Giovannini} \& {Althaus}}{{Kepler} et~al.}{2007}]{kepler2007}
{Kepler} S.~O.,  {Kleinman} S.~J.,  {Nitta} A.,  {Koester} D.,  {Castanheira}
  B.~G.,  {Giovannini} O.,    {Althaus} L.,  2007, in {Napiwotzki} R.,
  {Burleigh} M.~R.,  eds, 15th European Workshop on White Dwarfs Vol.~372 of
  Astronomical Society of the Pacific Conference Series, {The White Dwarf Mass
  Distribution}.
p.~35

\bibitem[\protect\citeauthoryear{{Kippenhahn} \& {Weigert}}{{Kippenhahn} \&
  {Weigert}}{1990}]{kippenhahn1990}
{Kippenhahn} R.,  {Weigert} A.,  1990, {Stellar Structure and Evolution}

\bibitem[\protect\citeauthoryear{{Kitamura}}{{Kitamura}}{2000}]{kitamura2000}
{Kitamura} H.,  2000, ApJ, 539, 888

\bibitem[\protect\citeauthoryear{{Kitamura} \& {Ichimaru}}{{Kitamura} \&
  {Ichimaru}}{1995}]{kitamura1995}
{Kitamura} H.,  {Ichimaru} S.,  1995, ApJ, 438, 300

\bibitem[\protect\citeauthoryear{{Lindemann}}{{Lindemann}}{1910}]{lindemann1910}
{Lindemann} F.~A.,  1910, Z. Phys., 11, 609

\bibitem[\protect\citeauthoryear{{Mannucci}, {Della Valle} \&
  {Panagia}}{{Mannucci} et~al.}{2006}]{Mannucci2006}
{Mannucci} F.,  {Della Valle} M.,    {Panagia} N.,  2006, MNRAS, 370, 773

\bibitem[\protect\citeauthoryear{{Marigo}}{{Marigo}}{2001}]{marigo2001}
{Marigo} P.,  2001, A\&A, 370, 194

\bibitem[\protect\citeauthoryear{{Mestel}}{{Mestel}}{1952a}]{mestel_1952_b}
{Mestel} L.,  1952a, MNRAS, 112, 583

\bibitem[\protect\citeauthoryear{{Mestel}}{{Mestel}}{1952b}]{mestel_1952_a}
{Mestel} L.,  1952b, MNRAS, 112, 598

\bibitem[\protect\citeauthoryear{{Miller Bertolami}, {Althaus} \&
  {Garc{\'{\i}}a-Berro}}{{Miller Bertolami} et~al.}{2013}]{millerbertolami2013}
{Miller Bertolami} M.~M.,  {Althaus} L.~G.,    {Garc{\'{\i}}a-Berro} E.,  2013,
  ApJL, 775, L22

\bibitem[\protect\citeauthoryear{{Napiwotzki}, {Christlieb}, {Drechsel},
  {Hagen}, {Heber}, {Homeier}, {Karl}, {Koester}, {Leibundgut}, {Marsh},
  {Moehler}, {Nelemans}, {Pauli}, {Reimers}, {Renzini} \&
  {Yungelson}}{{Napiwotzki} et~al.}{2003}]{Napi03}
{Napiwotzki} R.,  {Christlieb} N.,  {Drechsel} H.,  {Hagen} H.-J.,  {Heber} U.,
   {Homeier} D.,  {Karl} C.,  {Koester} D.,  {Leibundgut} B.,  {Marsh} T.~R.,
  {Moehler} S.,  {Nelemans} G.,  {Pauli} E.-M.,  {Reimers} D.,  {Renzini} A.,
   {Yungelson} L.,  2003, The Messenger, 112, 25

\bibitem[\protect\citeauthoryear{{Nomoto}}{{Nomoto}}{1982a}]{Nomoto1982b}
{Nomoto} K.,  1982a, ApJ, 257, 780

\bibitem[\protect\citeauthoryear{{Nomoto}}{{Nomoto}}{1982b}]{Nomoto1982a}
{Nomoto} K.,  1982b, ApJ, 253, 798

\bibitem[\protect\citeauthoryear{{Nomoto}, {Kobayashi} \& {Tominaga}}{{Nomoto}
  et~al.}{2013}]{Nomoto2013a}
{Nomoto} K.,  {Kobayashi} C.,    {Tominaga} N.,  2013, ARA\&A, 51, 457

\bibitem[\protect\citeauthoryear{{Ogata}, {Iyetomi} \& {Ichimaru}}{{Ogata}
  et~al.}{1991}]{ogata1991}
{Ogata} S.,  {Iyetomi} H.,    {Ichimaru} S.,  1991, ApJ, 372, 259

\bibitem[\protect\citeauthoryear{{Orio}}{{Orio}}{2013}]{Orio2013}
{Orio} M.,  2013, The Astronomical Review, 8, 010000

\bibitem[\protect\citeauthoryear{{Panei}, {Althaus}, {Chen} \& {Han}}{{Panei}
  et~al.}{2007}]{panei2007}
{Panei} J.~A.,  {Althaus} L.~G.,  {Chen} X.,    {Han} Z.,  2007, MNRAS, 382,
  779

\bibitem[\protect\citeauthoryear{{Renedo}, {Althaus}, {Miller-Bertolami},
  {Romero}, {C{\'o}rsico}, {Rohrmann} \& {Garc{\'{\i}}a-Berro}}{{Renedo}
  et~al.}{2010}]{renedo2010}
{Renedo} I.,  {Althaus} L.~G.,  {Miller-Bertolami} M.~M.,  {Romero} A.~D.,
  {C{\'o}rsico} A.~H.,  {Rohrmann} R.~D.,    {Garc{\'{\i}}a-Berro} E.,  2010,
  ApJ, 717, 183

\bibitem[\protect\citeauthoryear{{Salaris}, {Althaus} \&
  {Garc{\'{\i}}a-Berro}}{{Salaris} et~al.}{2013}]{salaris2013}
{Salaris} M.,  {Althaus} L.~G.,    {Garc{\'{\i}}a-Berro} E.,  2013, A\&A, 555,
  A96

\bibitem[\protect\citeauthoryear{{Salaris}, {Cassisi}, {Pietrinferni},
  {Kowalski} \& {Isern}}{{Salaris} et~al.}{2010}]{salaris2010}
{Salaris} M.,  {Cassisi} S.,  {Pietrinferni} A.,  {Kowalski} P.~M.,    {Isern}
  J.,  2010, ApJ, 716, 1241

\bibitem[\protect\citeauthoryear{{Salpeter} \& {van Horn}}{{Salpeter} \& {van
  Horn}}{1969}]{SalpeterVanHorn69}
{Salpeter} E.~E.,  {van Horn} H.~M.,  1969, ApJ, 155, 183

\bibitem[\protect\citeauthoryear{{Scalzo}, {Ruiter} \& {Sim}}{{Scalzo}
  et~al.}{2014}]{Scalzo2014}
{Scalzo} R.~A.,  {Ruiter} A.~J.,    {Sim} S.~A.,  2014, ArXiv e-prints,
  1408.6601

\bibitem[\protect\citeauthoryear{{Schramm} \& {Koonin}}{{Schramm} \&
  {Koonin}}{1990}]{schramm1990}
{Schramm} S.,  {Koonin} S.~E.,  1990, ApJ, 365, 296

\bibitem[\protect\citeauthoryear{{Shapiro} \& {Teukolsky}}{{Shapiro} \&
  {Teukolsky}}{1983}]{shapiro1983}
{Shapiro} S.~L.,  {Teukolsky} S.~A.,  1983, {Black holes, white dwarfs, and
  neutron stars: The physics of compact objects}

\bibitem[\protect\citeauthoryear{{Spergel}, {Verde}, {Peiris}, {Komatsu},
  {Nolta}, {Bennett}, {Halpern}, {Hinshaw}, {Jarosik}, {Kogut}, {Limon},
  {Meyer}, {Page}, {Tucker}, {Weiland}, {Wollack} \& {Wright}}{{Spergel}
  et~al.}{2003}]{spergel2003}
{Spergel} D.~N.,  {Verde} L.,  {Peiris} H.~V.,  {Komatsu} E.,  {Nolta} M.~R.,
  {Bennett} C.~L.,  {Halpern} M.,  {Hinshaw} G.,  {Jarosik} N.,  {Kogut} A.,
  {Limon} M.,  {Meyer} S.~S.,  {Page} L.,  {Tucker} G.~S.,  {Weiland} J.~L.,
  {Wollack} E.,    {Wright} E.~L.,  2003, ApJS, 148, 175

\bibitem[\protect\citeauthoryear{{Trimble} \& {Aschwanden}}{{Trimble} \&
  {Aschwanden}}{2004}]{Trimble04}
{Trimble} V.,  {Aschwanden} M.~J.,  2004, PASP, 116, 187

\bibitem[\protect\citeauthoryear{{Tsebrenko} \& {Soker}}{{Tsebrenko} \&
  {Soker}}{2014a}]{Tsebrenko2014a}
{Tsebrenko} D.,  {Soker} N.,  2014a, ArXiv e-prints, 1407.6231

\bibitem[\protect\citeauthoryear{{Tsebrenko} \& {Soker}}{{Tsebrenko} \&
  {Soker}}{2014b}]{Tsebrenko2014b}
{Tsebrenko} D.,  {Soker} N.,  2014b, ArXiv e-prints, 1409.0780

\bibitem[\protect\citeauthoryear{{van Horn}}{{van Horn}}{1968}]{VanHorn68}
{van Horn} H.~M.,  1968, ApJ, 151, 227

\bibitem[\protect\citeauthoryear{{Weidemann}}{{Weidemann}}{1967}]{weidemann1967}
{Weidemann} V.,  1967, Z. Astrophys., 67, 286

\bibitem[\protect\citeauthoryear{{Weidemann}}{{Weidemann}}{1977}]{weidemann1977}
{Weidemann} V.,  1977, A\&A, 59, 411

\bibitem[\protect\citeauthoryear{{Weidemann}}{{Weidemann}}{1990}]{weidemann1990}
{Weidemann} V.,  1990, ARA\&A, 28, 103

\bibitem[\protect\citeauthoryear{{Weidemann}}{{Weidemann}}{2000}]{weidemann2000}
{Weidemann} V.,  2000, A\&A, 363, 647

\bibitem[\protect\citeauthoryear{{Weiss}, {Hillebrandt}, {Thomas} \&
  {Ritter}}{{Weiss} et~al.}{2004}]{CoxGiuli2004}
{Weiss} A.,  {Hillebrandt} W.,  {Thomas} H.-C.,    {Ritter} H.,  2004, {Cox and
  Giuli's Principles of Stellar Structure}

\bibitem[\protect\citeauthoryear{{Yakovlev}, {Gasques}, {Afanasjev}, {Beard} \&
  {Wiescher}}{{Yakovlev} et~al.}{2006}]{yakovlev2006}
{Yakovlev} D.~G.,  {Gasques} L.~R.,  {Afanasjev} A.~V.,  {Beard} M.,
  {Wiescher} M.,  2006, PhRvC, 74, 035803

\end{thebibliography}

\end{document}